\documentclass[prx,twocolumn,unsortedaddress,superscriptaddress]{revtex4-2}
\usepackage[english]{babel}
\usepackage{graphicx}
\usepackage{epsfig,color}
\usepackage{amsmath,amssymb,eufrak,calc}
\usepackage{amsthm}
\usepackage{enumerate}
\usepackage{hyperref}
\usepackage{amsxtra,amsfonts,bm}
\usepackage{blkarray}
\usepackage{adjustbox}

\begin{document}

\title{Wilson Loops and Area Laws in Lattice Gauge Theory Tensor Networks}

\date{\today}

\author{Erez Zohar}
\address{Racah Institute of Physics, The Hebrew University of Jerusalem, Jerusalem 91904, Givat Ram, Israel.}

\begin{abstract}
Tensor network states have been a very prominent tool for the study of quantum many-body physics, thanks to their physically relevant entanglement properties and their ability to encode symmetries. In the last few years, the formalism has been extended and applied to theories with local symmetries too - lattice gauge theories. 
In order to extract physical properties (such as expectation values and correlation functions of physical observables)
out of such states, one has to use the so called transfer operators, whose local properties dictate the long-range behaviour of the state. In this work we study transfer operators of tensor network states (in particular, PEPS - projected entangled pair states)  of lattice gauge theories, and consider the implications of the local symmetry on their structure and properties.
In particular, we study the implications on the computation of the Wilson loop - a nonlocal, gauge-invariant observable which is central to pure gauge theories, whose long range decay behaviour probes the confinement or deconfinement of static charges. Using the symmetry, we show how to simplify the tensor contraction required for computing Wilson loop expectation values for such states, eliminate non-physical parts of the tensors and formulate conditions relating local properties (that is, of the tensors) to its decay fashion.
\end{abstract}

\maketitle

\section{Introduction}

In recent years, tensor network states \cite{orus_practical_2014} have been a very prominent tool, rooted in quantum information science, for the study of quantum many body systems and especially strongly correlated physics. In particular, Matrix Product States (MPS) \cite{white_density_1992,verstraete_matrix_2008} enable to study numerically and analytically physically relevant states, e.g. ground states of local many body Hamiltonians (that is, states exhibiting an entanglement entropy area law). In higher spatial dimensions, MPS generalize to PEPS - Projected Entangled Pair States \cite{verstraete_matrix_2008,cirac_matrix_2020}. These are useful for the description of strongly correlated physics in two spatial dimensions and more.

PEPS (and MPS) are constructed out of the contraction of local building blocks (tensors). They satisfy, by construction, the entanglement entropy area law (focusing on the physically relevant part of the Hilbert space) and allows the state to depend on very few local parameters and hence making it feasible for computations (compared with arbitrary states in the exponentially large Hilbert space). Furthermore, it also allows one to encode symmetries already on the single tensor level. By properly parametrizing the local tensors, a global symmetry of the whole PEPS under a symmetry group can be imposed \cite{sanz_matrix_2009,molnar_normal_2018}. This way, one can generate families of ansatz states in which the symmetry group of the studied model is encoded by construction.

While originally used mostly in the context of condensed matter physics, MPS and PEPS have recently been extended to the study of particle physics too - in particular, to lattice gauge theories, aiming at solving long standing open, non-perturbative questions of the standard model, such as the confinement of quarks \cite{wilson_confinement_1974}.
 Due to its running coupling \cite{gross_ultraviolet_1973} Quantum Chromodynamics (QCD)
 which allows one to use perturbation theory in high energy scales (collider physics) thanks to asymptotic freedom, 
 is strongly interacting in low energies, preventing the use of perturbative methods. Lattice Gauge Theories (LGTs) \cite{wilson_confinement_1974} have been introduced to overcome this difficulty, first as tools for lattice regularization of gauge invariant field theories. They quickly became a very successful numerical approach. Combined with quantum Monte Carlo it has been applied to nonperturbative QCD computations, such as the hadronic spectrum \cite{aoki_flag_2020}. However, quantum Monte Carlo does not allow for the direct observation of real time dynamics, and faces the fermionic sign problem in several important physical scenarios, not allowing one to probe some of the interesting exotic regions of the QCD phase diagram \cite{fukushima_phase_2011}, and this requires the use of other methods, with tensor networks being one such approach. The tensor network framework for lattice gauge theories has been rapidly growing in the last few years.

For $1+1d$ systems, MPS have already been extensively used. This does not only include abstract formalistic descriptions of MPS with a local symmetry \cite{silvi_lattice_2014,kull_classification_2017} or benchmarks of models that can be treated in other ways, such as, but not only, the Schwinger model \cite{banuls_mass_2013,buyens_matrix_2014,rico_tensor_2014,saito_temperature_2014,banuls_thermal_2015,banuls_chiral_2016,buyens_hamiltonian_2016,banuls_efficient_2017}. 
	 Successful numerical studies of lattice gauge theories in $1+1d$ have been carried out even in scenarios which face the sign problem when approached with conventional methods (such as real time evolution \cite{kuhn_non-abelian_2015,pichler_real-time_2016,buyens_real-time_2017} and finite density \cite{banuls_multi-flavor_2016,silvi_finite-density_2017,banuls_density_2017,banuls_mari_carmen_towards_2017}). This was done for both Abelian and non-Abelian models - see \cite{banuls_review_2020} and references therein for a discussion of that.
	  
The application of tensor networks to higher dimensional lattice gauge theories  has been discussed as well in the last few  years 
 \cite{tagliacozzo_entanglement_2011,tagliacozzo_tensor_2014,zohar_fermionic_2015,zohar_projected_2016,zohar_combining_2018,tschirsich_phase_2019,felser_two_2019,robaina_simulating_2020,emonts_variational_2020,magnifico_lattice_2020}.
 From the rather more abstract, or formalistic point of view, gauging mechanisms which lift globally invariant PEPS to locally invariant ones by adding a gauge field and entangling it to the matter properly were introduced and discussed \cite{haegeman_gauging_2015,zohar_building_2016}. 
   For a parallel approach in the action formalism - tensor field theory -  which uses tensor networks (but not tensor network states), see \cite{meurice_tensor_2020} and references therein.
   
   In this work, we will focus on a particular gauging mechanism - the one introduced in \cite{zohar_building_2016} and used mostly with fermionic matter, for creating gauged Gaussian fermionic PEPS \cite{zohar_fermionic_2015,zohar_projected_2016}: special PEPS constructions which allow for the description of fermionic matter coupled to dynamical gauge fields. Their construction may be seen as a minimal coupling procedure on the level of states, which is not possible in general but could be done in the context of PEPS \cite{emonts_gauss_2018}. While in general numerical computations are hard and challenging for PEPS in two spatial dimensions and more, it has been shown that, when this particular construction is used, the PEPS may be contracted efficiently (allowing one to extract physical information) when combining with Monte-Carlo methods which do not suffer from the sign problem \cite{zohar_combining_2018}. Variational Monte-Carlo then allows to find ground states of lattice gauge theory Hamiltonians when such states are used as ansatz states, which has already been demonstrated and benchmarked for a pure $\mathbb{Z}_3$ lattice gauge theory in $2+1$ dimensions \cite{emonts_variational_2020}.

A question that has to be asked when a PEPS is studied, is how physical information can be extracted from the contracted state - computation of expectation values of observables and correlation functions.
Thanks to the special structure of MPS, one may introduce a mathematical object called transfer matrix (or operator) \cite{fannes_finitely_1992} to compute efficiently expectation values of observables and correlation functions. This may be extended to two dimensional PEPS, by first contracting the rows, obtaining effectively a chain of the rows which is an MPS, and considering its transfer matrix \cite{yang_chiral_2015}. In this work, we will study such transfer operators of lattice gauge theory PEPS in two space dimensions.

Gauge theories are special in the sense that they exhibit a \emph{local symmetry}, responsible to mediating local interactions between the matter fields. This symmetry gives rise to many local constraints. All the physical states are invariant under \emph{gauge transformations} - local transformations parametrized by the elements of the so-called \emph{gauge group}. As a result, only gauge invariant observables and correlation functions - those which are invariant under local transformations -  give rise to a nonvanishing expectation values. Thus, LGT PEPS admit a very special structure manifested in the local tensors \cite{tagliacozzo_tensor_2014,haegeman_gauging_2015,zohar_building_2016}; in this work, we focus on the implications of the local symmetry on the transfer operators, and hence aim at using the symmetry to simplify the PEPS contraction, focusing on pure gauge theories (that is, without dynamical matter). 

In such scenarios, closed flux loops - usually referred to as the operators which create them, Wilson loops \cite{wilson_confinement_1974} - are perhaps the most important observables (and almost the only possible gauge invariant one). The decay rule of large Wilson loops in pure gauge theories serves as probes for confinement of static charges:  area law decay implies confinement, while a perimeter law - a deconfined (Coulomb) phase. Confinement implies a gapped, disordered phase, while deconfined phases are massless and ordered \cite{fradkin_order_1978}. The local ingredients of the Wilson loop are not gauge invariant - only their combination along the nonlocal path preserves the symmetry. This means, that when computing it for a gauge invariant PEPS, the transfer operator formalism must be extended and modified, requiring the inclusion of various types of transfer matrices which construct this nonlocal observable. The different building blocks will also have special properties \cite{zohar_projected_2016}, dictated by the special local symmetry, which will affect the behaviour of the Wilson loop and its decay. 

In this work, we will study the properties of transfer operators of gauge invariant PEPS. We will see how the symmetry affects the properties of the local tensors, and that thanks to it, some parts of the tensors may be excluded and ignored when a contraction is done (e.g. when combined with some numerical methods). We will also see
how that affects the  Wilson loop's decay - that is, how local properties of the tensors dictate the decay of large Wilson loops.

Note that PEPS have been previously used for the computation of Wilson loop expectation value in various cases - 
$\mathbb{Z}_2$ string nets  \cite{schotte_tensor_2019,ritz_wegner_2020},  as well as $U(1)$ \cite{zohar_fermionic_2015} and $SU(2)$ \cite{zohar_projected_2016} toy models; here we derive a general framework based on transfer matrix arguments and demonstrate with particular constructions

We begin with briefly reviewing important preliminaries from group theory and lattice gauge theory, in section II; move on to formulating gauge invariant PEPS and reviewing their symmetry properties, in section III; in section IV we 
introduce the transfer operators - after a brief review of their general properties, we formulate the flux-free transfer operators for LGT PEPS, study their properties and use them to calculate the norm; section V focuses on the contraction of Wilson loop expectation values for LGT PEPS, studying the relevant transfer operators and deriving conditions for area and perimeter decay laws; finally, in section VI, we give an explicit illustration, including both analytical and numerical arguments, for a $\mathbb{Z}_2$ lattice gauge theory. 

Throughout this work the Einstein summation convention (on doubly repeated indices) is assumed unless stated otherwise; with the only exception of irreducible representation indices, whose summation should not be assumed.

\section{Mathematical and Physical Preliminaries}

\subsection{Groups, transformations, representations}

Consider a group $G$, which is either a finite or a compact Lie one. Each group element $g\in G$ may be represented by different unitary matrices $D^j\left(g\right)$, labelled by the group's irreducible representations (irreps) $j$; the dimension of these matrices is referred to as the irrep dimension, $\text{dim}\left(j\right)$ (e.g. $2j+1$ for $SU(2)$).

In the Hilbert space $\mathcal{H}$ of some quantum mechanical theory, we would like to consider transformations parametrized by the elements of $G$. To do that, for each $g \in G$ we introduce a unitary operator $\theta_g$, and define it by its action on a basis states of the form $\left|jm\right\rangle$. $j$ labels the irreducible representations of $G$ and $m$ is an index labelling all states within this representation - that is, all the states that may be mixed by the transformations $\theta_g$ that act block-diagonally on the irreps:
	\begin{equation}
	\theta_g \left|jm\right\rangle =  \left|jm'\right\rangle D^{j}_{m'm}\left(g\right)
	\label{rt}
\end{equation}
We can hence express $\theta_g$ as
\begin{equation}
	\theta_g = \underset{j}{\bigoplus}D^j_{mn}\left(g\right) \left|jm \right\rangle \left\langle jn \right|
\end{equation}
- and therefore the dimension of the irrep $j$, $\text{dim}\left(j\right)$, is also the dimension of $\mathcal{H}_j$, the Hilbert subspace spanned by $\left|jm\right\rangle$, which we call a \emph{multiplet}. The Hilbert space may be seen as a direct sum of multiplet subspaces
\begin{equation}
	\mathcal{H} = \underset{j}{\bigoplus}\mathcal{H}_j
	\label{Hjj}
\end{equation}
In general, quantum Hilbert spaces may contain more than one multiplet carrying the same irreducible representation.

These transformations are sometimes referred to as \emph{right transformations}, since they mix the multiplet elements $\left|jm\right\rangle$, when seen as the components of a $\text{dim}\left(j\right)$ dimensional vector, via right matrix multiplication, as shown in (\ref{rt}). One can also introduce the \emph{left transformations}
\begin{equation}
	\tilde{\theta}_g \left|jm\right\rangle = D^{j}_{mm'}\left(g\right) \left|jm'\right\rangle 
	\label{lt}
\end{equation}
Note that the left transformations are not independent from the right ones: for each $g \in G$ one may find $h$ such that $\tilde{\theta}_g = \theta_h$. We introduce the left transformations separately nevertheless since they will be mathematically convenient later when the PEPS are constructed.

When $G$ is a compact Lie group, its elements may be uniquely identified in terms of group parameters or coordinates $\phi_a$; then, for each irrep $j$,
\begin{equation}
	D^{j}\left(g\right) = \text{exp}\left(i \phi_a\left(g\right) T^j_a\right)
\end{equation}
- the parameters $\phi_a\left(g\right)$ depend on the group element, while the generators $T^j_a$ depend on the representation. The latter form a set of matrices with dimension   $\text{dim}\left(j\right)$, satisfying the group's Lie algebra
\begin{equation}
	\left[T^j_a,T^j_b\right]=if_{abc}T^j_c
\end{equation}
where $f_{abc}$ are the group's structure constants. One may also introduce the abstract generators, $J_a$, which are block diagonal in the representations,
\begin{equation}
	J_a = \underset{j}{\bigoplus}\left(T_a^j\right)_{mn} \left|jm\right\rangle\left\langle jn\right|,
\end{equation}
satisfying the algebra
\begin{equation}
	\left[J_a,J_b\right]=if_{abc}J_c
\end{equation}
too.

The states $\left|jm\right\rangle$ are eigenstates of mutually commuting operators: the $j$ quantum numbers(s) labelling the irreducible representation (and hence the multiplet) are eigenvalues of the Casimir operators which commute with all the generators; within the representation, the states are labelled by the eigenvalues of a maximal set of mutually commuting generators (Cartan subalgebra) - $m$. Similarly, when the group is finite, $j$ labels the irreducible representation while the $m$ numbers are obtained from the simultaneous diagonalization of a maximal set of commuting transformations.

All the irreps of Abelian groups are one dimensional and thus no $m$ indices are required. In the $\mathbb{Z}_N$ case, the $N$ different irreps are labelled by the integers $j=0,...,N-1$, which label the group elements $g=0,...,N-1$ too, with $D^j\left(g\right) = \text{exp}\left(i \frac{2\pi}{N}jg\right)$. In the $U(1)$ case the group elements are labelled by one parameter as well, $\phi \in \left[0,2\pi\right)$, the representations are labelled by integers $j \in \mathbb{Z}$, and $D^j\left(\phi\right) = \text{exp}\left(i j \phi\right)$, and $T^j = j\mathbf{1}$. 

 As a non-Abelian example, consider $SU(2)$, whose irreps are labelled by $j$ that are non-negative integers and half-integers. The dimension of each representation is $\text{dim}\left(j\right) = 2j+1$, and the $2j+1$ within the multiplet are labelled by $m=-j,...,j$. There are three generators, satisfying the Lie algebra with $f_{abc}=\epsilon_{abc}$ - the anti-symmetric (Levi-Civita) symbol with $a,b,c=1,2,3$. The generators in this case are sometimes called the spin or angular momentum components, and then $a,b,c=x,y,z$. The $j=0$ (trivial) representation is one dimensional, with the singlet state $\left|00\right\rangle$. The next representation, $j=1/2$, is two dimensional ($m=\pm 1/2$ ), with generators proportional to the Pauli matrices, $T^{j=1/2}_a = \sigma_a/2$. In this case, there is a single Casimir operator, $\mathbf{J}^2 = J_aJ_a$, commuting with one generator at most (the Cartan subalgebra is of size one). Conventionally it is taken to be the $z$ or $3$ component of the angular momentum, and thus for $SU(2)$,
 \begin{equation}
 	\begin{aligned}
\mathbf{J}^2 \left|jm\right\rangle &= j\left(j+1\right)\left|jm\right\rangle \\
J_z \left|jm\right\rangle &= m\left|jm\right\rangle
 	\end{aligned}
  \end{equation}

\subsection{Lattice Gauge Theory Basics}

Just like  gauge theories in the continuum, LGTs 
describe the interaction of matter particles through gauge fields.
 In the lattice case, the matter fields reside on the lattice sites, while the gauge fields, mediating the interactions between matter particles, are on the links. One can either discretize both space and time \cite{wilson_confinement_1974}, as used for Euclidean, Monte-Carlo computations, or discretize only space while keeping time continuous \cite{kogut_hamiltonian_1975}. The latter corresponds to the Hamiltonian formulation widely used in the context of quantum simulation and tensor networks, including in this work. Since we consider Hamiltonian lattice gauge theory in $2+1$ dimensions, our lattice will be two dimensional. As this work focuses on the pure gauge case and matter fields are absent, all the degrees of freedom will reside on the links. We will review the basic ingredients of such models following the conventions of \cite{zohar_formulation_2015} and \cite{zohar_quantum_2016}.

\subsubsection{Local Hilbert Spaces}
Consider a two dimensional lattice, whose sites are labelled by vectors of integers $\mathbf{x} \in \mathbb{Z}^2$. $\hat{\mathbf{e}}_i$ denote the unit vectors pointing in directions $i=1,2$, and any link is classified by two numbers, $\left(\mathbf{x},i\right)$, standing for the beginning of the link and the direction to which it emanates, respectively.
Each link $\left(\mathbf{x},i\right)$ hosts a local gauge field Hilbert space $\mathcal{H}_{\text{gauge}}\left(\mathbf{x},i\right)$, which can be spanned by \emph{group element states} $\left\{\left|g\right\rangle\right\}_{g\in G}$ labelled by the gauge group elements.
 These states form a basis of  $\mathcal{H}_{\text{gauge}}$, with the orthogonality relation
\begin{equation}
	\left\langle g' | g \right\rangle = \delta\left(g',g\right),
\end{equation}
where $\delta\left(g',g\right)$ is the Kronecker delta if $G$ is finite, and a Dirac delta distribution in the compact lie case -  denoting the Haar measure of $G$ by $dg$, 
\begin{equation}
	 \int dg' f\left(g'\right) \delta\left(g',g\right) = f\left(g\right).
\end{equation}

Unlike in the multiplet case, here the right and left transformations are independent of one another, as group multiplications: we introduce two sets of unitary operators, $\Theta_g$ and $\tilde{\Theta}_g$, parametrized by the elements of the gauge group $G$, which implement right and left group multiplications (respectively) on the group element states,
\begin{equation}
\begin{aligned}
	&\Theta_g \left|h\right\rangle =  \left|hg^{-1}\right\rangle \\
	&\tilde{\Theta}_g \left|h\right\rangle =  \left|g^{-1}h\right\rangle
\end{aligned}
\end{equation} 

The space $\mathcal{H}_{\text{gauge}}$ can also be spanned by the dual  \emph{representation basis}, whose states are labelled by $\left|jmn\right\rangle$ - $j$ is an irrep and $m,n$ are identifiers within it. In a sense, using the multiplet states introduced previously,
\begin{equation}
	\left|jmn\right\rangle = \left|jm\right\rangle \otimes \left|jn\right\rangle
	\end{equation}
or,
\begin{equation}
\mathcal{H}_{\text{gauge}} = \underset{j}{\bigoplus}\mathcal{H}_j \otimes \mathcal{H}_j
\label{Hgauge}
	\end{equation}
where $\mathcal{H}_j$ is the $\text{dim}\left(j\right)$ dimensional subspace spanned by the $\left|jm\right\rangle$ multiplet states. We read this equation as a decomposition of the link's Hilbert space into a direct sum of products of multiplets of the groups on the left and right of the link, sharing the same irrep. Here one copy of each irreducible representation is used at most; one in the full, Kogut-Susskind case \cite{kogut_hamiltonian_1975}, but it is also possible to choose (for example, for reasons of feasibility of computation or experimental implementation) to truncate the sum and not include all the irreps in several ways \cite{brower_qcd_1999,zohar_formulation_2015} as we will discuss later.

In the non truncated case, using the Peter-Weyl theorem and the group's Fourier transform \cite{zohar_formulation_2015}, the transition between the two bases is given by
\begin{equation}
	\left\langle g | jmn \right\rangle = \sqrt{\frac{\text{dim}\left(j\right)}{|G|}} D^{j}_{mn}\left(g\right)
	\label{FourierT}
\end{equation}
where $|G|$ is the group's volume. In the representation basis,
\begin{equation}
	\begin{aligned}
		&\Theta_g \left|jmn\right\rangle =  \left|jmn'\right\rangle D^j_{n'n}\left(g\right) \\
		&\tilde{\Theta}_g \left|jmn\right\rangle =  D^j_{mm'}\left(g\right)\left|jm'n\right\rangle	.
		\label{Theta}
	\end{aligned}
\end{equation} 

In the compact Lie group case, one can introduce two sets of transformation generators, left and right - $L_a$ and $R_a$ respectively, such that
\begin{equation}
	\begin{aligned}
	\Theta_g = \text{exp}\left(i \phi_a\left(g\right) R_a\right) \\
	\tilde{\Theta}_g = \text{exp}\left(i \phi_a\left(g\right) L_a\right)	
	\end{aligned}
\end{equation}
satisfying the algebra
\begin{equation}
	\begin{aligned}
		\left[R_a,R_b\right]&=if_{abc}R_c \\
		\left[L_a,L_b\right]&=-if_{abc}L_c \\
		\left[R_a,L_b\right]&=0
	\end{aligned}
\end{equation}

Note that if the group is Abelian, there is no difference between left and right operations and the indices $m,n$ do not exist. Therefore, there $R=L\equiv E$. Thus, in the $U(1)$ case, for example, we have group states labelled by the single compact parameter $\left|\phi\right\rangle$ and representation states labelled by the single integer $\left|j\right\rangle$, related through the Fourier series formula
\begin{equation}
	\left\langle \phi | j \right\rangle = \frac{1}{\sqrt{2\pi}}e^{ij\phi}
\end{equation}
and the representation states $\left|j\right\rangle$ satisfy
\begin{equation}
	E\left|j\right\rangle=j\left|j\right\rangle
\end{equation}
For $\mathbb{Z}_N$, similarly, we obtain the discrete Fourier series formula
\begin{equation}
	\left\langle g | j \right\rangle = \frac{1}{\sqrt{N}}e^{i 2\pi j g / N}
\end{equation}

While in the $SU(2)$ case, since the group is non Abelian, the situation is more complicated. There are $(2j+1)^2$ $\left|jmn\right\rangle$ states for each $j$ - e.g. one singlet state $\left|000\right\rangle$ for $j=0$, and four $j=1/2$ states, $\left|\frac{1}{2},\pm\frac{1}{2},\pm\frac{1}{2}\right\rangle$. The group is parametrized by the three Euler angles $\alpha,\beta,\gamma$, and
\begin{equation}
	\left\langle \alpha,\beta,\gamma | jmn \right\rangle = \frac{\sqrt{2j+1}}{8\pi^2} D^{j}_{mn}\left(\alpha,\beta,\gamma\right)
\end{equation}
The Hilbert space in this case is that of a rigid rotator \cite{kogut_hamiltonian_1975,kasper_from_2020}. The right and left operators $R_a$ and $L_a$ correspond to the generators of its rotations in the space and body frames of reference. These two sets of generators commute, and give rise to the same total angular momentum (eigenvalue of the Casimir operator) since it is a rotation scalar quantity which does not depend on the frame of reference \cite{landau_quantum_1981,weinberg_lectures_2015}. Therefore,
\begin{equation}
	\begin{aligned}
	\mathbf{J}^2\left|jmn\right\rangle \equiv \mathbf{R}^2\left|jmn\right\rangle &= \mathbf{L}^2\left|jmn\right\rangle =j\left(j+1\right)\left|jmn\right\rangle \\
	L_z\left|jmn\right\rangle &= m\left|jmn\right\rangle \\
	R_z\left|jmn\right\rangle &= n\left|jmn\right\rangle
	\end{aligned}
\end{equation}

\subsubsection{Local Gauge Invariance}

At each site $\mathbf{x}$, and for each group element $g\in G$, we introduce the \emph{gauge transformation}
\begin{equation}
	\hat{\Theta}_g\left(\mathbf{x}\right) = 
	\tilde{\Theta}_g\left(\mathbf{x},1\right)
	\tilde{\Theta}_g\left(\mathbf{x},2\right)
	\Theta^{\dagger}_g\left(\mathbf{x}-\hat{\mathbf{e}}_1,1\right)
	\Theta^{\dagger}_g\left(\mathbf{x}-\hat{\mathbf{e}}_1,2\right)
	\label{ggg}
\end{equation}
which transforms all the four links intersecting at $\mathbf{x}$ with respect to the same group element - the outgoing links with the left transformation, and the ingoing ones with the inverse right one. The outgoing links, whose beginning (left) side connects to $\mathbf{x}$, undergo a left rotation, while the ingoing ones, connected through their end (right) side to $\mathbf{x}$, undergo an inverse right rotation.

\begin{figure}
	\includegraphics[width=\columnwidth]{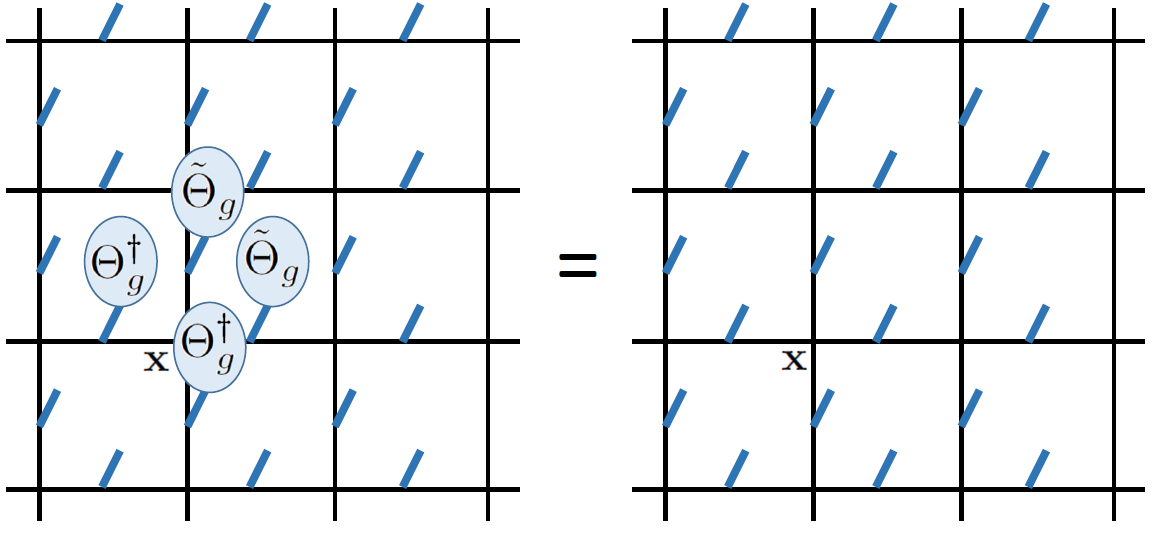}
	\caption{Gauge transformations act on the four links around a vertex with a particular set of unitary transformations parametrized by the same group element $g\in G$.}
	\label{Gauge_Inv}
\end{figure}

A gauge invariant state $\left|\psi\right\rangle$ satisfies
\begin{equation}
		\hat{\Theta}_g\left(\mathbf{x}\right) \left|\psi\right\rangle = \left|\psi\right\rangle
		\quad\quad\forall\mathbf{x}\in\mathbb{Z}^2,g\in G
		\label{gtrans}
\end{equation}
(see Fig. \ref{Gauge_Inv}) and similarly, for a gauge invariant operator $O$,
\begin{equation}
	\hat{\Theta}_g\left(\mathbf{x}\right) O \hat{\Theta}^{\dagger}_g\left(\mathbf{x}\right) = O
	\quad\quad\forall\mathbf{x}\in\mathbb{Z}^2,g\in G
	\label{ginv}
\end{equation}
(one can extend it to the case of static charges \cite{kasper_from_2020} which we do not discuss here).
In a lattice gauge theory, only gauge invariant states and operators are considered physical. 

If $G$ is a compact Lie group, we can formulate the gauge transformations $\hat{\Theta}_g\left(\mathbf{x}\right) $ in terms of their generators,
\begin{equation}
	\mathcal{G}_a\left(\mathbf{x}\right) = L_a\left(\mathbf{x},1\right)+
	L_a\left(\mathbf{x},2\right)-
	R_a\left(\mathbf{x}-\hat{\mathbf{e}}_1,1\right)-
	R_a\left(\mathbf{x}-\hat{\mathbf{e}}_2,2\right)
\end{equation} 
Gauge invariance is then formulated in terms of the \emph{Gauss laws}
\begin{equation}
	\mathcal{G}_a\left(\mathbf{x}\right) \left|\psi\right\rangle = 0
	\quad\quad\forall\mathbf{x}\in\mathbb{Z}^2,a
\end{equation}
(once again, excluding static charges \cite{kasper_from_2020}).

We call this eigenvalue equation the \emph{Gauss law}, since $\mathcal{G}_a\left(\mathbf{x}\right)$ can clearly be seen as the divergence of electric fields - $L_a$ and $R_a$ - on a site. For physical states - the one which satisfy the local constraints (\ref{gtrans}) - the divergence of electric fields is zero. It is very apparent in the $U(1)$ case, where it takes the explicit form
\begin{equation}
	\begin{aligned}
&\left(E\left(\mathbf{x},1\right)+
E\left(\mathbf{x},2\right)-
E\left(\mathbf{x}-\hat{\mathbf{e}}_1,1\right)-
E\left(\mathbf{x}-\hat{\mathbf{e}}_2,2\right)\right)\left|\psi\right\rangle \\
&\equiv\nabla \cdot \mathbf{E}\left(\mathbf{x}\right) \left|\psi\right\rangle =0
\quad\quad\forall\mathbf{x}\in\mathbb{Z}^2,a
\end{aligned}
\end{equation}
In non Abelian cases, the divergence involves left and right electric fields, which is related to the charge carried by non-Abelian gauge bosons \cite{kogut_hamiltonian_1975} (e.g., the colour charged gluon vs the electric neutral photon).

\subsubsection{Wilson Loops}

Since we deal with gauge invariant states, it is expected that the expectation values of non gauge invariant operators will vanish. Thus, when classifying the phases and behaviour of gauge theories one needs to consider only gauge invariant observables and correlation functions.

One option, for compact Lie group, is to compute expectation values of electric field operators and functions thereof (and only of Casimir operators if the group is non-Abelian). Another possible gauge invariant observable is the loop variable, and in particular Wilson Loops \cite{wilson_confinement_1974}.

On the local link Hilbert spaces we introduce the \emph{group element operators}:
\begin{equation}
	U^{j}_{mn} = \int dg D^j_{mn}\left(g\right) \left|g\right\rangle\left\langle g\right|
	\label{Udef}
\end{equation}
$U^j$ is a matrix of dimension $\text{dim}\left(j\right)\times\text{dim}\left(j\right)$, whose elements are operators acting on the link's gauge field Hilbert space, $\mathcal{H}_{\text{gauge}}$ (on each link $\ell$ we can define such operators $U^j_{mn}\left(\ell\right)$). Even though they are Hilbert space operators, all the elements of $U^j$ commute - one can see in the definition above that they are all diagonal in the same basis. The matrix elements of $U^j$ mix with respect to the transformation properties of the $j$ representation,
\begin{equation}
	\begin{aligned}
		\Theta_g U^j_{mn} \Theta^{\dagger}_g &= U^j_{mn'} D^{j}_{n'n}\left(g\right) \\
		\tilde{\Theta}_g U^j_{mn} \tilde{\Theta}^{\dagger}_g &= D^{j}_{mm'}\left(g\right) U^j_{m'n}
\end{aligned}
\label{Utrans}
\end{equation}	
and, in the compact Lie case,
\begin{equation}
	\begin{aligned}
		\left[R_a,U^j_{mn}\right]&=  U^j_{mn'} \left(T^{j}_a\right)_{n'n} \\
		\left[L_a,U^j_{mn}\right] &= \left(T^{j}_a\right)_{mm'} U^j_{m'n}
		\label{RU}
	\end{aligned}
	\end{equation}

\begin{figure}
	\includegraphics[width=0.7\columnwidth]{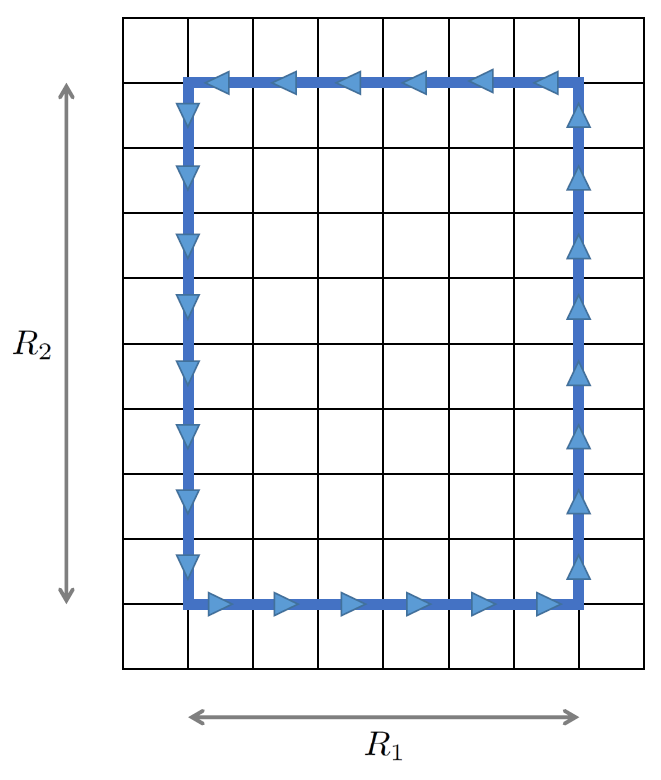}
	\caption{A Wilson loop: rectangular loop of electric flux. On links in the positive directions (pointing rightwards and upwards, here on the lower and right edges of the loop) the group element operator $U$ is used; on links in the negative directions (pointing leftwards and downwards, here on the left and upper edges of the loop), $U^{\dagger}$ is used.}
	\label{Wilson_Loop}
\end{figure}

Let us take some closed path $\mathcal{C}$ on the lattice. We define the Wilson loop operator $\mathcal{W}\left(\mathcal{C}\right)$ as the ordered contraction of group element operators along this closed path, that is
\begin{equation}
\begin{aligned}
\mathcal{W}^j\left(\mathcal{C}\right) &= \text{Tr}\left(\underset{\ell\in\mathcal{C}}{\prod}U^j\left(\ell\right)\right) =\\
&=U^j\left({\ell_1}\right)_{m_1 m_2} U^j\left({\ell_2 }\right)_{m_2 m_3} \cdots U^j\left({\ell_L }\right)_{m_L m_1}
\end{aligned}
\label{Wdef}
\end{equation}
It is a trace over the product of the group element operators $U$, seen as matrices, ordered along the closed path $\mathcal{C}$ with length $L$ (which is simply the number of links along the path). Depending on the orientation of the path, one may have to use $U^{\dagger}$ instead of $U$, on half of the links along the path - those pointing leftwards or downwards (see Fig. \ref{Wilson_Loop}). For simplicity, we will omit the $j$ indices below, but obviously the same irrep must be used along the path, otherwise the matrix product is ill defined. Consider $U(1)$ with $j=1$ as an example; there,
\begin{equation}
	\mathcal{W}\left(\mathcal{C}\right) = \text{exp}\left(i\underset{\ell \in \mathcal{C}}{\sum}\phi\left(\ell\right)\right)
\end{equation}
with half of the phases with a minus sign, according to their orientation.

In order to consider the action of the group element operators on representation states, we use the Clebsch-Gordan series and coefficients $\left\langle J M j m | K N \right\rangle$ \cite{rose_elementary_1995} and obtain 
\begin{equation}
	\begin{aligned}
	U^{j}_{mm'}\left|JMM'\right\rangle &=\\ \sqrt{\frac{\text{dim}\left(J\right)}{\text{dim}\left(K\right)}}
	&\left\langle J M j m | K N \right\rangle \left\langle K N' | J M' j m' \right\rangle \left| K N N' \right\rangle
	\end{aligned}
\label{Urep}
\end{equation}
- that is, the action of the group element operator $U^j$ on a state with representation $J$ yields states with all representations which are obtained by combining $j$ and $J$ (more precisely, fusing the two irreps together). Acting with a loop operator hence excites the representations along the loop with respect to that rule. One may truncate the Hilbert space in the representation basis: as long as all the irreducible representations used are taken completely and connected by nonzero Clebsch Gordan coefficients when $j$ is added, one may use (\ref{Urep}) to define a $U^j$ operator acting on that truncated space. The transformation properties (\ref{Theta}), (\ref{Utrans}) and (\ref{RU}) will still hold \cite{zohar_formulation_2015}, which may make it convenient for some numerical approaches (or quantum simulation implementations \cite{zohar_quantum_2021})
but, since the group structure will be lost, the group element basis will no longer be defined, making, in particular, (\ref{Udef}) and the Fourier transform (\ref{FourierT}) invalid.

In most cases, rectangular Wilson loops are considered. We denote by $W\left(R_1,R_2\right)$ a rectangular loop sized $R_1 \times R_2$ (see Fig. \ref{Wilson_Loop}). Very large Wilson loops of pure gauge theories are a probe for confinement (or deconfinement) of static charges, as introduced by Wilson in \cite{wilson_confinement_1974} (see also \cite{fradkin_order_1978,kogut_introduction_1979,polyakov_gauge_1987}). In a confining phase, 
\begin{equation}
-\log\left\langle W\left(R_1,R_2\right) \right\rangle \propto R_1 R_2
\end{equation}
for $R_1,R_2 \gg 1$ (area law), while in a deconfined phase
\begin{equation}
-\log\left\langle W\left(R_1,R_2\right) \right\rangle \propto R_1 + R_2
\end{equation}
for $R_1,R_2 \gg 1$ (perimeter law).

In \cite{creutz_asymptotic_1980}, Creutz introduced the  parameter
\begin{equation}
\chi\left(R_1,R_2\right) = -\log \left(
\frac{W\left(R_1,R_2\right)W\left(R_1-1,R_2-1\right)}{W\left(R_1-1,R_2\right)W\left(R_1,R_2-1\right)}
\right)
\label{Creutz}
\end{equation}
for the detection of static charge confinement. In the general case of
\begin{equation}
\left\langle W\left(R_1,R_2\right)\right\rangle = W_0 e^{-\kappa_A R_1R_2 -\kappa_P\left(R_1+R_2\right)}
\label{Wlarge}
\end{equation}
For large $R_1,R_2$, the area factor $\kappa_A$ (called the \emph{string tension}), should it exist, is the most dominant one. The Creutz parameter $\chi$ filters out the contributions of the constant prefactor $W_0$ and the perimeter coefficient $\kappa_P$, and thus within a confining phase, $\chi\left(R_1,R_2\right) \rightarrow \kappa_A > 0$ for $R_1,R_2 \gg 1$, while in a deconfining one it converges to zero.

\section{Gauge Invariant PEPS}

In this work, we use the lattice gauge theory PEPS formalism of \cite{zohar_building_2016,zohar_combining_2018}, with slightly different notations (and restricted to the pure gauge case). First of all, let us review it.

\subsection{Review of the PEPS construction}
Each site $\mathbf{x} \in \mathbb{Z}^2$ of our square, periodic lattice is at the intersection of four legs. The outgoing ones are in the right and up directions, while the left and down directed legs are considered ingoing. 
We wish, as usual with PEPS, to construct a physical lattice state describing different physical degrees of freedom located on different sites. Each such degree of freedom is described by a local physical Hilbert space: if we had matter, we would fix a physical matter Hilbert space to each lattice site. Here, however, the gauge fields are our only physical degrees of freedom, and they reside on the links. Thus, with each lattice site $\mathbf{x}$ we associate two  physical Hilbert spaces, located on the outgoing legs. We refer to them as the side ($\mathcal{H}_s$) and top ($\mathcal{H}_t$) \emph{physical} Hilbert spaces.

These are local gauge field Hilbert spaces (note that the word local here has to do with being defined on a single link, not with the gauge symmetry being local) - that is, either the full $\mathcal{H}_{\text{gauge}}$  spaces introduced in Eq.  (\ref{Hgauge}), or truncated versions thereof containing only some representations. When truncating, it is important to make sure that all the $\left|jmn\right\rangle$ for an included $j$ are present, otherwise no gauge invariance can be imposed, as explained above \cite{zohar_formulation_2015,zohar_building_2016}.

When constructing a PEPS, in order to connect the local physical building blocks to one physical quantum state, one has to introduce \emph{auxiliary} or \emph{virtual} degrees of freedom, on top of the physical ones given by the model we study. These are
 used merely for the purpose of contraction. 
On each of the four legs we introduce an auxiliary or \emph{virtual Hilbert space}, $\mathcal{H}_r,\mathcal{H}_u,\mathcal{H}_l,\mathcal{H}_d$ for the right,  up, left and down going legs, respectively. They are spanned by group multiplet states of the form $\left|jm\right\rangle$, as defined in Eq. (\ref{Hjj}). One may include all such multiplets, truncate, or include several copies of the same multiplet, which allows to increase the number of variational parameters; but once again, all the states within a multiplet included must be present, and the representations used in the physical spaces must be included (though possibly with a higher multiplicity). For more details about that, refer to \cite{zohar_building_2016} where the general construction of such states is discussed.

\begin{figure}
	\includegraphics[width=\columnwidth]{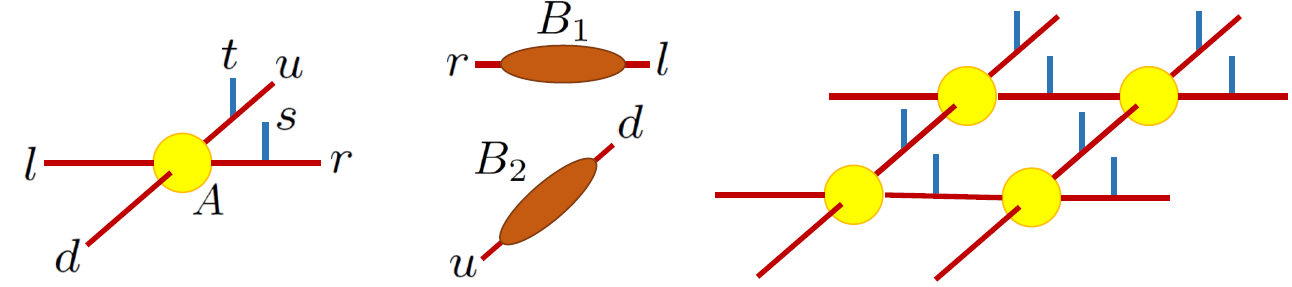}
	\caption{The building blocks of the PEPS: on the left, the site tensors $A$ (\ref{Aket}), with the physical legs $s,t$ and the virtual ones $r,u,l,d$. In the middle, the link projectors, $B_{1,2}$ (\ref{Bket}), connecting the outgoing legs $r,u$ with the ingoing legs $l,d$ of the next sites, to the right and above respectively. On the right, the contracted PEPS $\left|\psi\right\rangle$ (\ref{psi}), obtained after the projection.}
	\label{PEPS_Blocks}
\end{figure}

On each site, we construct the physical-virtual state
\begin{widetext}
\begin{equation}
\left|A\right\rangle = A^{j_s m_s n_s; j_t m_t n_t}_{j_r m_r; j_u m_u;j_l m_l;j_d m_d}\left|j_s m_s n_s; j_t m_t n_t\right\rangle \left|j_r m_r; j_u m_u;j_l m_l;j_d m_d\right\rangle \in \mathcal{H}_s \times \mathcal{H}_t \times \mathcal{H}_r \times \mathcal{H}_u \times \mathcal{H}_l \times \mathcal{H}_d
\label{Aket}
\end{equation}
\end{widetext}
where the first ket refers to the physical states and the second to the virtual ones (see Fig. \ref{PEPS_Blocks}). The coordinate $\mathbf{x}$ was omitted for simplicity, but the Hilbert spaces are all associated with particular sites and, in general, the tensors 
$ A^{j_s m_s n_s; j_t m_t n_t}_{j_r m_r; j_u m_u;j_l m_l;j_d m_d}$ may depend on the position, although we will focus on translationally invariant PEPS and thus they will be independent of $\mathbf{x}$.

To contract the PEPS, on each link we introduce the maximally entangled states
\begin{equation}
	\begin{aligned}
	\left|B_1\left(\mathbf{x}\right)\right\rangle &= \underset{j}{\sum}
	\left|jm\right\rangle_{r,\mathbf{x}} \left|jm\right\rangle_{l,\mathbf{x}+\hat{\mathbf{e}}_1} \\
	\left|B_2\left(\mathbf{x}\right)\right\rangle &= \underset{j}{\sum}
	\left|jm\right\rangle_{u,\mathbf{x}} \left|jm\right\rangle_{d,\mathbf{x}+\hat{\mathbf{e}}_2} 
	\label{Bket}
	\end{aligned}
\end{equation}
As usual, we construct our PEPS $\left|\psi\right\rangle$ by projecting the virtual states on the legs onto the maximally entangled states, 
\begin{equation}
	\left|\psi\right\rangle = \underset{\mathbf{x},i}{\bigotimes} \left\langle B_i \left(\mathbf{x}\right) \right| 
	\underset{\mathbf{x}}{\bigotimes} \left|A\left(\mathbf{x}\right)\right\rangle
	\label{psi}
\end{equation}
Note that $\underset{\mathbf{x}}{\bigotimes}\left|A\left(\mathbf{x}\right)\right\rangle$ in both the physical and virtual spaces, while 
$\underset{\mathbf{x},i}{\bigotimes} \left\langle B_i \left(\mathbf{x}\right) \right| $ is only virtual. Thus the result of this projection, $\left|\psi\right\rangle$, is still a quantum state, including only physical degrees of freedom - the virtual, or auxiliary ones, are all contracted: this is the standard way to contract PEPS, and a frequently used notation. The physical degrees of freedom are now correlated, and in particular, thanks to maximally entangling nearest neighbours, this guarantees the entanglement entropy area law.

\begin{figure}
	\includegraphics[width=\columnwidth]{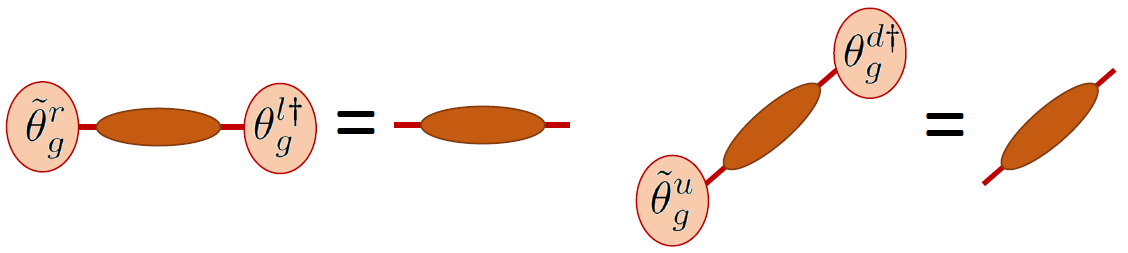}
	\caption{The invariance properties of the links.}
	\label{Links_Inv}
\end{figure}

One still has some freedom to choose which maximally entangled states to use; the ones that we picked here 
are invariant under the following group transformations:
\begin{equation}
	\begin{aligned}
		\tilde{\theta}^{r}_g\left(\mathbf{x}\right) \theta^{l\dagger}_g\left(\mathbf{x}+\hat{\mathbf{e}}_1\right)
		\left|B_1\left(\mathbf{x}\right)\right\rangle &= \left|B_1\left(\mathbf{x}\right)\right\rangle \\
		\tilde{\theta}^{u}_g\left(\mathbf{x}\right) \theta^{d\dagger}_g\left(\mathbf{x}+\hat{\mathbf{e}}_2\right)
		\left|B_2\left(\mathbf{x}\right)\right\rangle &= \left|B_2\left(\mathbf{x}\right)\right\rangle
		\label{Btrans}
	\end{aligned}
\end{equation}
(with $\theta_g,\tilde\theta_g$ defined in (\ref{rt}), (\ref{lt}) respectively) as depicted in Fig. \ref{Links_Inv}. This allows to construct states with a global or local symmetry, as we shall now see.

\subsection{Imposing the local symmetry}

\begin{figure*}
	\includegraphics[width=0.8\textwidth]{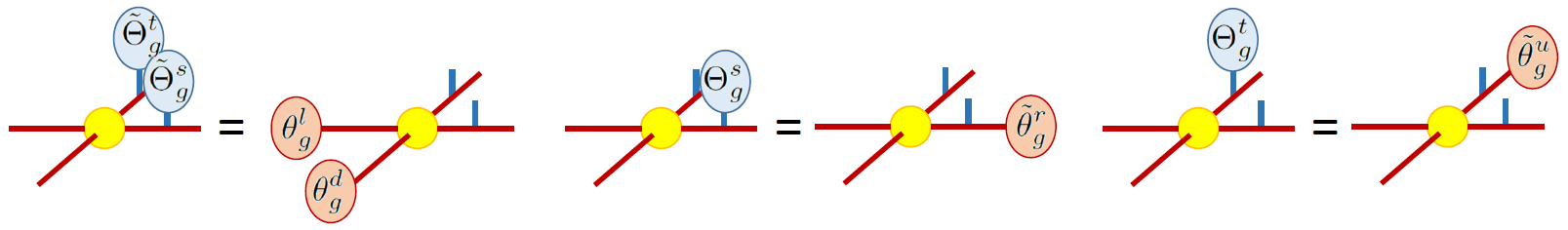}
	\caption{The invariance properties of the tensors (\ref{Atrans}), allowing for a physical local (gauge) symmetry.}
	\label{PEPS_inv}
\end{figure*}

We want our PEPS $\left|\psi\right\rangle$ (\ref{psi}) to be gauge invariant as in (\ref{gtrans}) with respect to the local gauge transformations defined in (\ref{ggg}). If the local physical-virtual states on each site satisfy (\cite{zohar_building_2016})
\begin{equation}
	\begin{aligned}
	&\tilde{\Theta}^s_g\left(\mathbf{x}\right)\tilde{\Theta}^t_g\left(\mathbf{x}\right)\left|A\left(\mathbf{x}\right)\right\rangle = \theta^l_g\left(\mathbf{x}\right) \theta^d_g\left(\mathbf{x}\right)\left|A\left(\mathbf{x}\right)\right\rangle, \\
	&\Theta^s_g\left(\mathbf{x}\right)\left|A\left(\mathbf{x}\right)\right\rangle = \tilde{\theta}^r_g\left(\mathbf{x}\right) \left|A\left(\mathbf{x}\right)\right\rangle, \\
	&\Theta^t_g\left(\mathbf{x}\right)\left|A\left(\mathbf{x}\right)\right\rangle = \tilde{\theta}^u_g\left(\mathbf{x}\right) \left|A\left(\mathbf{x}\right)\right\rangle	\quad \quad \forall g\in G
	\label{Atrans}
	\end{aligned}
\end{equation}
where the physical Hilbert spaces are transformed using $\Theta,\tilde\Theta$ defined in (\ref{Theta}), and the virtual ones using $\Theta,\tilde\Theta$ defined in (\ref{rt}) and (\ref{lt}) respectively; see Fig. \ref{PEPS_inv}. Using the transformation properties of the maximally entangled states (\ref{Btrans}) one obtains that $\left|\psi\right\rangle$ is gauge invariant.
 
In order to get a more intuitive picture of the symmetry conditions (\ref{Atrans}), let us consider the compact Lie group case again. Omitting the coordinate, since we deal we a single coordinate $\mathbf{x}$, let us denote the right and left generators of the physical degrees of freedom by $R^{s/t}_a$ and $L^{s/t}_a$. For the virtual degrees of freedom we can also define such operators, but in their case note that they do not commute, since they do not act on separate degrees of freedom ($\left|jm\right\rangle$ states, unlike the physical $\left|jmn\right\rangle$ states). The conditions (\ref{Atrans}) can be expressed, using these notations, as Gauss laws:
\begin{equation}
	\begin{aligned}
		&\left(L_a^s +L_a^t\right)\left|A\right\rangle = \left(R_a^l + R_a^d\right)\left|A\right\rangle, \\
		&R_a^s\left|A\right\rangle = L_a^r \left|A\right\rangle, \\
		&R_a^t\left|A\right\rangle = L_a^u \left|A\right\rangle	\quad \quad \forall a
		\label{AtransG}
	\end{aligned}
\end{equation}

The first condition looks like the familiar physical Gauss law. It
 implies that the two ingoing representations of the virtual indices must combine to the same representation to which the two physical representations combine: $j_s \otimes j_t \sim j_l \otimes j_d$ . Therefore, the tensor 
 $A^{j_s m_s n_s; j_t m_t n_t}_{j_r m_r; j_u m_u;j_l m_l;j_d m_d}$ should be proportional to the appropriate Clebsch-Gordan coefficients, $\left\langle j_l m_l j_d m_d | j_1 m_1 \right\rangle \left\langle j_1 m_1 | j_s m_s j_t m_t \right\rangle$.
 
 The other two conditions, are different identifying the right constituents of the physical degrees of freedom with the virtual states on the same legs. This implies that $j_r = j_s$, $j_u = j_t$, $m_r = n_s$ and $m_u = n_t$;
   $A^{j_s m_s n_s; j_t m_t n_t}_{j_r m_r; j_u m_u;j_l m_l;j_d m_d}$ must  be proportional to 
 $\delta_{j_s j_r} \delta_{j_t j_t} \delta_{n_s,m_r} \delta_{n_t,m_u}$.
 Combining the first condition with the other two, we can obtain a condition on the four virtual legs: the elements of  $A^{j_s m_s n_s; j_t m_t n_t}_{j_r m_r; j_u m_u;j_l m_l;j_d m_d}$ must vanish, unless  
 \begin{equation}
j_r \otimes j_u \sim j_l \otimes j_d.
\label{jgauss}
 \end{equation}
 
 Examples for constructions satisfying that have been previously given \cite{zohar_building_2016,zohar_fermionic_2015,zohar_projected_2016,emonts_gauss_2018}; let us just briefly comment on some special cases. When the group is Abelian, only the irrep indices remain and the Clebsch-Gordan coefficients are simply Kronecker deltas. One can then formulate $j_r \otimes j_u \sim j_l \otimes j_d$ in a very simple way. For $U(1)$, for example, $\left\langle j_1 j_2 | J\right\rangle = \delta_{j_1+j_2,J}$, and  the $\mathbb{Z}_N$ is the appropriate modular modification, $\left\langle j_1 j_2 | J\right\rangle = \delta_{j_1+j_2,J \text{mod} N}$. We thus obtain, in the $U(1)$ case, only tensor elements for which
 $j_r + j_u - j_l - j_d =0$ may be nonzero ($j_r + j_u - j_l - j_d = N\mathbb{Z}$ for $\mathbb{Z}_N$). 
  
 The same applies to non-Abelian groups as well, but since physical states contain the (generally different) $m,n$ quantum numbers it is less simple. For $SU(2)$, e.g., if we choose to include only the $j=0,1/2$ representations, the only non-vanishing tensor elements will be those with an even number of virtual legs (ingoing or outgoing) with $j=1/2$, such that a singlet can be formed by combining the contributions of all four legs.
 
 The only freedom left in the definition of $A^{j_s m_s n_s; j_t m_t n_t}_{j_r m_r; j_u m_u;j_l m_l;j_d m_d}$ is to introduce some parameters $f^{j}_{j_r,j_u,j_l,j_d}$ which only depend on the representations, and we obtain \cite{zohar_building_2016}:
 \begin{widetext}
 	\begin{equation}
 		A^{j_s m_s n_s; j_t m_t n_t}_{j_r m_r; j_u m_u;j_l m_l;j_d m_d} = 
 		\underset{j}{\sum} f^{j}_{j_r,j_u,j_l,j_d}
 		\left\langle j_l m_l j_d m_d |j m\right\rangle \left\langle j m | j_s m_s j_t m_t \right\rangle	
 		\delta_{j_s j_r} \delta_{j_t j_t} \delta_{n_s,m_r} \delta_{n_t,m_u}
 	\end{equation}
 \end{widetext}

In the following, we will focus on PEPS satisfying the above symmetry properties, with no more than one copy of each irrep in the virtual spaces. This may seem restrictive when attempting to apply the states to real, physically relevant Hamiltonians; here, however, we wish to consider the most minimal constructions which capture the relevant symmetry properties, allowing us to demonstrate our claims and results as accurately as possible. When applied to Hamiltonians as variational ansatz states the states may have to be generalized indeed but in a straight forward way that does not affect the properties we discuss here. For example, as was demonstrated already in the $\mathbb{Z}_3$ case \cite{emonts_variational_2020}, several copies of the virtual representations are required in order to use such PEPS in order to variationally find the ground states of the $\mathbb{Z}_3$ Hamiltonian. 

 One could also consider a more general PEPS construction, in which such properties are only satisfied after blocking, for effective sites and effective links. The symmetry conditions described above will hold in this case too - for the blocked tensor network, rather than the original, microscopic" one, and thus what we study here could easily be applied to such cases too. A more general scenario would be with local MPO symmetries \cite{cirac_matrix_2020}, but this is out of the scope of this work and requires its own, separate discussion.

\subsection{Tensor notation}
 
 The projection (\ref{psi}) which generates the PEPS $\left|\psi\right\rangle$ can simply be seen as a set of contraction rules for the virtual indices of the tensors $A^{j_s m_s n_s; j_t m_t n_t}_{j_r m_r; j_u m_u;j_l m_l;j_d m_d}$,  associating the indices of $r$ at $\mathbf{x}$ with those of $l$ at $\mathbf{x}+\hat{\mathbf{e}}_1$, as well as $u$ at $\mathbf{x}$ with $d$ at $\mathbf{x}+\hat{\mathbf{e}}_2$. Hence instead of looking at the local states $\left|A\right\rangle$ and their projection onto the link stats $\left|B_i\left(x\right)\right\rangle$, we may use, as our basic local building block, 
 \begin{equation}
 	A = A^{s_1 s_2 ; t_1 t_2}_{ruld} \left|l\right\rangle\left\langle r\right| \otimes  \left|d\right\rangle\left\langle u\right| \left|s_1 s_2 t_1 t_2\right\rangle
 	\label{Aten}
 \end{equation}
where, for the sake of notation simplicity,  $r\equiv \left\{j_r,m_r\right\}$,$u\equiv \left\{j_u,m_u\right\}$,$l\equiv \left\{j_l,m_l\right\}$,$d\equiv \left\{j_d,m_d\right\}$,$s_1\equiv \left\{j_s,m_s\right\}$,$s_2\equiv \left\{j_s,n_s\right\}$, $t_1\equiv \left\{j_t,m_t\right\}$,$t_2\equiv \left\{j_t,n_t\right\}$ and
$A^{s_1 s_2 ; t_1 t_2}_{ruld}\equiv A^{j_s m_s n_s; j_t m_t n_t}_{j_r m_r; j_u m_u;j_l m_l;j_d m_d}$.
 In (\ref{Aten})  the virtual states and their projection are replaced by the matrix products of $\left|l\right\rangle\left\langle r\right|$ along horizontal lines (with the positive direction from the left to the right) and $\left|d\right\rangle\left\langle u\right|$ on the vertical lines (positive direction - upwards). This sets the contraction rules of the tensors $A^{s_1 s_2 ; t_1 t_2}_{ruld}$.
 
 To illustrate, let us reduce to one space dimension and one dimensional PEPS - an MPS \cite{fannes_finitely_1992}. Each local tensor along the one dimensional system includes one physical leg, spanned by states $\left|p\right\rangle$, and two virtual ones, on the left and right direction. The state is thus parametrized by the tensors $A^p_{lr}$, and their contraction is simply a matrix multiplication of the virtual indices along the system. For a periodic system with $\mathcal{N}$ sites (the modification for open boundaries is straightforward) the state takes the form
 \begin{equation}
 	\begin{aligned}
 	\left|\psi_0\right\rangle &=
 	\text{Tr}\left[A^{p_1} A^{p_2}\cdots A^{p_{\mathcal{N}}}\right]\left|p_1,p_2,...,p_{\mathcal{N}}\right\rangle \\&= \begin{gathered}
 		\includegraphics[scale=0.15]{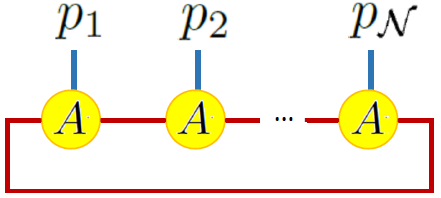}\end{gathered} \left|p_1,p_2,...,p_{\mathcal{N}}\right\rangle 
 	\label{MPS}
 	\end{aligned}
 \end{equation}
 The PEPS contraction rules in two space dimensions are simply a two dimensional generalization of the trace contraction in the one dimensional case.
 
The symmetry conditions (\ref{Atrans}) may also be expressed as properties of the tensor $A^{j_s m_s n_s; j_t m_t n_t}_{j_r m_r; j_u m_u;j_l m_l;j_d m_d}$. 
For that, we introduce the (reducible) representation matrices $\mathcal{D}\left(g\right)$ which are direct sums of the irreducible unitaries $D^j\left(g\right)$; using them, the symmetry condition (\ref{Atrans})  may be reformulated as
\begin{equation}
\begin{aligned}
&A^{s'_1 s_2; t'_1 t_2}_{ruld} \mathcal{D}_{s'_1 s_1} \left(g\right) \mathcal{D}_{t'_1 t_1} \left(g\right) = 
\mathcal{D}_{ll'} \left(g\right) \mathcal{D}_{dd'} \left(g\right) A^{s_1 s_2 ; t_1 t_2}_{rul'd'} \\
&\mathcal{D}_{s_2 s'_2} \left(g\right)  A^{s_1 s'_2; t_1 t_2}_{ruld} = 
  A^{s_1 s_2 ; t_1 t_2}_{r'uld} \mathcal{D}_{r'r}\left(g\right)  \\
 &\mathcal{D}_{t_2 t'_2} \left(g\right) A^{s_1 s_2 ;t'_1 t_2}_{ruld} = 
 A^{s_1 s_2 ; t_1 t_2}_{ru'ld} \mathcal{D}_{u'u}\left(g\right) \quad \quad \forall g\in G
 \label{tensortrans}
\end{aligned}
\end{equation}

\section{Transfer Operators and Norms of PEPS}

Before turning to the study of the transfer operator of our gauge invariant PEPS, let us recall what the transfer operator of a PEPS is. First, we briefly review the one dimensional, MPS case \cite{fannes_finitely_1992}. We strictly focus on the translationally invariant case, since this work is aimed at translational invariant systems; however, the general transfer matrix discussion may be (and has been) generalized to the non-translationally invariant case.

\subsection{Brief review of MPS transfer matrices}

The transfer matrix of the MPS $\left|\psi_0\right\rangle$ from (\ref{MPS}) is defined as
\begin{equation}
	E_{ll',rr'}=\text{Tr}_{\text{phys}}\left[A^p_{lr}\left|p\right\rangle \left\langle p'\right|\bar{A}^{p'}_{l'r'}\right]=A^p_{lr}\bar{A}^p_{l'r'}
	= \begin{gathered}
		\includegraphics[scale=0.2]{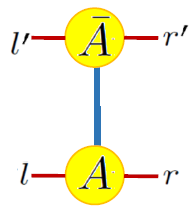}\end{gathered}
\end{equation}
- it is a matrix with double valued indices, $ll'$ to the left and $rr'$ to the right. Thus, if $l,r$ take $D$ values each (i.e. the virtual Hilbert spaces used for contracting the MPS are $D$ dimensional), $E$ is a $D^2 \times D^2$ matrix, acting on the $D^2$ space formed by the product of two copies if the virtual Hilbert space. Using $E$, we can first write down the norm of the state, 
\begin{equation}
	\left\langle\psi_0 | \psi_0 \right\rangle = \text{Tr}\left[E^{\mathcal{N}}\right]	= \begin{gathered}
		\includegraphics[scale=0.2]{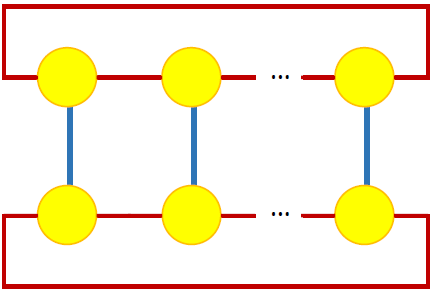}\end{gathered}
\end{equation}

For the computation of an expectation value of some operator $O$ at site $x$, we will need to define
\begin{widetext}
\begin{equation}
\left(E_O\right)_{ll',rr'}=\text{Tr}_{\text{phys}}\left[A^p_{lr}O\left|p\right\rangle \left\langle p'\right|\bar{A}^{p'}_{l'r'}\right] = 
\underset{p,p'}{\sum} A^p_{lr}\bar{A}^{p'}_{l'r'} \left\langle p' \right| O \left| p \right\rangle = \begin{gathered}
	\includegraphics[scale=0.2]{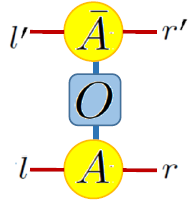}\end{gathered}
\end{equation}
using which we may write
\begin{equation}
	\left\langle O\left(x\right)\right\rangle = \frac{\left\langle \psi_0 \right| O\left(x\right) \left|\psi_0\right\rangle}{\left\langle\psi_0 | \psi_0 \right\rangle}
	=\frac{\text{Tr}\left[E_OE^{{\mathcal{N}}-1}\right]}{\text{Tr}\left[E^{\mathcal{N}}\right]}=\frac{\begin{gathered}
			\includegraphics[scale=0.2]{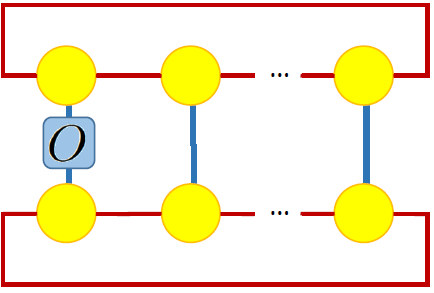}\end{gathered}}{\begin{gathered}
			\includegraphics[scale=0.2]{MPS_Norm.png}\end{gathered}}
\end{equation}

Suppose we wish to compute the two-point correlator of $O_1\left(x_1\right)$ and $O_2\left(x_2\right)$ (assuming for simplicity that $x_2-x_1=R>0$),
\begin{equation}
	\begin{aligned}
	F\left(x_1,x_2\right)&=\left\langle O_1\left(x_1\right)O_2\left(x_2\right)\right\rangle
	- \left\langle O_1\left(x_1\right)\right\rangle \left\langle O_2\left(x_2\right)\right\rangle 
	 = \frac{\text{Tr}\left[E_{O_1}E^{R-1}E_{O_2}E^{{\mathcal{N}}-R-1}\right]}{\text{Tr}\left[E^{\mathcal{N}}\right]}
	 -\frac{\text{Tr}\left[E_{O_1}E^{{\mathcal{N}}-1}\right]\text{Tr}\left[E_{O_2}E^{{\mathcal{N}}-1}\right]}{\text{Tr}^2\left[E^{\mathcal{N}}\right]}
	\\& = \frac{\begin{gathered}
			\includegraphics[scale=0.2]{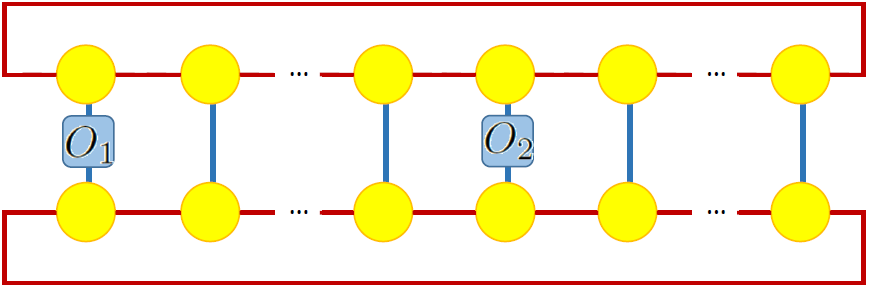}\end{gathered}}{\begin{gathered}
			\includegraphics[scale=0.2]{MPS_Norm.png}\end{gathered}} - 
		\frac{\begin{gathered}
				\includegraphics[scale=0.2]{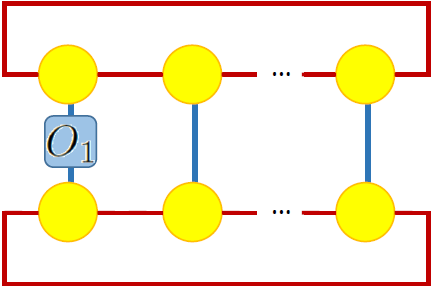}\end{gathered}\times\begin{gathered}\includegraphics[scale=0.2]{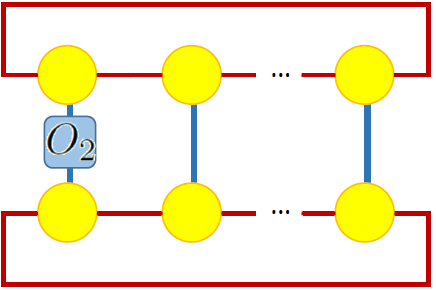}\end{gathered}}
			{\left(\begin{gathered}
					\includegraphics[scale=0.2]{MPS_Norm.png}\end{gathered}\right)^2}
	 \end{aligned}
\label{MPScorr}
\end{equation}
\end{widetext}

We introduce the left and right eigenvectors of $E$, $\left\langle w_i \right| E = \left\langle w_i \right|\rho_i$ and 
 $ E \left|v_i\right\rangle = \rho_i \left|v_i\right\rangle$, sharing the same eigenvalues $\rho_i$ and satisfying the orthonormality relation $\left\langle w_i | v_j \right\rangle = \delta_{ij}$, and expand $E$ as
 \begin{equation}
 	E=\underset{i}{\sum}\rho_i \left|v_i\right\rangle \left\langle w_i \right|
 \end{equation}
Let us sort the eigenvalues in a descending order and assume that the largest one is non-degenerate, that is
$\left|\rho_1\right| > \left|\rho_2\right| \geq \left|\rho_3\right|  \geq ...$. Then, for $N \gg 1$, $R \gg 1$, one obtains that
\begin{equation}
F\left(x_1,x_2\right) \approx \rho_1^{-2}\underset{i>1}{\sum}\left(\frac{\rho_i}{\rho_1}\right)^{R-1}
\left\langle w_1 \right| E_{O_1} \left| v_i \right\rangle 
\left\langle w_i \right| E_{O_2} \left| v_1 \right\rangle 
\end{equation}
- the correlations decay exponentially, with a finite correlation length $\xi = -1/\log\left|\frac{\rho_2}{\rho_1}\right|$.

\subsection{Transfer operators of PEPS}

The transfer matrix approach can be generalized to two dimensional PEPS, such as the ones we consider here, constructed in (\ref{psi}). We assume the system has periodic boundary conditions - a torus of size $\mathcal{N} \times \mathcal{N}$ (generalizations to other boundary conditions are straightforward).
The local transfer operator of a PEPS on a site is a map from two double virtual Hilbert spaces, associated with the ingoing (left and down) legs, to other two double virtual spaces, directed to the outgoing directions (right and up):
\begin{equation}
\hat{T}=T_{ll',rr',dd',uu'} \left|ll'\right\rangle \left\langle rr' \right| \otimes  \left|dd'\right\rangle \left\langle uu' \right|
\label{Tdef}
\end{equation}
(note that we use again a convention in which the input vectors are denoted by bras, in accordance with matrix product ordered from left to right in the positive system directions).

In full analogy with the one dimensional case, the elements of the transfer tensor $T_{ll',rr',dd',uu'}$ are given by 
\begin{widetext}
\begin{equation}
T_{ll',rr',dd',uu'}=\text{Tr}_{s,t}\left[A^{st}_{ruld} \left|st\right\rangle \left\langle s't' \right| \bar{A}^{s't'}_{r'u'l'd'}\right]
=A^{st}_{ruld}  \bar{A}^{st}_{r'u'l'd'} = \begin{gathered}
	\includegraphics[scale=0.15]{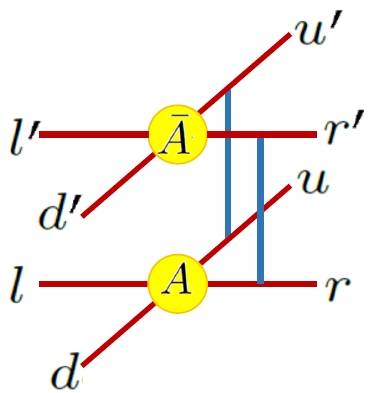}\end{gathered} \equiv \begin{gathered}
	\includegraphics[scale=0.15]{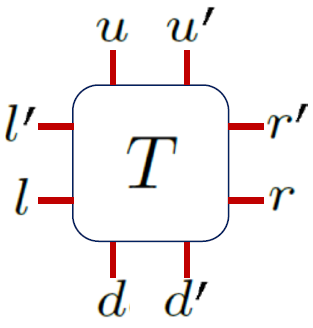}\end{gathered}
\label{Tform}
\end{equation}

The norm may be computed by properly contracting products of $\hat{T}$ on all the lattice sites; in expectation values of observables, the numerator may be computed by replacing $\hat{T}$ at the relevant sites by
\begin{equation}
\left(T_O\right)_{ll',rr',dd',uu'}=\text{Tr}_{\text{phys}}\left[A^{st}_{ruld} O \left|st\right\rangle \left\langle s't' \right| \bar{A}^{s't'}_{r'u'l'd'}\right]
= A^{st}_{ruld}  \bar{A}^{s't'}_{r'u'l'd'} \left\langle s't' \right| O \left| st \right\rangle =  \begin{gathered}
	\includegraphics[scale=0.15]{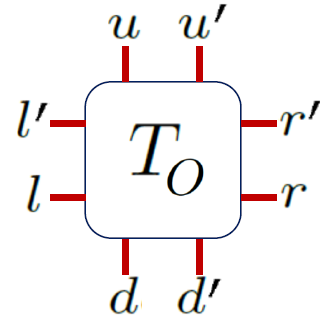}\end{gathered}
\label{TOcomp}
\end{equation} 

To compute correlations, we will first contract all the tensors along one dimension of the PEPS, converting it effectively to an MPS \cite{yang_chiral_2015,zohar_fermionic_2015,zohar_projected_2016} whose transfer matrix can be defined as above.
For example, the transfer matrix of a row of length $\mathcal{N}$ is obtained by contraction along the horizontal direction, 
\begin{equation}
	\begin{aligned}
\hat{E} &= T_{l_1l'_1,r_1r'_1,d_1d'_1,u_1u'_1} T_{l_2l'_2,r_2r'_2,d_2d'_2,u_2u'_2} \cdots T_{l_{\mathcal{N}}l'_{\mathcal{N}},r_{\mathcal{N}}r'_{\mathcal{N}},d_{\mathcal{N}}d'_{\mathcal{N}},u_{\mathcal{N}}u'_{\mathcal{N}}} \text{Tr}\left[\left|l_1l'_1\right\rangle\left\langle r_1 r'_1 | l_2 l'_2 \right\rangle \left\langle r_2 r'_2 \right| \cdots \left|l_{\mathcal{N}}l'_{\mathcal{N}}\right\rangle\left\langle r_{\mathcal{N}} r'_{\mathcal{N}} \right|\right] \\
& \times \left|d_1 d'_1\right\rangle \left\langle u_1 u'_1 \right|  \otimes ...
\otimes  \left|d_{\mathcal{N}} d'_{\mathcal{N}}\right\rangle \left\langle u_{\mathcal{N}} u'_{\mathcal{N}} \right| =
 E_{d_1d,_1,...,d_{\mathcal{N}},d'_{\mathcal{N}};u_1u'_1,...,u_{\mathcal{N}},u'_{\mathcal{N}}} 
\left|d_1 d'_1\right\rangle \left\langle u_1 u'_1 \right|  \otimes ...
 \otimes  \left|d_{\mathcal{N}} d'_{\mathcal{N}}\right\rangle \left\langle u_{\mathcal{N}} u'_{\mathcal{N}} \right| \\&=
  \begin{gathered}
 	\includegraphics[scale=0.3]{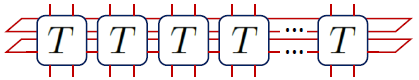}\end{gathered}
\end{aligned}
\end{equation}
where
\begin{equation}
E_{d_1d,_1,...,d_{\mathcal{N}},d'_{\mathcal{N}};u_1u'_1,...,u_{\mathcal{N}},u'_{\mathcal{N}}}  = T_{i_1,i'_1 , i_2 i'_2,d_1 d'_1 ,u_1 u'_1}T_{i_2,i'_2 , i_3 i'_3,d_2 d'_2,u_2 u'_2}\cdots T_{i_{\mathcal{N}} i'_{\mathcal{N}}, i_1 i'_1 ,d_{\mathcal{N}} d'_{\mathcal{N}} ,u_{\mathcal{N}} u'_{\mathcal{N}}}
\end{equation}
\end{widetext}

Using $E$ and similar transfer matrices which include observables, one may use the entire MPS machinery for computations of norms, expectation values and correlation functions. Naively, one may deduce that correlations in this case decay exponentially as in the MPS case \cite{fannes_finitely_1992}. However, unlike in the one dimensional, MPS case, here the transfer matrix is a composite object with some internal structure, which can lead to different results. It was shown in \cite{verstraete_criticality_2006}, for example, that two dimensional PEPS can describe critical physics, exhibiting power law contributions.

\subsection{Flux free transfer operators}

\begin{figure}
	\includegraphics[width=\columnwidth]{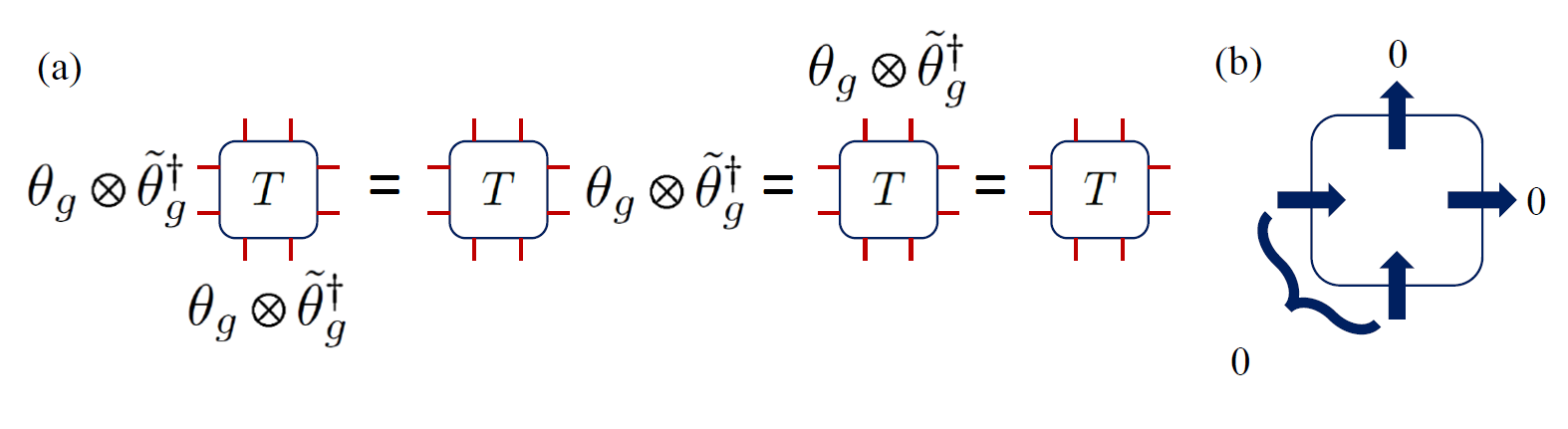}
	\caption{The invariance properties of the transfer operator $\hat T$ (a) and its map interpretation (b).}
	\label{Trules}
\end{figure}

Let us apply the above to the computation of the norm. For that, consider the flux-free transfer operator, that is, the local building block of the transfer matrix on a single site, with no string (group element operator $U^j$), $\hat{T}$, as defined in (\ref{Tdef}). We calculate its elements using (\ref{Tform}), and thanks to the symmetry conditions  (\ref{tensortrans}) we obtain that for every $g \in G$,
\begin{equation}
	\left(\theta^l_g \otimes \tilde{\theta}^{\dagger l'}_g\right) \otimes \left(\theta^d_g \otimes \tilde{\theta}^{\dagger d'}_g\right) \hat{T} = \hat{T} \theta^r_g \otimes \tilde{\theta}^{\dagger r'}_g
	= \hat{T} \theta^u_g \otimes \tilde{\theta}^{\dagger u'}_g = \hat{T} 
\end{equation}
(see Fig. \ref{Trules}(a)).
This implies, that in (\ref{Tdef}), the outgoing vectors
$\left|rr'\right\rangle$ and $\left|uu'\right\rangle$ are both separately singlets under the action of $\left(\theta_g \otimes \tilde{\theta}^{\dagger }_g\right)$ - that is, they are 
\emph{on-leg singlets} , denoted by $\left\langle0\left(j_r\right)\right|$ and $\left\langle0\left(j_u\right)\right|$
and defined as
\begin{equation}
 \left|0\left(j\right)\right\rangle = \left| jm jm\right\rangle
\label{ols}
\end{equation}
 The ingoing legs $\left|ll'\right\rangle \otimes \left|dd'\right\rangle$, on the other hand, combine \emph{together} to a singlet under $\left(\theta^l_g \otimes \tilde{\theta}^{\dagger l'}_g\right) \otimes \left(\theta^d_g \otimes \tilde{\theta}^{\dagger d'}_g\right)$:
$\left\langle j_l m_l j_d m_d | j'_l m'_l j'_d m'_d \right\rangle
\left|j_l m_l, j'_l m'_l\right\rangle  \otimes 
\left|j_d m_d, j'_d m'_d\right\rangle $.
 We can therefore conclude that the general structure of $\hat{T}$ is
\begin{widetext}
\begin{equation}
	\hat{T} = \underset{\left\{j\right\}}{\sum}T_{j_l,j_l',j_r;j_d,j_d',j_u}
	\left\langle j_l m_l j_d m_d | j'_l m'_l j'_d m'_d \right\rangle
	\left|j_l m_l, j'_l m'_l\right\rangle \left\langle 0\left(j_r\right) \right| \otimes 
	\left|j_d m_d, j'_d m'_d\right\rangle \left\langle 0\left(j_u\right) \right|
	\label{Tstruct}
\end{equation}
\end{widetext}
- it is a map with two inputs and two outputs, which takes a joint singlet (on both the ingoing legs) into two separate on-leg singlets, on each outgoing leg alone (see Fig. \ref{Trules}(b)).

\subsection{The row transfer matrix and the norm}

Suppose we wish to compute the norm, which involves contracting the tensor product of $\hat{T}$ everywhere. Each  $\hat{T}$ obtains its inputs from the neighbouring $\hat{T}$ operators on its left and bottom, whose outputs are on-leg singlets: that is, when the norm is computed, the inputs 
$\left|j_l m_l, j'_l m'_l\right\rangle \left\langle 0\left(j_r\right) \right|$ on the left leg and $\left|j_d m_d, j'_d m'_d\right\rangle \left\langle 0\left(j_r\right) \right|$ on the lower one are being contracted with the outputs from neighbouring sites - $\left\langle 0\left(j_r\right)\right|$ and  $\left\langle 0\left(j_u\right)\right|$ respectively. 
 Thus, for the norm contraction it is enough to focus only on a subset of the $T$ elements, where only on-leg singlets are allowed as input. Denoting by $\Pi_0 = \underset{j}{\sum}\left|0\left(j\right)\right\rangle \left\langle 0\left(j\right)\right|$ the projection operator onto on-leg singlets $\left|0\left(j\right)\right\rangle = \left|jm jm\right\rangle$, we define
\begin{equation}
	\hat{\tau}_0 = \Pi_0 \otimes \Pi_0 \hat{T} \equiv \begin{gathered}
	\includegraphics[scale=0.25]{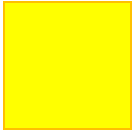}\end{gathered}
	\label{tau0def}
\end{equation}
(introducing a new notation which will be used for tiling diagrams below, in which the legs are implicit). It takes the simple form
\begin{equation}
	\hat{\tau}_0 = \underset{\left\{j\right\}}{\sum}\left(\tau_0\right)_{j_l,j_r;j_d,j_u}
	\left|0\left(j_l\right)\right\rangle \left\langle 0\left(j_r\right) \right| \otimes 
	\left|0\left(j_d\right)\right\rangle \left\langle 0\left(j_u\right) \right|
		\label{tau0}
\end{equation}
where $\left(\tau_0\right)_{j_l,j_r;j_d,j_u} = T_{j_l,j_l,j_r;j_d,j_d,j_u}$.

To see how this simplifies the contraction, let us consider some illustrative examples. First, consider the $\mathbb{Z}_N$ case, in which (disregarding multiplicities) the virtual Hilbert spaces are spanned by $D=N$ basis states, corresponding to the $j=0,...,N-1$ irreps. Thus we will have $N$ on-leg singlets, of the form
\begin{equation}
\left|0\left(j\right)\right\rangle = \left|j j\right\rangle
\label{Z0}
\end{equation}
The tensor $\hat{\tau}_0$ will thus contain $N^4$ elements; having considered $T$ without taking the symmetry into account, with two $N$ dimensional legs per direction, we would have instead $N^8$ tensor elements! That is, the number of elements that actually need to be used for contraction is $N^4$ times smaller. Next, generalize to $U(1)$, and suppose we truncate and allow for the $\left|j\right|  \leq J$ for some $J > 0$. Then we will have once again on-leg singlets of the form (\ref{Z0}). There are $D=2J+1$ irreps in the virtual Hilbert space, we  have $D$ on-leg singlets and, similarly to the $\mathbb{Z}_N$ case, we obtain a reduction of $D^4$: $D^4$ elements in $\hat{\tau}_0$ which we need for the contraction, rather than the $D^8$ in the most general case.

The simplification is even bigger when we consider non-Abelian groups, because the tensors $\hat{\tau}_0$ only see the representations and not the different $m$ values within them. For example, consider $SU(2)$, with the smallest truncation, containing the $j=0,1/2$ representations. This implies that each virtual Hilbert space has dimension $3$. Naively speaking, $T$ would be a tensor with $3^8 = 6561$ elements. Reducing to $\hat{\tau}_0$, with only two on-leg singlets for the two irreps used, the number of relevant elements decreases to $2^4 = 16$, that is, approximately 410 times less! If we wish to consider, a little more generally, all the irreps of $SU(2)$ between $0$ to some $J$, the dimension of the virtual Hilbert spaces would be 
$D\left(J\right) =  \underset{j=0}{\overset{J}{\sum}} \left(2j+1\right) = \left(J+1\right)\left(2J+1\right)$ (note that the sum runs on both integer and half-integer values). Thus $T$ has $D^8\left(J\right) = \left(J+1\right)^8\left(2J+1\right)^8$ elements. However the number of on-leg singlets is as the number of irreps, $2J+1$, and hence $\hat \tau_0$ is a tensor with $\left(2J+1\right)^4$ elements: the reduction factor is $\left(J+1\right)^8 \left(2J+1\right)^4$, which scales as $J^{-12}$ for large cutoffs - a very significant reduction!

To examine further the properties of $\hat \tau_0$, let us consider $\left(\tau_0\right)_{j_l,j_r;j_d,j_u}$ as a matrix with the multivalued indices $j_l,j_r$ and  $j_u,j_d$. If we assume horizontal-vertical reflection symmetry, we find that it is a symmetric matrix,
\begin{equation}
	\left(\tau_0\right)_{j_l,j_r;j_d,j_u} = \left(\tau_0\right)_{j_d,j_u;j_l,j_r}
\end{equation}
Furthermore, it is a real matrix, since using (\ref{Tform}), with the restriction (\ref{tau0}), we obtain that
\begin{equation}
	\begin{aligned}
		\left(\tau_0\right)_{j_l,j_r;j_d,j_u} & = T_{j_l,j_l,j_r;j_d,j_d,j_u} \\&=
		\underset{\left\{j,m,n\right\}}{\sum}\left|A^{j_s m_s n_s; j_t m_t n_t}_{j_r m_r; j_u m_u;j_l m_l;j_d m_d}\right|^2
		\end{aligned}
\end{equation}
Therefore, there exists an orthogonal matrix $V$, such that
\begin{equation}
	\tau_0 = V \Lambda V^{\dagger}
	\label{tau0diag}
\end{equation}
where $\Lambda$ is a diagonal matrix with eigenvalues $\lambda_{\mu}$. This allows us to bring $\hat  \tau_0$ to the convenient form
\begin{equation}
	\hat{\tau}_0 = \underset{\mu}{\sum}\lambda_{\mu} \hat{M}_{\mu} \otimes \hat{M}_{\mu}
\end{equation}
where
\begin{equation}
	\hat{M}_{\mu} = \underset{j_1,j_2}{\sum} V_{j_1j_2,\mu}\left| 0\left(j_1\right) \right\rangle \left\langle 0\left(j_2\right) \right|
	\label{Mdef}
\end{equation}
- one copy of which acts on the horizontal direction and the other on the virtual one.

The real matrices $\left\{\hat{M}_{\mu}\right\}$ form an orthonormal set with respect to the trace inner product. Since $V$ is orthogonal, it is straightforward to show that
\begin{equation}
	\text{Tr}\left[\hat{M}_{\mu}\hat{M}^T_{\nu}\right] = \delta_{\mu \nu}
	\label{ortho}
\end{equation}
Suppose our tensor includes $D$ irreps, all the $j$s take $D$ different values. Then there are $D$ different on-leg singlets, and the matrix $\tau_0$ is $D^{2} \times D^{2}$; thus, $\mu = 1,...,D^2$ and we have $D^2$ $\hat{M}_{\mu}$ matrices. They act on the $D$ dimensional space spanned by the $D$ linearly independent on-leg singlets $\left|0\left(j\right)\right\rangle$. These matrices form a $D^2$ linear space; we have shown that $\hat{M}_{\mu}$ is an orthonormal set of $D^2$ matrices within this space, and thus it is an orthonormal basis and the $\hat{M}_{\mu}$ span the whole space of $D \times D$ real matrices.

The row transfer matrix and the norm thus take the forms
\begin{widetext}
\begin{equation}
	\begin{aligned}
	\hat{E} &\equiv 
	\begin{gathered}
		\includegraphics[scale=0.3]{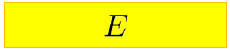}\end{gathered}
	=
	\text{Tr}_{\text{row}}\left[\begin{gathered}
	 	\includegraphics[scale=0.3]{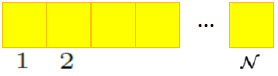}\end{gathered}\right]
	\\&=
	\underset{\left\{\mu\right\}}{\sum} 
	\lambda_{\mu_1}\lambda_{\mu_2}\cdots\lambda_{\mu_{\mathcal{N}}}
	\text{Tr}\left[\hat M_{\mu_1}\hat M_{\mu_2}\cdots \hat M_{\mu_{\mathcal{N}}}\right]\hat M_{\mu_1} \otimes \hat M_{\mu_2} \otimes \cdots \otimes \hat M_{\mu_{\mathcal{N}}}
	\label{Edef}
	\end{aligned}
\end{equation}
and 
\begin{equation}
	\begin{aligned}
	\left\langle \psi | \psi \right\rangle = \text{Tr}\left[\hat E^{\mathcal{N}}\right]&=
	\text{Tr}\left[\begin{gathered}
		\includegraphics[scale=0.3]{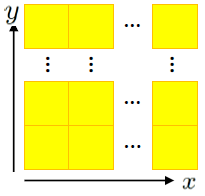}\end{gathered}\right]
	=
	\text{Tr}\left[\begin{gathered}
		\includegraphics[scale=0.3]{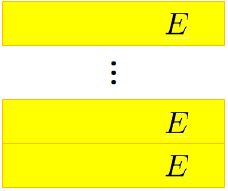}\end{gathered}\right]
	=
	\text{Tr}\left[\begin{gathered}
		\includegraphics[scale=0.3]{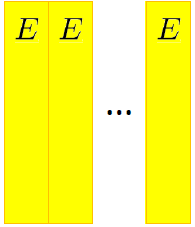}\end{gathered}\right]=
	\\&=
	\underset{\left\{\mu\left(x,y\right)\right\}}{\sum}\underset{x,y}{\prod}\lambda_{\mu\left(x,y\right)}
	\underset{y}{\prod} \text{Tr}\left[\hat M_{\mu\left(1,y\right)}\hat M_{\mu\left(2,y\right)}\cdots \hat M_{\mu\left({\mathcal{N}},y\right)}\right]
	\underset{x}{\prod} \text{Tr}\left[\hat M_{\mu\left(x,1\right)}\hat M_{\mu\left(x,2\right)}\cdots \hat M_{\mu\left(x,{\mathcal{N}}\right)}\right]
	\end{aligned}
\end{equation}
\end{widetext}

\subsection{Spectrum of the flux-free transfer matrix}

We have used the fact that each leg of $\hat\tau_0$ forms a singlet $\left|0\left(j\right)\right\rangle$; however, recall the symmetry properties of the tensor $A$ out of which the transfer operators were constructed, and the Gauss law satisfied by its four legs (\ref{jgauss}): $j_r \otimes j_u \sim j_l \otimes j_d$. 
This implies that 
$\left(j_l \otimes j_r\right) \otimes \left(j_d \otimes j_u\right)$ must contain the single representation: the horizontal representations and the vertical ones must be such that can fuse to a singlet together. As a consequence of that, elements of $\left(\tau_0\right)_{j_l,j_r;j_d,j_u}$ whose indices do not satisfy it must vanish. This splits the matrix $\left(\hat{\tau}_0\right)_{j_l,j_r;j_d,j_u}$ into separate blocks which can be separately diagonalized, implying similar block structure of the $V$ matrices as well, splitting the $\hat{M}_{\mu}$ operators defined in (\ref{Mdef}) into different sets.

First, consider the so-called zero block $\hat{B}_0$ in which $j_l=j_r$ as well as $j_d=j_u$. The elements of this block will be linear combinations of products of horizontal and vertical on-leg singlet projectors,
\begin{equation}
	\hat{B}_0 = \left(\tau_0\right)_{jj;j'j'} \left|0\left(j\right)\right\rangle \left\langle0\left(j\right) \right| \otimes \left|0\left(j'\right)\right\rangle \left\langle 0\left(j'\right) \right| 
\end{equation}
The $\hat{M}_{\mu}$ operators derived from this block will be diagonal in the space of singlets; the block $\left(\tau_0\right)_{jj;j'j'}$ is a simple symmetric matrix, diagonalizable by the orthogonal block $V^{(0)}_{j\mu}$, using which we obtain the diagonal operators
\begin{equation}
	\hat{M}^{(0)}_{\mu} = \underset{j}{\sum} V^{(0)}_{j\mu} \left|0\left(j\right)\right\rangle \left\langle0\left(j\right) \right| 
\end{equation}

The next blocks are responsible to $\hat{M}_{\mu}$ which are off-diagonal in the singlet space. In the $U(1)$ case, for example, we will have blocks 
for which $j_l - j_r = j_u - j_d = \pm k$ (for any integer $k$ allowed by our tensors)
\begin{equation}
	\hat{B}_{\pm k} = \left(\tau_0\right)_{j,j \mp k;j',j'\pm k} \left|0\left(j\right)\right\rangle \left\langle 0\left(j\mp k\right) \right| \otimes \left|0\left(j'\right)\right\rangle \left\langle 0\left(j'\pm k\right) \right| 
\end{equation}

Let us choose, in our $U(1)$ example, to include one copy of each irrep $\left|j\right| \leq J$ ($J$ may also be infinite). The matrix $\left(\tau_0\right)_{j_l,j_r;j_u,j_d}$ will have dimension of $\left(2J+1\right)^2$.
 The zeroth block of $\hat{\tau}$, $B_0 = \left(\tau_0\right)_{j,j k;j',j'}$, will be a $2J+1$ dimensional matrix (since there are $2J+1$ possible on-leg singlet states). The blocks $B_k = \left(\tau_0\right)_{j,j \mp k;j',j'\pm k}$ will each be $2J+1-\left|k\right|)$ dimensional (counting the number of $j$ values allowing for $j-k$ and $j+k$ values which agree with $\left|j\right| \leq J$), from $k=\pm 1$ until $k=\pm 2J$ - altogether $2J+1$ blocks whose dimensions add up, properly, to the right matrix dimension, $\underset{k=-2J}{\overset{2J}{\sum}}\left(2J+1-\left|k\right|\right)=\left(2J+1\right)^2$. Finally, since  $\left(\hat{\tau}_0\right)_{j_l,j_r;j_d,j_u}$ is a symmetric matrix, we obtain that
 $B_k = B_{-k}^T$, and write down the matrix in the block form
 \begin{widetext}
\begin{equation}
	\renewcommand\arraystretch{1.2}
	\tau_0 =
	\begin{blockarray}{cccccccccc}
		\text{lr / du}&&& \left|0\left(j\right)\right\rangle\left\langle 0\left(j\right) \right|   &  \cdots & 
		\left|0\left(j\right)\right\rangle\left\langle 0\left(j-k\right) \right| & \left|0\left(j-k\right)\right\rangle\left\langle 0\left(j\right) \right| &  \cdots & \left|0\left(J\right)\right\rangle\left\langle 0\left(-J\right) \right| & \left|0\left(-J\right)\right\rangle\left\langle 0\left(J\right) \right| &
		\\
		\begin{block}{cc(c@{}cccccc@{}c)}
			\left|0\left(j\right)\right\rangle\left\langle 0\left(j\right) \right|  &  && B_0 & \cdots & 0 & 0& \cdots & 0 & 0 \\
			\vdots  &  && \vdots &  & \vdots &  \vdots &  & \vdots & \vdots \\
			\left|0\left(j\right)\right\rangle\left\langle 0\left(j-k\right) \right|  & && 0 & \cdots & 0 & B_k& \cdots & 0 & 0 \\ 
			\left|0\left(j-k\right)\right\rangle\left\langle 0\left(j\right) \right|  & && 0 & \cdots &  B_k^T & 0 & \cdots & 0 & 0 \\ 
			\vdots  &  && \vdots &  & \vdots &  \vdots &  & \vdots & \vdots \\
			\left|0\left(J\right)\right\rangle\left\langle 0\left(-J\right) \right|  &  && 0 & \cdots & 0 & 0& \cdots & 0 & B_J \\ 
			\left|0\left(-J\right)\right\rangle\left\langle 0\left(J\right) \right|  & && 0 & \cdots & 0 & 0& \cdots & B_J^T & 0 \\ \\
		\end{block}
	\end{blockarray}
\end{equation}
 \end{widetext}
(where the headers of the rows and columns denote the type of operators they connect with). This matrix can be easily blockwise diagonalized, involving the diagonalization of $J+1$ different blocks. Similar forms can be written also for other gauge groups (later on, we will work out a detailed example for the $\mathbb{Z}_2$ case).

Before moving on to the contraction of Wilson loops, we shall consider some simple illustrative cases of norm computation, regardless of the gauge group. First, assume that all the blocks but the zeroth one vanish, and, on top of that, that the zeroth block is diagonal, that is 
\begin{equation}
	\hat{\tau}_0 = \underset{j}{\sum}\lambda_j \left|0\left(j\right)\right\rangle \left\langle0\left(j\right) \right| \otimes \left|0\left(j\right)\right\rangle \left\langle0\left(j\right) \right| 
\end{equation}
- all the relevant $\hat{M}_{\mu}$ operators are projectors (the other ones do not contribute since they are associated with zero eigenvalues). Then, it is easy to see that the transfer matrix is
\begin{equation}
	\hat{E} = \underset{j}{\sum} 
	\lambda_{j}^{\mathcal{N}}
	\left|0\left(j\right)\right\rangle \left\langle0\left(j\right) \right| \otimes  \left|0\left(j\right)\right\rangle \left\langle0\left(j\right)\right| \otimes \cdots \otimes \left|0\left(j\right)\right\rangle \left\langle0\left(j\right)\right|
\end{equation}
Then, the eigenvectors are product vectors of the same representation, $\left\langle w_j \right| = \left\langle 0\left(j\right) \right| \otimes \cdots \otimes \left\langle 0\left(j\right)\right|$ with eigenvalues $\rho_j = \lambda_j^{\mathcal{N}}$, 
and the norm is
\begin{equation}
	\left\langle \psi | \psi \right\rangle = \text{Tr}\left[\hat{E}^{\mathcal{N}}\right] = \underset{j}{\sum}\lambda_j^{{\mathcal{N}}^2}
\end{equation}

Next, we keep the off-diagonal terms of the zeroth block zero, but allow for very small nonzero elements in the other blocks - that is, significantly smaller (in absolute value) than the diagonal terms of the zeroth block. If $D$ irreps participate in our state, we have $\hat{M}_{\mu} = \left|0\left(j_{\mu}\right)\right\rangle \left\langle0\left(j_{\mu}\right) \right|$ for $\mu = 1,...,D$, with eigenvalues $\left|\lambda_{1}\right| \geq  ... \geq \left|\lambda_{D}\right| > 0$; while for some $K>D$, $\left|\lambda_{\mu+1}\right| \geq  ... \geq \left|\lambda_{\mu+K}\right| > 0$ and there is some $1 \leq L \leq D$ for which $\left|\lambda_{\mu+1}\right| \ll  \left|\lambda_{L}\right|$.
Then one may use perturbation theory to find the spectrum of $\hat E$. The nonperturbed part is 
$
\underset{\mu \leq L}{\sum} 
\lambda_{\mu}^N
\left|0\left(j_{\mu}\right)\right\rangle \left\langle0\left(j_{\mu}\right) \right| \otimes  \left|0\left(j_{\mu}\right)\right\rangle \left\langle0\left(j_{\mu}\right)\right| \otimes \cdots \otimes \left|0\left(j_{\mu}\right)\right\rangle \left\langle0\left(j_{\mu}\right)\right|
$
giving rise to zeroth order eigenvectors as before, with corrections which are product vectors as well.

Now allow for nonzero weak off diagonal elements in the zeroth block. Perturbation theory is still valid, keeping our eigenvectors close to product states along the row. In fact, as long as the diagonal terms of the zeroth block are significantly stronger (in absolute value) than the rest of the $\tau_0$ elements, this argument holds. As these other terms get larger and larger, the perturbative description loses its validity and the eigenvectors get farther from being product states along the row. 

This may be interpreted as the lack or the presence of long-range order: the farther we are from product states along the row, the longer ranged order we have. Since confinement has to do with disorder \cite{fradkin_order_1978}, we find here the first hint to detecting area law from the transfer operators. As we shall see later on, indeed, the closer the transfer matrix eigenvectors are to product states, the closer we are to an area law of the Wilson loop.

\section{A tale of tiling: contracting Wilson Loops}

After having computed the norms, we move further to the contraction of Wilson Loop expectation values, which first requires studying further local ingredients: the flux-carrying transfer operators.  

\subsection{Flux carrying transfer operators}

\begin{figure*}
	\includegraphics[width=0.8\textwidth]{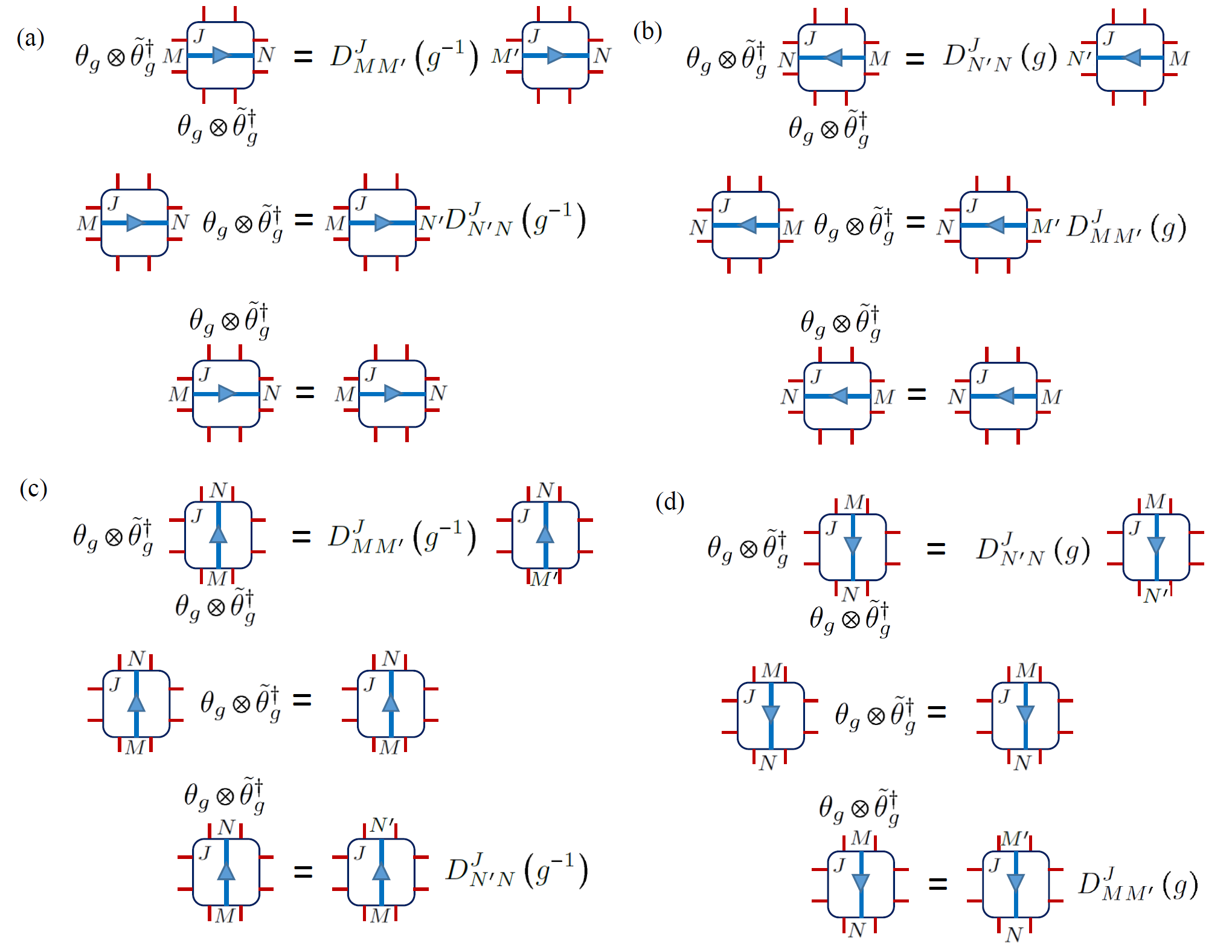}
	\caption{Transformation rules of the straight flux carrying transfer operators: (a) $\left[\hat T_{\rightarrow}\right]^{J}_{MN}$ - Eq. (\ref{Th});
	(b) $\left[\hat T_{\leftarrow}\right]^{J}_{MN}$ - Eq. (\ref{Thd});
(c) $\left[\hat T_{\uparrow}\right]^{J}_{MN}$ - Eq. (\ref{Tv});
(d) $\left[\hat T_{\downarrow}\right]^{J}_{MN}$ - Eq. (\ref{Tvd});}
	\label{Tst_rules}
\end{figure*}

\begin{figure}
	\includegraphics[width=\columnwidth]{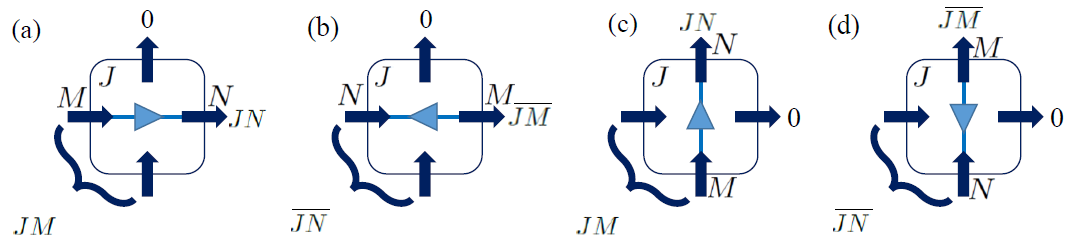}
	\caption{The straight flux line transfer operators as maps.}
	\label{Tst_map}
\end{figure}

Consider the transfer operators associated with sites carrying a straight flux line - that is, a group element operator $U^j$ (or $U^{j\dagger}$) acting on either the horizontal or vertical direction, computed using (\ref{TOcomp}):
\begin{widetext}
\begin{equation}
	\begin{aligned}
\left(\left[T_{\rightarrow}\right]^{J}_{MN}\right)_{ll',rr';dd',uu'} & = \text{Tr}_{\text{phys}}\left[A^{st}_{ruld} U^{sJ}_{MN} \left|st\right\rangle \left\langle s't' \right| \bar{A}^{s't'}_{r'u'l'd'}\right]
= A^{st}_{ruld}  \bar{A}^{s't}_{r'u'l'd'} \left\langle s' \right| U^{J}_{MN} \left| s \right\rangle \equiv \begin{gathered}\includegraphics[scale=0.1]{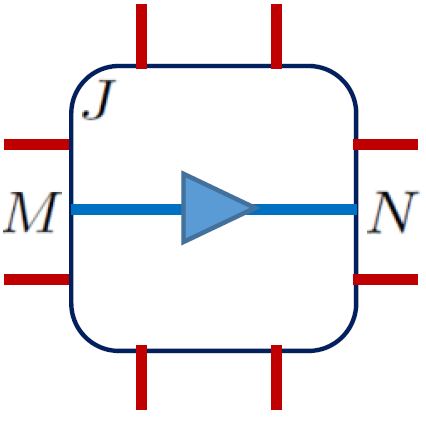}\end{gathered} \\
\left(\left[T_{\uparrow}\right]^{J}_{MN}\right)_{ll',rr';dd',uu'} & = \text{Tr}_{\text{phys}}\left[A^{st}_{ruld} U^{tJ}_{MN} \left|st\right\rangle \left\langle s't' \right| \bar{A}^{s't'}_{r'u'l'd'}\right]
= A^{st}_{ruld}  \bar{A}^{st'}_{r'u'l'd'} \left\langle t' \right| U^{J}_{MN} \left|t \right\rangle \equiv \begin{gathered}\includegraphics[scale=0.1]{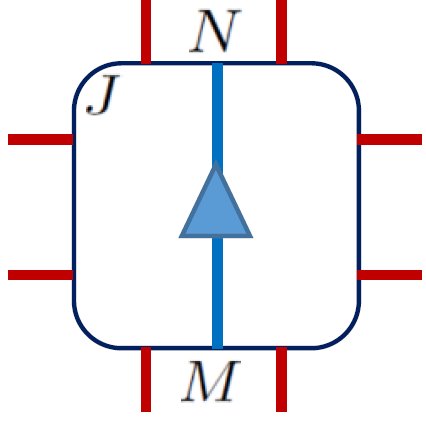}\end{gathered} \\
\left(\left[T_{\leftarrow}\right]^{J}_{MN}\right)_{ll',rr';dd',uu'} & = \text{Tr}_{\text{phys}}\left[A^{st}_{ruld} U^{sJ\dagger}_{MN} \left|st\right\rangle \left\langle s't' \right| \bar{A}^{s't'}_{r'u'l'd'}\right]
= A^{st}_{ruld}  \bar{A}^{s't}_{r'u'l'd'} \left\langle s' \right| U^{J\dagger}_{MN} \left| s \right\rangle \equiv \begin{gathered}\includegraphics[scale=0.1]{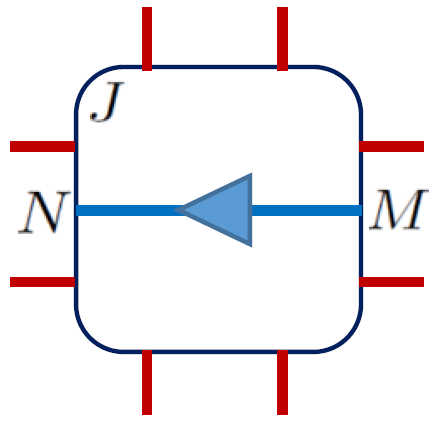}\end{gathered}\\
\left(\left[T_{\downarrow}\right]^{J}_{MN}\right)_{ll',rr';dd',uu'} & = \text{Tr}_{\text{phys}}\left[A^{st}_{ruld} U^{tJ\dagger}_{MN} \left|st\right\rangle \left\langle s't' \right| \bar{A}^{s't'}_{r'u'l'd'}\right]
= A^{st}_{ruld}  \bar{A}^{st'}_{r'u'l'd'} \left\langle t' \right| U^{J\dagger}_{MN} \left|t \right\rangle \equiv \begin{gathered}\includegraphics[scale=0.1]{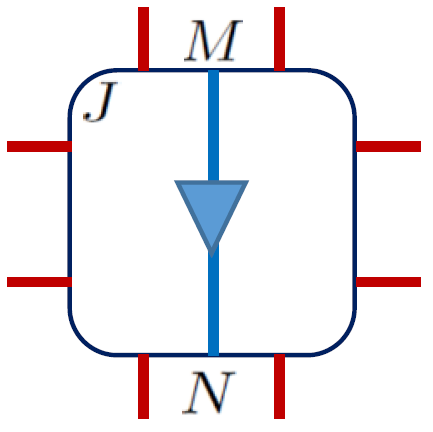}\end{gathered} 
	\end{aligned}
\end{equation}

Using the symmetry conditions (\ref{Atrans}) as well as the transformation properties of the group element operators (\ref{Utrans}), we obtain that for every $g \in G$,
\begin{equation}
	\begin{aligned}
		\left(\theta^l_g \otimes \tilde{\theta}^{\dagger l'}_g\right) \otimes \left(\theta^d_g \otimes \tilde{\theta}^{\dagger d'}_g\right) \left[\hat T_{\rightarrow}\right]^{J}_{MN} &= D^J_{MM'}\left(g^{-1}\right)\left[\hat T_{\rightarrow}\right]^{J}_{M'N}  \\
		\left[\hat T_{\rightarrow}\right]^{J}_{MN}\left(\theta^r_g \otimes \tilde{\theta}^{\dagger r'}_g\right)  &= \left[\hat T_{\rightarrow}\right]^{J}_{MN'}	D^J_{N'N}\left(g^{-1}\right) \\
		\left[\hat T_{\rightarrow}\right]^{J}_{MN}\left(\theta^u_g \otimes \tilde{\theta}^{\dagger u'}_g\right)  &= \left[\hat T_{\rightarrow}\right]^{J}_{MN}	
		\label{Th}
	\end{aligned}
\end{equation}
\end{widetext}
(see Fig. \ref{Tst_rules}(a)). That is, $\left[\hat T_{\rightarrow}\right]^{J}_{MN}$ maps from a total $\left\langle JM\right|$ on both ingoing legs (with respect to $\left(\theta^l_g \otimes \tilde{\theta}^{\dagger l'}_g\right) \otimes \left(\theta^d_g \otimes \tilde{\theta}^{\dagger d'}_g\right)$) onto $\left\langle JN\right|$ with respect to $\left(\theta^r_g \otimes \tilde{\theta}^{\dagger r'}_g\right)$ on the outgoing horizontal leg and a singlet with respect to $\left(\theta^u_g \otimes \tilde{\theta}^{\dagger u'}_g\right)$ on the outgoing vertical leg (see Fig. \ref{Tst_map}(a)). As in the flux-free case, that will have implications on the structure of the $\left[\hat T_{\rightarrow}\right]^{J}_{MN}$ operators.

Furthermore, the transfer operators  $\left[\hat T_{\rightarrow}\right]^{J}_{MN}$ form a multiplet for each $J$, whose elements are mixed by the transformations. There is no problem with that, because in the contraction of the Wilson loop we sum over the $M,N$ indices (matrix product and tracing of the $U$ matrices). As usual, in the Abelian case the multiplets are trivial and contain one operator only, allowing us to give an intuitive illustration. For example, let us consider $U(1)$ with the fundamental representation $j=1$; there, the transformations take the simple form
\begin{equation}
	\begin{aligned}
		e^{i\phi\left(E^l-E^{l'}+E^d-E^{d'}\right)} \hat T_{\rightarrow} &= e^{-i\phi}\hat T_{\rightarrow}  \\
		\hat T_{\rightarrow}	e^{i\phi\left(E^r-E^{r'}\right)} &= 	 e^{-i\phi}\hat T_{\rightarrow} \\
		\hat T_{\rightarrow}e^{i\phi\left(E^u-E^{u'}\right)}  &= \hat T_{\rightarrow}	
	\end{aligned}
\end{equation}

For the inverse horizontal flux line, one obtains
\begin{widetext}
\begin{equation}
	\begin{aligned}
		\left(\theta^l_g \otimes \tilde{\theta}^{\dagger l'}_g\right) \otimes \left(\theta^d_g \otimes \tilde{\theta}^{\dagger d'}_g\right) \left[\hat T_{\leftarrow}\right]^{J}_{MN} &= \left[\hat T_{\leftarrow}\right]^{J}_{MN'} D^J_{N'N}\left(g\right)  \\
		\left[\hat T_{\leftarrow}\right]^{J}_{MN}\left(\theta^r_g \otimes \tilde{\theta}^{\dagger r'}_g\right)  &= D^J_{MM'}\left(g\right)\left[\hat T_{\leftarrow}\right]^{J}_{M'N}	 \\
		\left[\hat T_{\leftarrow}\right]^{J}_{MN}\left(\theta^u_g \otimes \tilde{\theta}^{\dagger u'}_g\right)  &= \left[\hat T_{\leftarrow}\right]^{J}_{MN}	
	\end{aligned}
\label{Thd}
\end{equation}
(see Fig. \ref{Tst_rules}(b)) - the difference from the right going flux is not very big, and has to do mainly on the opposite flux orientation: $g$ instead of $g^{-1}$ appears in the transformation, and the beginning index $M$ is now associated with the right side rather than the left (similarly, $N$ with the left rather than the right), since the flux goes backwards. This corresponds to transposition, and since the representations are unitary, $D^{j}_{nm}\left(g\right)=\overline{D^{j}_{mn}\left(g^{-1}\right)}$ - i.e., the conjugate representation $J$. As a result, we denote the input of both legs as $\left\langle \overline{JN}\right|$ and the output of the right leg as $\left\langle \overline{JM}\right|$ - vectors with a conjugate transformation rule (see Fig. \ref{Tst_map}(b)).

In the vertical direction, we have
\begin{equation}
	\begin{aligned}
		\left(\theta^l_g \otimes \tilde{\theta}^{\dagger l'}_g\right) \otimes \left(\theta^d_g \otimes \tilde{\theta}^{\dagger d'}_g\right) \left[\hat T_{\uparrow}\right]^{J}_{MN} &= D^J_{MM'}\left(g^{-1}\right)\left[\hat T_{\uparrow}\right]^{J}_{M'N}  \\
		\left[\hat T_{\uparrow}\right]^{J}_{MN}\left(\theta^r_g \otimes \tilde{\theta}^{\dagger r'}_g\right)  &= \left[\hat T_{\uparrow}\right]^{J}_{MN}	 \\
		\left[\hat T_{\uparrow}\right]^{J}_{MN}\left(\theta^u_g \otimes \tilde{\theta}^{\dagger u'}_g\right)  &= \left[\hat T_{\uparrow}\right]^{J}_{MN'}	D^J_{N'N}\left(g^{-1}\right)
	\end{aligned}
\label{Tv}
\end{equation}
(Fig. \ref{Tst_rules}(c)) and
\begin{equation}
	\begin{aligned}
		\left(\theta^l_g \otimes \tilde{\theta}^{\dagger l'}_g\right) \otimes \left(\theta^d_g \otimes \tilde{\theta}^{\dagger d'}_g\right) \left[\hat T_{\downarrow}\right]^{J}_{MN} &= \left[\hat T_{\downarrow}\right]^{J}_{MN'} D^J_{N'N}\left(g\right)  \\
		\left[\hat T_{\downarrow}\right]^{J}_{MN}\left(\theta^r_g \otimes \tilde{\theta}^{\dagger r'}_g\right)  &= \left[\hat T_{\downarrow}\right]^{J}_{MN}	 \\
		\left[\hat T_{\downarrow}\right]^{J}_{MN}\left(\theta^u_g \otimes \tilde{\theta}^{\dagger u'}_g\right)  &= D^J_{MM'}\left(g\right)\left[\hat T_{\downarrow}\right]^{J}_{M'N}	
	\end{aligned}
\label{Tvd}
\end{equation}
\end{widetext}
(Fig. \ref{Tst_rules}(d)). The input/output pictures, when looking at these operators as maps, are shown in Fig. \ref{Tst_map}(c,d).
Note that when plugging the trivial representation into any of the results for straight flux lines, that is $J=M=N=0$,  $\hat T$ is obtained.

\begin{figure}
	\includegraphics[width=\columnwidth]{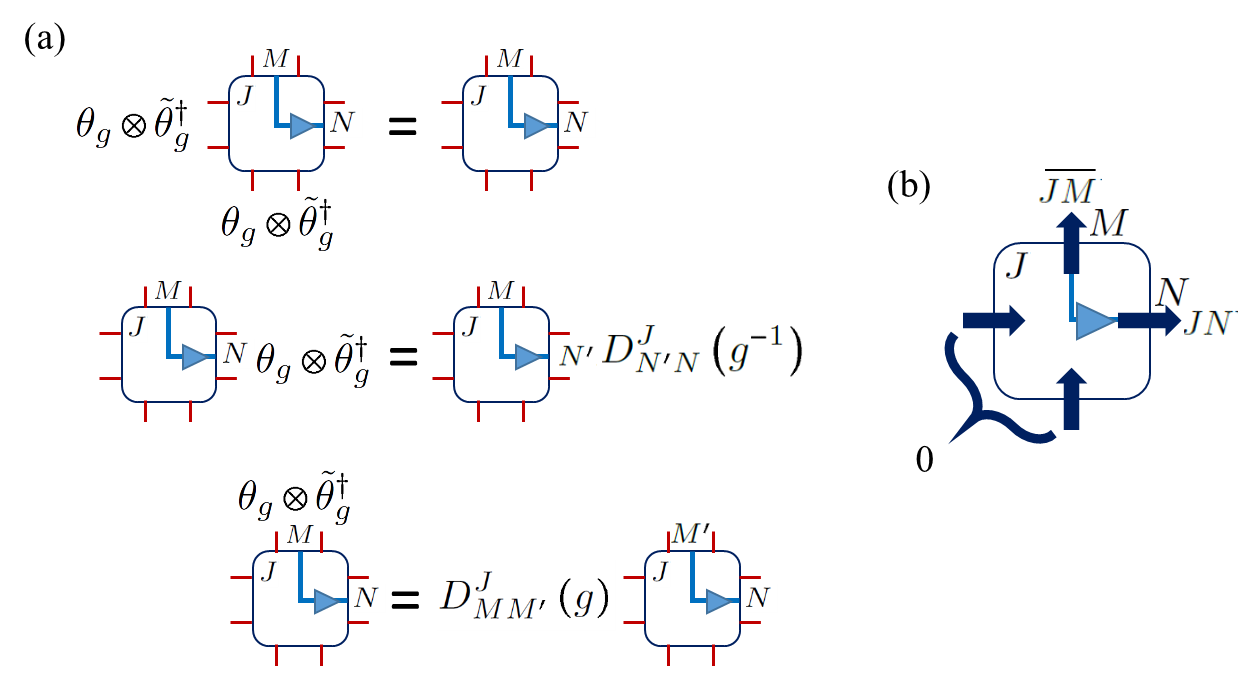}
	\caption{The lower left corner transfer operator: (a) transformation rules and (b) as a map.}
	\label{Tcorner}
\end{figure}

There are many other options to consider, in which flux line(s) go through a site. Here we only look at the ones required for our counter-clockwise Wilson loop contraction, which implies naively that four further types of transfer operators, for the corners, are required. However, we only need one, as we shall see shortly when tiling the loop,
\begin{equation}
	\begin{aligned}
		&\left(\left[\hat T_{\searrow}\right]^{J}_{MN}\right)_{ll',rr';dd',uu'}  \\&= \text{Tr}_{\text{phys}}\left[A^{st}_{ruld} U^{tJ\dagger}_{MK}U^{sJ}_{KN} \left|st\right\rangle \left\langle s't' \right| \bar{A}^{s't'}_{r'u'l'd'}\right]
		\\&= A^{st}_{ruld}  \bar{A}^{s't'}_{r'u'l'd'} \left\langle t' \right| U^{J\dagger}_{MK} \left| t \right\rangle \left\langle s' \right| U^{J}_{KN} \left| s \right\rangle   \equiv \begin{gathered}\includegraphics[scale=0.1]{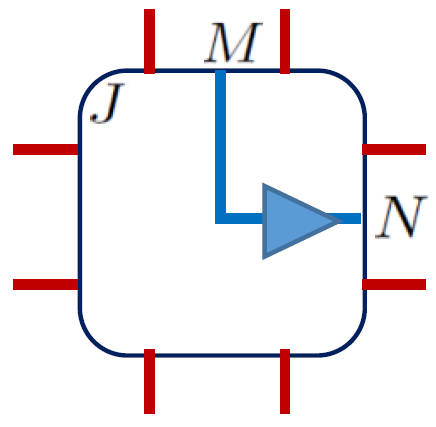}\end{gathered} 
		\end{aligned}
\end{equation}
Its transformation properties may be similarly derived, resulting in
\begin{widetext}
\begin{equation}
	\begin{aligned}
		\left(\theta^l_g \otimes \tilde{\theta}^{\dagger l'}_g\right) \otimes \left(\theta^d_g \otimes \tilde{\theta}^{\dagger d'}_g\right) \left[\hat T_{\searrow}\right]^{J}_{MN} &= \left[\hat T_{\searrow}\right]^{J}_{MN} \\
		\left[\hat T_{\searrow}\right]^{J}_{MN}\left(\theta^r_g \otimes \tilde{\theta}^{\dagger r'}_g\right)  &= \left[\hat T_{\searrow}\right]^{J}_{MN'}	D^J_{N'N}\left(g^{-1}\right)\\
		\left[\hat T_{\searrow}\right]^{J}_{MN}\left(\theta^u_g \otimes \tilde{\theta}^{\dagger u'}_g\right)  &= D^J_{MM'}\left(g\right)\left[\hat T_{\searrow}\right]^{J}_{M'N}		
	\end{aligned}
\end{equation}
\end{widetext}
- the ingoing legs form a combined singlet, while both the outgoing legs, separately, belong to the $J$ representation (one regular, one conjugate) - see Fig. \ref{Tcorner}. 

\subsection{Tiling the loop and projecting onto smaller spaces}

Do we need to use all the elements of the transfer operators for the Wilson loop contraction? The answer is no; we can ignore some of them in the computation, while tiling the different building blocks together, thanks to the local symmetry and the special properties it enforces on the states and the transfer operators, just like we did in the case of the norm.
As discussed, each of the local transfer operators used for the contraction, either with or without flux, can be seen as a map between the two ingoing legs to the two outgoing ones. While the ingoing legs form \emph{together} a multiplet vector of the group, the output is a product of two separate multiplet vectors on the two outgoing legs (see Fig. \ref{Trules}(b), \ref{Tst_map} and \ref{Tcorner}(b)). The numerator of the Wilson loop expectation value requires a particular tiling of the transfer operators, closing the loop. Since the output to each direction forms a multiplet vector, this will also be the input of the neighbouring transfer operators in the outgoing directions, and we can restrict all our transfer operators by cutting off all the input options that could not be realized within the Wilson loop tiling. This is done in a very similar way to what did in the norm computation, where we defined $\hat{\tau}_0$ (\ref{tau0}) instead of $\hat T$. 

Since our system is translationally invariant, let us identify the lower left corner of the loop with the lower left corner of our system. Let us consider the numerator of the expectation value of the Wilson loop:
\begin{widetext}
\begin{equation}
	\left\langle \psi \right| W \left|\psi\right\rangle  =
	\left\langle \psi \right| \begin{gathered}\includegraphics[scale=0.2]{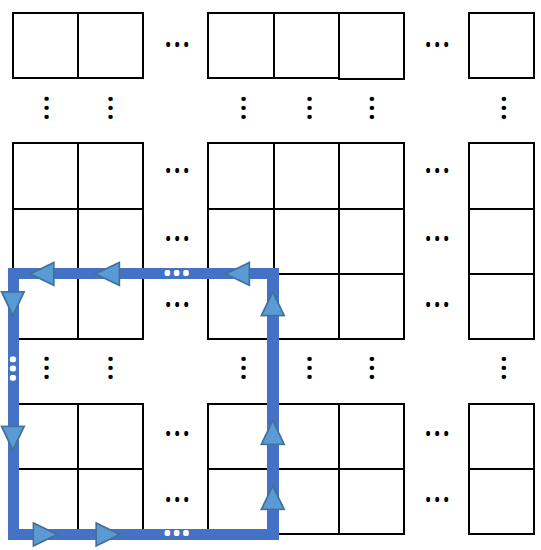}\end{gathered}  \left|\psi\right\rangle
	 =
	\text{Tr}\left[ \begin{gathered}\includegraphics[scale=0.3]{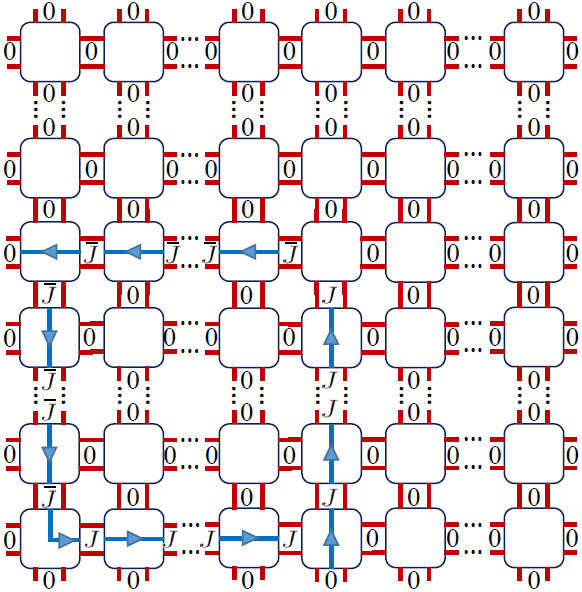}\end{gathered}  \right]
\end{equation}
\end{widetext}
Where the trace is on both directions, assuming periodic boundaries (similar results may be easily derived for open boundary conditions); the $J,M,N$ indices of the flux-carrying transfer operators have been omitted for simplicity, but it is assumed that they all carry the same irrep $J$ (otherwise it would make no physical sense) and that the $M,N$ indices are properly connected and summed over along the loop. Using the mapping properties summarized in figures \ref{Trules}(b), \ref{Tst_map} and \ref{Tcorner}(b), we can write on each of the outgoing legs its output representation - $0$, $J$ or $\overline{J}$ for the conjugate representations used in the backwards fluxes cases. This immediately determines onto which inputs the transfer operators should be projected.
Note that the lower right and both upper corners do not seem right in the equation above; nevertheless these are the right ingredients to be used, as explained below.

The tiling is composed of the following ingredients:
\begin{itemize}
	\item  Outside of the loop and within it, on sites through which no flux lines pass, we  use the flux-free transfer operator $\hat T$. They only receive $0$ as inputs, and thus may be replaced by $\hat{\tau}_0$  from Eq. (\ref{tau0def}) in all these places.
	\item On the lower left corner, we use the $\left[\hat T_{\searrow}\right]^{J}$, which, thanks to receiving $0$ inputs on both directions from $\hat{\tau}_0$ operators, may be replaced by
\begin{equation}
	\left[\hat \tau_{\searrow}\right]^{J}_{MN}=\Pi_0 \otimes \Pi_0 \left[\hat T_{\searrow}\right]^{J}_{MN}
	\equiv \begin{gathered}\includegraphics[scale=0.3]{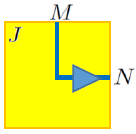}\end{gathered} 
\end{equation}

	\item Along the lower edge, until the next corner, we  use $\left[\hat T_{\rightarrow}\right]^{J}$. As the input of these operators is $J$ from the left and $0$ from below, they may be replaced by
\begin{equation}
	\left[\hat \tau_{\rightarrow}\right]^{J}_{MN}=\Pi_{JM} \otimes \Pi_0 \left[\hat T_{\rightarrow}\right]^{J}_{MN}	
	\equiv \begin{gathered}\includegraphics[scale=0.3]{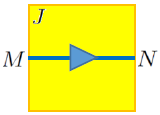}\end{gathered} 
\end{equation}	
using the on-leg projector $\Pi_{JM} = \underset{j,j'}{\sum}\left|JM\left(j,j'\right)\right\rangle\left\langle JM\left(j,j'\right)\right|$ (no summation on $M$) and defining
	\begin{equation}
		\left|JM \left(j,j'\right)\right\rangle = \left\langle J M j m | j' m' \right\rangle \left| j m j' m' \right\rangle 
\end{equation}

	\item When turning upwards, in the lower right corner, we  use $\left[\hat T_{\uparrow}\right]^J$: our tensors only contain physical degrees of freedom on the outgoing links; at this site the only physical leg carrying flux is the one pointing upwards  and therefore this is the relevant transfer operator. Its input allows us to restrict it to
\begin{equation}
	\left[\hat \tau_{\nearrow}\right]^{J}_{MN}=\Pi_{JM} \otimes \Pi_0 \left[\hat T_{\uparrow}\right]^{J}_{MN}
	\equiv \begin{gathered}\includegraphics[scale=0.3]{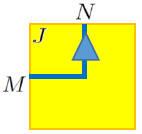}\end{gathered} 
\end{equation}
	
	\item We  go along with $\left[\hat T_{\uparrow}\right]^J$ all the way up until the top right corner, but with different input, introducing 
	\begin{equation}
		\left[\hat \tau_{\uparrow}\right]^{J}_{MN}=\Pi_0 \otimes \Pi_{JM} \left[\hat T_{\uparrow}\right]^{J}_{MN}
		\equiv \begin{gathered}\includegraphics[scale=0.3]{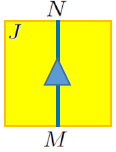}\end{gathered} 
	\end{equation}

	\item At the upper right corner, the fluxes only come from the ingoing legs, and therefore the relevant transfer operator is once again $\hat{T}$, projected this time onto
	\begin{equation}
		\left[\hat \tau_{\nwarrow}\right]^{J}_{MN}=\Pi_{\overline{JN}} \otimes \Pi_{JM} \hat T
	\equiv \begin{gathered}\includegraphics[scale=0.3]{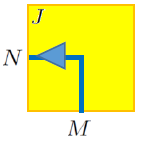}\end{gathered} 
\end{equation}
	where $\Pi_{\overline{JN}} = \underset{j,j'}{\sum}\left|\overline{JN}\left(j,j'\right)\right\rangle\left\langle \overline{JN}\left(j,j'\right)\right|$ (no summation on $N$) and
		\begin{equation}
			\left|\overline{JM} \left(j,j'\right)\right\rangle  = \left\langle J M j' m' | j m \right\rangle \left| j m j' m' \right\rangle
	\end{equation}

	\item All the way to the left  we proceed with $\left[\hat T_{\leftarrow}\right]^J$ along the upper edge. Until the next corner - and without including it, it can be replaced by
	\begin{equation}
	\left[\hat \tau_{\leftarrow}\right]^{J}_{MN}=\Pi_{\overline{JN}} \otimes \Pi_0 \left[\hat T_{\leftarrow}\right]^{J}_{MN}
	\equiv \begin{gathered}\includegraphics[scale=0.3]{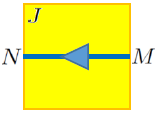}\end{gathered} 
\end{equation}

	\item At the upper left corner we still use $\left[\hat T_{\leftarrow}\right]^J$ but with different inputs, projecting it to
	\begin{equation}
	\left[\hat \tau_{\swarrow}\right]^{J}_{MN}=\Pi_0 \otimes \Pi_{\overline{JN}}\left[\hat T_{\leftarrow}\right]^J_{MN}
\equiv \begin{gathered}\includegraphics[scale=0.3]{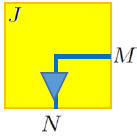}\end{gathered} 
\end{equation}
	
	\item Finally, we  go down with $\left[\hat T_{\downarrow}\right]^J$ all the way to the starting point, restricting it to
		\begin{equation}
			\left[\hat \tau_{\downarrow}\right]^{J}_{MN}=\Pi_0 \otimes \Pi_{\overline{JN}} \left[\hat T_{\downarrow}\right]^{J}_{MN}
		\equiv \begin{gathered}\includegraphics[scale=0.3]{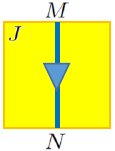}\end{gathered} 
	\end{equation}
\end{itemize}

Just like in the case of $\hat \tau_0$ compared with $\hat T$, these newly introduced operators contain less tensor elements and simplify the contraction of the Wilson loop, 
\begin{equation}
	\left\langle \psi \right| W \left|\psi\right\rangle  =
	\text{Tr}\left[ \begin{gathered}\includegraphics[scale=0.4]{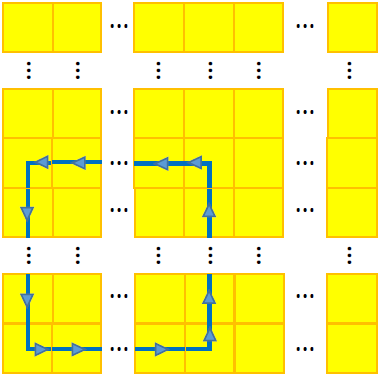}\end{gathered}  \right]
\end{equation}

\subsection{The decay of Wilson loops: is an area law possible?} \label{ssc}

Now we have all the ingredients required for the computation of a Wilson loop whose dimensions are $R_1 \times R_2$, and compute it using row transfer matrices, by contracting first in the horizontal direction, within an ${\mathcal{N}} \times {\mathcal{N}}$ system with periodic boundary conditions (torus).

We denote the transfer matrix corresponding to the first row we contract (the one containing the lower edge of the loop) by 
	\begin{equation}
	\begin{aligned}
		\left[\hat{E}_b\right]^J_{MN}\left(R\right) &\equiv  
		\begin{gathered}\includegraphics[scale=0.35]{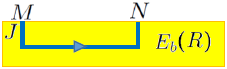}\end{gathered}
		\\&=	\text{Tr}_{\text{row}}\left[ \begin{gathered}\includegraphics[scale=0.35]{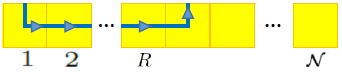}\end{gathered}  \right]	
	\end{aligned}
\end{equation}
on top of it, there will be $R_2-1$ rows with parallel vertical flux lines, represented by
	\begin{equation}
	\begin{aligned}
		\left[\hat{E}_{\parallel}\right]^J_{M_l N_l M_r N_r}\left(R\right) &\equiv  
		\begin{gathered}\includegraphics[scale=0.35]{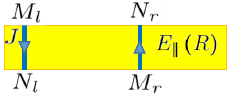}\end{gathered}
		\\&=	\text{Tr}_{\text{row}}\left[ \begin{gathered}\includegraphics[scale=0.35]{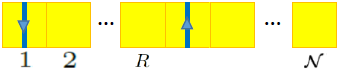}\end{gathered}  \right]		
		\end{aligned}
\end{equation}
 the top of the loop is represented the row transfer matrix we define by
 	\begin{equation}
 	\begin{aligned}
 		\left[\hat{E}_t\right]^J_{MN}\left(R\right)  &\equiv  
 		\begin{gathered}\includegraphics[scale=0.35]{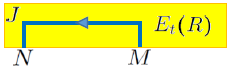}\end{gathered}
 		\\&=	\text{Tr}_{\text{row}}\left[ \begin{gathered}\includegraphics[scale=0.35]{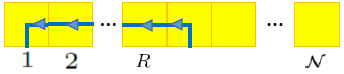}\end{gathered}  \right]	
 	\end{aligned}
 \end{equation}
 
 and all the remaining rows simply contribute $\hat{E}$ (we omit the $J,M,N$ indices for simplicity, assuming some given $J$ for the Wilson loop, and implicitly contracting over the $M,N$ indices).
The expectation value of the Wilson loop may then be written as 
\begin{equation}
\begin{aligned}
&\left\langle W\left(R_1,R_2\right)\right\rangle
=  
	\frac{\text{Tr}\left[\begin{gathered}
		\includegraphics[scale=0.5]{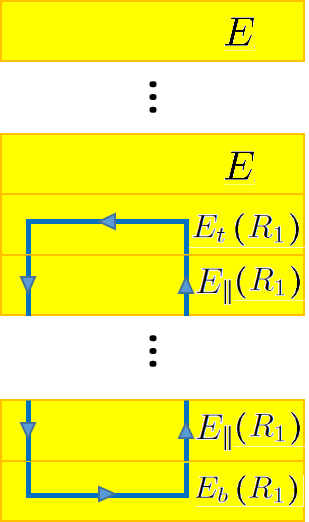}\end{gathered}\right]}
	{\text{Tr}\left[\begin{gathered}
	\includegraphics[scale=0.3]{Norm_E.png}\end{gathered}\right]}
	 = 
	\\&\frac{\text{Tr}\left[\hat E_b\left(R_1\right)\hat E^{R_2-1}_{\parallel}\left(R_1\right)\hat E_t\left(R_1\right)\hat E^{{\mathcal{N}}-R_2-1}\right]}
	{\text{Tr}\left[E^{\mathcal{N}}\right]}
\label{W12}
\end{aligned}
\end{equation}
It is very similar to the MPS expression used for computing correlation functions (\ref{MPScorr}) with one major difference. Due to the local symmetry, in between the two rows closing the loop, we need to use a different transfer matrix, $\hat E_{\parallel}$: the long range decay properties depend now two different transfer matrices, instead of one. 

As stated in the beginning of this subsection, we have omitted the $M,N$ indices and we assume implicit summation over them when contracting the loop. The Wilson loop contraction consists of the contraction of $2\left(R_1+R_2\right)$ indices, each taking $\text{dim}\left(J\right)$ values - naively speaking, we would have to consider $\text{dim}^{2\left(R_1+R_2\right)}\left(J\right)$ different contractions; however, the singular values are independent of these indices and depend only on the irrep $J$. Thanks to this symmetry, all the $\text{dim}^{2\left(R_1+R_2\right)}\left(J\right)$ are equal, so it is enough to make one choice of the indices and multiply the result by  $\text{dim}^{2\left(R_1+R_2\right)}\left(J\right)$. This will be a perimeter-law term, however in the presence of an area law term it will not contribute in the large loop limit. Hence we focus below on computing for one particular choice of the indices. 

Consider the diagonalization of the two transfer matrices which matter for the long range properties,
\begin{equation}
\begin{aligned}
\hat E &= \underset{i}{\sum}\rho_i \left|v_i\right\rangle  \left\langle w_i\right| \\
\hat E_{\parallel}\left(R\right) &= \underset{i}{\sum}\rho'_i\left(R\right) \left|v'_i\left(R\right)\right\rangle  \left\langle w'_i\left(R\right)\right|
\end{aligned}
\end{equation}
Once again, we sort the eigenvalues in decreasing order, but in this case we do not care if the highest one is degenerate (but assume the existence of a spectral gap): for some integers $K,K'\geq 1$,
$\left|\rho_1\right| = ... = \left|\rho_K\right| > \left|\rho_{K+1}\right| \geq \left|\rho_{K+2}\right| \geq ...$ and
$\left|\rho'_1\left(R\right)\right| = ... = \left|\rho'_{K'}\left(R\right)\right| > \left|\rho'_{K'+1}\left(R\right)\right| \geq \left|\rho'_{K'+2}\left(R\right)\right| \geq ...$ 

Let us use this to compute the expectation value of the Wilson loop (\ref{W12})  in the thermodynamic limit ${\mathcal{N}} \gg R_2$:
\begin{widetext}
\begin{equation}
\begin{aligned}
\left\langle W\left(R_1,R_2\right)\right\rangle &= \text{dim}^{2\left(R_1+R_2\right)}\left(J\right)\frac
{\rho_1^{N-R_2-1}\text{Tr}\left[\left(\underset{i=1}{\overset{K}{\sum}}\left|v_i\right\rangle  \left\langle w_i\right|
	+\underset{i>K}{\sum}\left(\frac{\rho_i}{\rho_1}\right)^{N-R_2-1} \left|v_i\right\rangle  \left\langle w_i\right|\right)E_b\left(R_1\right)E^{R_2-1}_{\parallel}\left(R_1\right)E_t\left(R_1\right)\right]}
{\rho_1^{N}\text{Tr}\left[\underset{i=1}{\overset{K}{\sum}}\left|v_i\right\rangle  \left\langle w_i\right|
	+\underset{i>K}{\sum}\left(\frac{\rho_i}{\rho_1}\right)^{{\mathcal{N}}} \left|v_i\right\rangle  \left\langle w_i\right|\right]}\\
&\rightarrow \frac{\text{dim}^{2\left(R_1+R_2\right)}\left(J\right)}{K\rho_1^{R_2+1}}\underset{i=1}{\overset{K}{\sum}}\left\langle w_i\right|
\hat E_b\left(R_1\right)\hat E^{R_2-1}_{\parallel}\left(R_1\right)\hat E_t\left(R_1\right)\left|v_i\right\rangle 
\end{aligned}
\end{equation}
(we assumed that $\rho_1 = ... = \rho_K$; the generalization for the case of different phases is straightforward).

We further assume that the loop is large, that is - $R_1,R_2 \gg 1$, allowing us to perform a similar simplification for $E_{\parallel}$, and obtain that in the thermodynamic limit, for large loops,
\begin{equation}
\left\langle W\left(R_1,R_2\right)\right\rangle
\rightarrow \text{dim}^{2\left(R_1+R_2\right)}\left(J\right)
\frac{\rho'^{R_2-1}_1\left(R_1\right)}{K\rho_1^{R_2+1}}
\underset{i=1}{\overset{K}{\sum}}
\underset{j=1}{\overset{K'}{\sum}}
\left\langle w_i\right| \hat E_b \left(R_1\right) \left|v'_j\left(R_1\right)\right\rangle
\left\langle w'_j \left(R_1\right) \right| \hat E_t \left(R_1\right)\left| v_i \right\rangle
\label{formula}
\end{equation}
(This holds only if $\underset{i=1}{\overset{K}{\sum}}
\underset{j=1}{\overset{K'}{\sum}}
\left\langle w_i\right| \hat E_b \left(R_1\right) \left|v'_j\left(R_1\right)\right\rangle
\left\langle w'_j \left(R_1\right) \right| \hat E_t \left(R_1\right)\left| v_i \right\rangle \neq 0$; if this condition is not fulfilled, the vectors $\left|v'_j\right\rangle$ and $\left\langle w'_i\right|$ should not be seen as those corresponding to the highest eigenvalues, but rather as those with the highest eigenvalues for which this condition is satisfied. We assumed here that $\rho'_1 = ... = \rho'_{K'}$; the generalization for the case of different phases is straightforward).

Assuming rotational invariance, we could repeat the same procedure by contracting the columns first, to obtain
\begin{equation}
	\left\langle W\left(R_1,R_2\right)\right\rangle=  
	\frac{\text{Tr}\left[\begin{gathered}
			\includegraphics[scale=0.3]{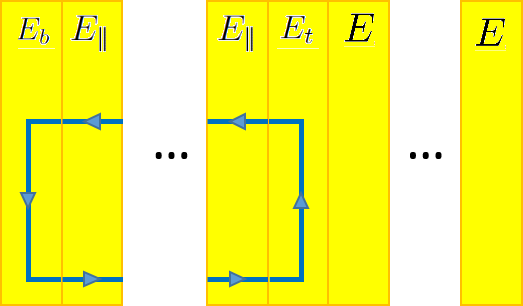}\end{gathered}\right]}
	{\text{Tr}\left[\begin{gathered}
			\includegraphics[scale=0.3]{Norm_E_rot.png}\end{gathered}\right]}
	\rightarrow \text{dim}^{2\left(R_1+R_2\right)}\left(J\right)
	\frac{\rho'^{R_1-1}_1\left(R_2\right)}{K\rho_1^{R_1+1}}
	\underset{i=1}{\overset{K}{\sum}}
	\underset{j=1}{\overset{K'}{\sum}}
	\left\langle w_i\right|\hat  E_b \left(R_2\right) \left|v'_j\left(R_2\right)\right\rangle
	\left\langle w'_j \left(R_2\right) \right| \hat E_t \left(R_2\right)\left| v_i \right\rangle
	\label{Wexp1}
\end{equation}

Both expressions must be equal; therefore, we deduce that
\begin{equation}
	\rho'^{R_2-1}_1\left(R_1\right) \underset{i=1}{\overset{K}{\sum}}
	\underset{j=1}{\overset{K'}{\sum}}
	\left\langle w_i\right| \hat E_b \left(R_1\right) \left|v'_j\left(R_1\right)\right\rangle
	\left\langle w'_j \left(R_1\right) \right|\hat  E_t \left(R_1\right)\left| v_i \right\rangle
	\propto
	\frac{1}{\rho_1^{R_1+1}}
	\label{Wexp2}
	\end{equation}

But the more interesting question is whether $\partial \rho'_1\left(R\right) / \partial R =0$ or not. If the largest eigenvalue of $\hat E_{\parallel}\left(R\right)$ does not depend on $R$, we obtain that
\begin{equation}
\left\langle W\left(R_1,R_2\right)\right\rangle
\rightarrow
\tilde C \left(\text{dim}^2\left(J\right)\frac{\rho'_1}{\rho_1}\right)^{R_1+R_2}
\end{equation}
with some constant $\tilde C$: perimeter law decay of the Wilson loop (unless $\rho_1 = \rho'_1$). On the other hand, an area law is possible if
\begin{equation}
\rho'_1\left(R\right) \sim \Gamma e^{-\kappa R}
\label{rhogammakappa}
\end{equation}
with $\kappa > 0$. Let us plug this expression into (\ref{Wexp1}) and (\ref{Wexp2}). We will obtain the equation
\begin{equation}
	\begin{aligned}
	\left\langle W\left(R_1,R_2\right)\right\rangle
	&\rightarrow
	\frac{\text{dim}^{2\left(R_1+R_2\right)}\left(J\right)}{K\Gamma\rho_1}\left(\frac{\Gamma}{\rho_1 e^{-\kappa}}\right)^{R_1} e^{-\kappa R_1 R_2} 
		\underset{i=1}{\overset{K}{\sum}}
	\underset{j=1}{\overset{K'}{\sum}}
	\left\langle w_i\right|\hat  E_b \left(R_2\right) \left|v'_j\left(R_2\right)\right\rangle
	\left\langle w'_j \left(R_2\right) \right| \hat E_t \left(R_2\right)\left| v_i \right\rangle \\
	&=
		\frac{\text{dim}^{2\left(R_1+R_2\right)}\left(J\right)}{K\Gamma\rho_1}\left(\frac{\Gamma}{\rho_1 e^{-\kappa}}\right)^{R_2} e^{-\kappa R_1 R_2} 
	\underset{i=1}{\overset{K}{\sum}}
	\underset{j=1}{\overset{K'}{\sum}}
	\left\langle w_i\right|\hat  E_b \left(R_1\right) \left|v'_j\left(R_1\right)\right\rangle
	\left\langle w'_j \left(R_1\right) \right|\hat  E_t \left(R_1\right)\left| v_i \right\rangle
	\end{aligned}
\end{equation}
Rotation invariance guarantees that
\begin{equation}
\underset{i=1}{\overset{K}{\sum}}
\underset{j=1}{\overset{K'}{\sum}}
\left\langle w_i\right| \hat  E_b \left(R\right) \left|v'_j\left(R\right)\right\rangle
\left\langle w'_j \left(R\right) \right| \hat E_t \left(R\right)\left| v_i \right\rangle \sim C \left(\frac{\Gamma}{\rho_1 e^{-\kappa}}\right)^{R}
\label{eq71}
\end{equation}
\end{widetext}
for some constant $C$, and we and obtain, finally, for large Wilson loops, that if $\rho'_1\left(R\right) \sim \Gamma e^{-\kappa R}$,
\begin{equation}
\left\langle W\left(R_1,R_2\right)\right\rangle
\rightarrow
\frac{C}{K\Gamma\rho_1}\left(\frac{\Gamma \text{dim}^{2}\left(J\right)}{\rho_1 e^{-\kappa}}\right)^{R_1+R_2} e^{-\kappa R_1 R_2} 
\end{equation}
- exactly the same form of (\ref{Wlarge}), with $W_0 = \frac{C}{K\Gamma\rho_1}$, $\kappa_A=\kappa$ and $\kappa_P=\log\left(\frac{\rho_1}{\Gamma\text{dim}^{2}\left(J\right)}\right)-\kappa$.

Therefore, we conclude that a perimeter law will be obtained if the largest relevant 
(in terms of accessible through $\hat{E}_{b}$ and $\hat{E}_{t}$) eigenvalue of $\hat E_{\parallel}\left(R\right)$ is independent of $R$; an area law is \emph{possible} if it depends on $R$ exponentially. Why only possible? To see why this condition is necessary but not sufficient for the area law to hold, let us consider the following scenario.

Previously, we made the assumption that the eigenvectors of the flux-free transfer matrix should be close to product vectors in order to make an area law possible. We also know that the expectation value of the Wilson loop depends on the zeroth flux transfer operators $\hat{\tau}_0$ inside and outside the loop, and some other, flux-carrying transfer operators along the loop. 
Let us assume that we are, indeed, in a scenario in which the eigenvalues of the transfer matrix are close to product states. Denote as usual the highest eigenvalue of the transfer operator by $\lambda_1$ . Then the norm, for a large system, will roughly scale as $\lambda_{1}^{\mathcal{N}^2}$: each site contributes a single power of $\lambda_1$. This is the denominator of the expectation value formula. In the numerator, we will have a contribution of  $\lambda_1$ for each site outside the loop; within the loop, it depends. 

If the flux carrying transfer operators along the loop take us from the singlet subspace corresponding to $\lambda_1$ to that of another eigenvalue - denote it by $\lambda'$ - we will have a contribution of $\lambda'$ for each of the sites within the loop, and the Wilson loop's expectation value will scale as $\left(\lambda' / \lambda_1\right)^A$ where $A$ is the area of the loop ($E_{\parallel}\left(R\right) \propto \lambda'^R$). However, if the flux carrying transfer operators do not take us to another singlet subspace \emph{with a different eigenvalue}, we will not have an area dependent contribution. In this case, the largest eigenvalue of $\hat E_{\parallel}\left(R\right)$ depends exponentially on $R$ (through $\lambda'^R$) but an area law is not obtained, which shows us why this condition is necessary but not sufficient.

On the other hand, if the eigenvectors of $\hat{E}$ are far from product vectors, which means they are governed by some collective, long range effect, we cannot have area-dependent contributions at all.

\section{Illustration: the $\mathbb{Z}_2$ case}

To conclude and illustrate our discussion, we will show an explicit example, where the gauge group is $\mathbb{Z}_2$. In this case, the group Hilbert space on each link is two dimensional, with representations labelled by $j=+,-$, which can be simply seen as spins. The group element operators are Hermitian, $U=U^{\dagger}=X$, and invert the spin,
\begin{equation}
	X\left|\pm\right\rangle = \left|\mp\right\rangle
\end{equation}
and the group operations $\Theta$ (no difference between left and right in Abelian groups) are the identity operator as well as
\begin{equation}
	Z\left|\pm\right\rangle = \pm \left|\pm\right\rangle
\end{equation}
Gauge transformations are given by
\begin{equation}
	\hat{\Theta}\left(\mathbf{x}\right) = 
	Z\left(\mathbf{x},1\right)
	Z\left(\mathbf{x},2\right)
	Z\left(\mathbf{x}-\hat{\mathbf{e}}_1,1\right)
	Z\left(\mathbf{x}-\hat{\mathbf{e}}_1,2\right)
	\label{Theta2}
\end{equation}

We would like to consider the most general PEPS with translational and rotational invariance, with physical spaces containing all the irreps and virtual ones containing a single copy of each irrep (minimal construction - as explained above, to consider real physical scenarios one will most likely have to generalize in a straight forward manner and add more copies, as was necessary in the $\mathbb{Z}_3$ demonstration of Ref. \cite{emonts_gauss_2018}). Thus, the physical and virtual spaces will be the same, two dimensional spin-like spaces spanned by the representation states $\left|\pm\right\rangle$. The state will be parametrized by the tensors $A^{st}_{lrdu}$, with $s,t,l,r,d,u = \pm$. The most general construction satisfying these conditions is given by
\begin{equation}
	\begin{aligned}
	&A^{++}_{++++}=
	\begin{gathered}\includegraphics[scale=0.3]{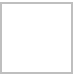}\end{gathered}=\alpha\\
	&A^{++}_{-+-+}=\begin{gathered}\includegraphics[scale=0.12]{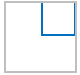}\end{gathered}=\beta\\
	&A^{-+}_{+--+}=\begin{gathered}\includegraphics[scale=0.12]{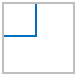}\end{gathered}=\beta\\
	&A^{--}_{+-+-}=\begin{gathered}\includegraphics[scale=0.12]{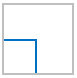}\end{gathered}=\beta\\
	&A^{+-}_{-++-}=\begin{gathered}\includegraphics[scale=0.12]{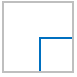}\end{gathered}=\beta\\
	&A^{+-}_{++--}=\begin{gathered}\includegraphics[scale=0.3]{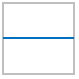}\end{gathered}=\gamma\\
	&A^{-+}_{--++}=\begin{gathered}\includegraphics[scale=0.3]{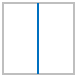}\end{gathered}=\gamma\\
	&A^{--}_{----}=\begin{gathered}\includegraphics[scale=0.3]{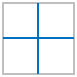}\end{gathered}=\delta
	\label{Atenel}
	\end{aligned}
\end{equation}
and the rest of the elements, which violate the symmetry, vanish. If we consider the $\left|+\right\rangle$ states as flux free states, and the $\left|-\right\rangle$ as flux carrying, we can interpret $\alpha$ as the amplitude of having no fluxes going through the site, $\beta$ as the amplitude of corner flux, $\gamma$ - of straight line fluxes and $\delta$ - two intersecting flux lines.

Here we will be interested in the properties of the transfer operators constructed for such states, and the computation of the Wilson loop expectation value. 

\subsection{The transfer operators}
The transfer operator $\hat{T}$ may be simply built using (\ref{Tdef}) and (\ref{Tform}).

Let us identify the elements of the vector space spanned by the double legs of the transfer matrix. The on-leg transformations here admit the simple form $\theta \otimes \tilde \theta^{\dagger} = Z \otimes Z$ for the only group element which is not the identity; Since there are two irreps, we will have two on-leg singlets,
\begin{equation}
	\begin{aligned}
	\left|0\left(+\right)\right\rangle &= \left|++\right\rangle \equiv \left| \uparrow \right\rangle \otimes \left| s\right\rangle \\
	\left|0\left(-\right)\right\rangle &= \left|--\right\rangle \equiv \left| \downarrow \right\rangle \otimes \left|s\right\rangle 
	\label{not1}
	\end{aligned}
\end{equation}
as well as two non-singlets,
\begin{equation}
	\begin{aligned}
		 \left|1\left(+,-\right)\right\rangle &=\left|+-\right\rangle \equiv \left|\uparrow\right\rangle \otimes \left| n\right\rangle\\
		 \left|1\left(-,+\right)\right\rangle &= \left|-+\right\rangle \equiv \left|\downarrow \right\rangle \otimes \left|n\right\rangle
		 	\label{not2}
	\end{aligned}
\end{equation}
Where the new notation introduced in the two equations above factorizes the on-leg Hilbert space into the product of two spin spaces; one detects whether the state is an on-leg singlet ($s$) or not ($n$) and the other labels the two states within each of these options by $\uparrow$ and $\downarrow$.

Using these states, we can write down all the relevant transfer operators and their reductions. For example, 
\begin{widetext}
	\begin{equation}
		\begin{aligned}
			\hat\tau_0 = &\left|\alpha\right|^2 \left|0\left(+\right)\right\rangle \left\langle 0\left(+\right) \right|
			\otimes \left|0\left(+\right)\right\rangle \left\langle 0\left(+\right) \right|  +
			\left|\gamma\right|^2 \left(\left|0\left(+\right)\right\rangle \left\langle 0\left(+\right) \right|\otimes \left|0\left(-\right)\right\rangle \left\langle 0\left(-\right) \right| 
			+ \left|0\left(-\right)\right\rangle \left\langle 0\left(-\right) \right| \otimes \left|0\left(+\right)\right\rangle \left\langle 0\left(+\right) \right| \right) \\+&
			\left|\delta\right|^2 \left|0\left(-\right)\right\rangle \left\langle 0\left(-\right) \right|
			\otimes \left|0\left(-\right)\right\rangle \left\langle 0\left(-\right) \right| +
			\left|\beta\right|^2 \left(\left|0\left(+\right)\right\rangle \left\langle 0\left(-\right) \right|
			+ \left|0\left(-\right)\right\rangle \left\langle 0\left(+\right) \right|\right)\otimes \left(\left|0\left(+\right)\right\rangle \left\langle 0\left(-\right) \right|
			+ \left|0\left(-\right)\right\rangle \left\langle 0\left(+\right) \right|\right) 	
		\end{aligned}
	\end{equation}
We can simplify by writing it in the matrix form, as well as adopting the new notation introduced in (\ref{not1}) and (\ref{not2}),
\begin{equation}
	\renewcommand\arraystretch{1.2}
	\hat{\tau}_0 =
	\begin{blockarray}{rcccccccc}
		\text{lr / du}&&& \left|\uparrow\right\rangle \left\langle \uparrow \right| \otimes \left|s\right\rangle\left\langle s\right| & \left|\downarrow\right\rangle \left\langle \downarrow \right| \otimes \left|s\right\rangle\left\langle s\right| & \left|\uparrow\right\rangle \left\langle \downarrow \right| \otimes \left|s\right\rangle\left\langle s\right| &  \left|\downarrow\right\rangle \left\langle \uparrow \right| \otimes \left|s\right\rangle\left\langle s\right| & \\
		\begin{block}{rc(c@{}ccccc@{}c)}
			\left|\uparrow\right\rangle \left\langle \uparrow \right| \otimes \left|s\right\rangle\left\langle s\right|  &  && \left|\alpha\right|^2 & \left|\gamma\right|^2 & 0 & 0& \\
			\left|\downarrow\right\rangle \left\langle \downarrow \right| \otimes \left|s\right\rangle\left\langle s\right|  &  && \left|\gamma\right|^2 & \left|\delta\right|^2  & 0 & 0 & \\
		\left|\uparrow\right\rangle \left\langle \downarrow \right| \otimes \left|s\right\rangle\left\langle s\right| &  && 0 &0 & \left|\beta\right|^2& \left|\beta\right|^2 & \\
		\left|\downarrow\right\rangle \left\langle \uparrow \right| \otimes \left|s\right\rangle\left\langle s\right|&  && 0 & 0 &\left|\beta\right|^2  & \left|\beta\right|^2& \vphantom{\smash[t]{\bigg|}}\\
		\end{block}
	\label{tau0z2}
	\end{blockarray}
\end{equation}
where the block structure is clearly seen; the first one is the zeroth block, mixing only projection operators. It depends on $\alpha,\gamma,\delta$ - the amplitudes for which fluxes do not change directions, and thus the representations are not changed horizontally and vertically on the state, and the on-leg singlets are not flipped on the transfer operators. The second block, where the representation / singlet change, depends on $\beta$ - the turning (corner) flux amplitude.
Furthermore, as the parameter $\gamma$ has to do with straight flux lines going through the site, we expect that the larger it gets, the farther the $\hat{M}_{\mu}$ operators derived from the zeroth block are from projection operators, and the farther we are from an area law; indeed, as we see, it appears on the off-diagonal terms of the zeroth block, and when $\gamma=0$ the  $\hat{M}_{\mu}$ operators of the zeroth blocks are projectors.

This matrix can be easily diagonalized as in (\ref{tau0diag}), with the eigenvalues (not necessarily in descending order - this depends on the values of the parameters):
\begin{equation}
	\begin{aligned}
	&\lambda_{1,2} = \frac{1}{2}\left(\left|\alpha\right|^2 +\left|\delta\right|^2 \pm \sqrt{\left(\left|\alpha\right|^2 -\left|\delta\right|^2\right)^2 + 4 \left|\gamma\right|^2}\right),
	\\&
	\lambda_3 = 2\left|\beta\right|^2,
		\quad
	\lambda_4 = 0
	\end{aligned}
\end{equation}
with the diagonalizing matrix
\begin{equation}
	\renewcommand\arraystretch{1.2}
V =
\begin{blockarray}{rcccccccc}
	&&& \mu = 1 & \mu = 2 & \mu=3 &  \mu=4 & \\
	\begin{block}{rc(c@{}ccccc@{}c)}
		\left|\uparrow\right\rangle \left\langle \uparrow \right| \otimes \left|s\right\rangle\left\langle s\right|  &  && u_{11}\left(\alpha,\gamma,\delta\right)& u_{12}\left(\alpha,\gamma,\delta\right) & 0 & 0& \\
		\left|\downarrow\right\rangle \left\langle \downarrow \right| \otimes \left|s\right\rangle\left\langle s\right|  &  && u_{21}\left(\alpha,\gamma,\delta\right) & u_{22}\left(\alpha,\gamma,\delta\right)  & 0 & 0 & \\
		\left|\uparrow\right\rangle \left\langle \downarrow \right| \otimes \left|s\right\rangle\left\langle s\right| &  && 0 &0 & \frac{1}{\sqrt{2}}& -\frac{1}{\sqrt{2}} & \\
		\left|\downarrow\right\rangle \left\langle \uparrow \right| \otimes \left|s\right\rangle\left\langle s\right|&  && 0 & 0 &\frac{1}{\sqrt{2}} & \frac{1}{\sqrt{2}} & \vphantom{\smash[t]{\bigg|}}\\
	\end{block}
\end{blockarray}
\end{equation}

Using all that, we obtain the operators $\hat{M}_{\mu}$ as defined in (\ref{Mdef}),
\begin{equation}
	\begin{aligned}
	\hat{M}_{1} &= \left( u_{11}\left(\alpha,\gamma,\delta\right) \Pi_{\uparrow} + u_{21}\left(\alpha,\gamma,\delta\right) \Pi_{\downarrow}\right),\\
	\hat{M}_{2} &= \left( u_{12}\left(\alpha,\gamma,\delta\right) \Pi_{\uparrow} + u_{22}\left(\alpha,\gamma,\delta\right) \Pi_{\downarrow}\right),\\
	\hat{M}_{3} &= \frac{1}{\sqrt{2}}\sigma_x,\\
	\hat{M}_{4} &= -\frac{i}{\sqrt{2}}\sigma_y.
	\label{Mz2}
\end{aligned}
\end{equation}
where $\Pi_{\uparrow}=\left|\uparrow\right\rangle \left\langle \uparrow \right|$ and $\Pi_{\downarrow} = \left|\downarrow\right\rangle \left\langle \downarrow \right|$.
Since $V$ is orthogonal, they form an orthonormal basis as in (\ref{ortho}). $\hat{M}_4$ is irrelevant, since $\lambda_4=0$; the $\left|s\right\rangle\left\langle s \right|$ is also irrelevant since it multiplies everything, and hence we will omit it and refer to the operators $\hat{M}_{\mu}$ as two dimensional. Note that as expected the first two ones, $\hat{M}_{1,2}$, having to do with the zeroth block, are diagonal, while the other ones are not.

Similarly, we can compute and write down the other relevant matrices.  Note that since the fluxes have no orientation in our case, $\hat\tau_{\rightarrow} = \hat\tau_{\leftarrow} \equiv \hat{\tau}_{-}$ and $\hat\tau_{\uparrow} = \hat\tau_{\downarrow} \equiv \hat{\tau}_{|}$. We thus require only six rather than eight further matrices. The first is
\begin{equation}
	\renewcommand\arraystretch{1.2}
	\hat{\tau}_- =
	\begin{blockarray}{rcccccccc}
		\text{lr / du}&&& \Pi_{\uparrow} \otimes \left|s\right\rangle\left\langle s\right| & \Pi_{\downarrow} \otimes \left|s\right\rangle\left\langle s\right| &\sigma_{+} \otimes \left|s\right\rangle\left\langle s\right|&  \sigma_{-} \otimes \left|s\right\rangle\left\langle s\right| & \\
		\begin{block}{rc(c@{}ccccc@{}c)}
			\Pi_{\uparrow} \otimes \left|n\right\rangle\left\langle n\right|  &  && \alpha\overline\gamma & \gamma\overline\delta & 0 & 0& \\
			\Pi_{\downarrow} \otimes \left|n\right\rangle\left\langle n\right|  &  && \gamma\overline\alpha & \delta\overline\gamma  & 0 & 0 & \\
			\sigma_{+} \otimes \left|n\right\rangle\left\langle n\right| &  && 0 &0 & \left|\beta\right|^2& \left|\beta\right|^2 & \\
			\sigma_{-} \otimes \left|n\right\rangle\left\langle n\right| &  && 0 & 0 &\left|\beta\right|^2  & \left|\beta\right|^2& \vphantom{\smash[t]{\bigg|}}\\
		\end{block}
	\label{tauh}
	\end{blockarray}
\end{equation}
connecting operators acting on the non-singlet subspace in the horizontal direction with ones acting on the singlet space in the vertical one. The same block structure is apparent; the first block is a generalization of the zeroth block - still only connecting projection operators, though acting on different spaces, and the second block changes the representations. As in the $\tau_0$ case, the parameter $\gamma$ is the one "spoiling" the area law: all the amplitudes of $\hat{L}_{\mu}$ operators which do not change the on-leg singlet eigenvalue subspace are proportional to it. One it is set to zero, when crossing a flux line the subspace will change.

We can formally perform a horizontal-vertical singular value decomposition and obtain an expression of the form $\hat{\tau}_-  = \underset{\mu}{\sum}\eta_{\mu}\hat K_{\mu}\otimes \hat L_{\mu}$. Since the horizontal operators act only within the non-singlet subspace and the vertical ones only within the singlet subspace, we can represent $\hat K_{\mu}$ and $\hat L_{\mu}$ by two dimensional matrices.

 $\hat{\tau}_|$ is simply obtained by transposition,
\begin{equation}
	\renewcommand\arraystretch{1.2}
	\hat{\tau}_| =
	\begin{blockarray}{rcccccccc}
		\text{lr / du}&&& \Pi_{\uparrow} \otimes \left|n\right\rangle\left\langle n\right| & \Pi_{\downarrow} \otimes \left|n\right\rangle\left\langle n\right| &\sigma_{+} \otimes \left|n\right\rangle\left\langle n\right|&  \sigma_{-} \otimes \left|n\right\rangle\left\langle n\right| & \\
		\begin{block}{rc(c@{}ccccc@{}c)}
			\Pi_{\uparrow} \otimes \left|s\right\rangle\left\langle s\right|  &  && \alpha\overline\gamma & \gamma\overline\alpha & 0 & 0& \\
			\Pi_{\downarrow} \otimes \left|s\right\rangle\left\langle s\right|  &  && \gamma\overline\delta & \delta\overline\gamma  & 0 & 0 & \\
			\sigma_{+} \otimes \left|s\right\rangle\left\langle s\right| &  && 0 &0 & \left|\beta\right|^2& \left|\beta\right|^2 & \\
			\sigma_{-} \otimes \left|s\right\rangle\left\langle s\right| &  && 0 & 0 &\left|\beta\right|^2  & \left|\beta\right|^2& \vphantom{\smash[t]{\bigg|}}\\
		\end{block}
		\label{tauv}
	\end{blockarray}
\end{equation}
and $\hat{\tau}_|  = \underset{\mu}{\sum}\eta_{\mu}\hat L_{\mu}\otimes \hat K_{\mu}$.

Finally, let us consider the transfer operators of the four corners. We begin with the lower left corner
\begin{equation}
	\renewcommand\arraystretch{1.2}
	\hat{\tau}_{\llcorner} =
	\begin{blockarray}{rcccccccc}
		\text{lr / du}&&& \Pi_{\uparrow} \otimes \left|s\right\rangle\left\langle n\right| & \Pi_{\downarrow} \otimes \left|s\right\rangle\left\langle n\right| &\sigma_{+} \otimes \left|s\right\rangle\left\langle n\right|&  \sigma_{-} \otimes \left|s\right\rangle\left\langle n\right| & \\
		\begin{block}{rc(c@{}ccccc@{}c)}
			\Pi_{\uparrow} \otimes \left|s\right\rangle\left\langle n\right|  &  && \alpha\overline\beta & \gamma\overline\beta & 0 & 0& \\
			\Pi_{\downarrow} \otimes \left|s\right\rangle\left\langle n\right|  &  && \gamma\overline\beta & \delta\overline\beta  & 0 & 0 & \\
			\sigma_{+} \otimes \left|s\right\rangle\left\langle n\right| &  && 0 &0 & \beta\overline\alpha& \beta\overline\gamma & \\
			\sigma_{-} \otimes \left|s\right\rangle\left\langle n\right| &  && 0 & 0 &\beta\overline\gamma  & \beta\overline\delta& \vphantom{\smash[t]{\bigg|}}\\
		\end{block}
	\end{blockarray}
\end{equation}
where in both dimensions we get a singlet input and obtain a non-singlet output. Here, after performing the singular value decomposition, we will also use two dimensional operators acting only on the "spin space" since this corner operator connects to the right $s/n$ subspaces.
The other corner operators are
\begin{equation}
	\renewcommand\arraystretch{1.2}
	\hat{\tau}_{\lrcorner} =
	\begin{blockarray}{rcccccccc}
		\text{lr / du}&&& \Pi_{\uparrow} \otimes \left|s\right\rangle\left\langle n\right| & \Pi_{\downarrow} \otimes \left|s\right\rangle\left\langle n\right| &\sigma_{+} \otimes \left|s\right\rangle\left\langle n\right|&  \sigma_{-} \otimes \left|s\right\rangle\left\langle n\right| & \\
		\begin{block}{rc(c@{}ccccc@{}c)}
			\Pi_{\uparrow} \otimes \left|n\right\rangle\left\langle s\right|  &  && \alpha\overline\beta & \gamma\overline\beta & 0 & 0& \\
			\Pi_{\downarrow} \otimes \left|n\right\rangle\left\langle s\right|  &  && \gamma\overline\beta & \delta\overline\beta  & 0 & 0 & \\
			\sigma_{+} \otimes \left|n\right\rangle\left\langle s\right| &  && 0 &0 & \beta\overline\gamma& \beta\overline\delta & \\
			\sigma_{-} \otimes \left|n\right\rangle\left\langle s\right| &  && 0 & 0 &\beta\overline\alpha  & \beta\overline\gamma& \vphantom{\smash[t]{\bigg|}}\\
		\end{block}
	\end{blockarray}
\end{equation}
\begin{equation}
	\renewcommand\arraystretch{1.2}
	\hat{\tau}_{\urcorner} =
	\begin{blockarray}{rcccccccc}
		\text{lr / du}&&& \Pi_{\uparrow} \otimes \left|n\right\rangle\left\langle s\right| & \Pi_{\downarrow} \otimes \left|n\right\rangle\left\langle s\right| &\sigma_{+} \otimes \left|n\right\rangle\left\langle s\right|&  \sigma_{-} \otimes \left|n\right\rangle\left\langle s\right| & \\
		\begin{block}{rc(c@{}ccccc@{}c)}
			\Pi_{\uparrow} \otimes \left|n\right\rangle\left\langle s\right|  &  && \alpha\overline\beta & \gamma\overline\beta & 0 & 0& \\
			\Pi_{\downarrow} \otimes \left|n\right\rangle\left\langle s\right|  &  && \gamma\overline\beta & \delta\overline\beta  & 0 & 0 & \\
			\sigma_{+} \otimes \left|n\right\rangle\left\langle s\right| &  && 0 &0 & \beta\overline\delta& \beta\overline\gamma & \\
			\sigma_{-} \otimes \left|n\right\rangle\left\langle s\right| &  && 0 & 0 &\beta\overline\gamma  & \beta\overline\alpha& \vphantom{\smash[t]{\bigg|}}\\
		\end{block}
	\end{blockarray}
\end{equation}
and
\begin{equation}
	\renewcommand\arraystretch{1.2}
	\hat{\tau}_{\ulcorner} =
	\begin{blockarray}{rcccccccc}
		\text{lr / du}&&& \Pi_{\uparrow} \otimes \left|n\right\rangle\left\langle s\right| & \Pi_{\downarrow} \otimes \left|n\right\rangle\left\langle s\right| &\sigma_{+} \otimes \left|n\right\rangle\left\langle s\right|&  \sigma_{-} \otimes \left|n\right\rangle\left\langle s\right| & \\
		\begin{block}{rc(c@{}ccccc@{}c)}
			\Pi_{\uparrow} \otimes \left|s\right\rangle\left\langle n\right|  &  && \alpha\overline\beta & \gamma\overline\beta & 0 & 0& \\
			\Pi_{\downarrow} \otimes \left|s\right\rangle\left\langle n\right|  &  && \gamma\overline\beta & \delta\overline\beta  & 0 & 0 & \\
			\sigma_{+} \otimes \left|s\right\rangle\left\langle n\right| &  && 0 &0 & \beta\overline\gamma& \beta\overline\alpha & \\
			\sigma_{-} \otimes \left|s\right\rangle\left\langle n\right| &  && 0 & 0 &\beta\overline\delta  & \beta\overline\gamma& \vphantom{\smash[t]{\bigg|}}\\
		\end{block}
	\end{blockarray}
\end{equation}
Note that all the elements of the corner operators are proportional to either $\beta$ or $\overline{\beta}$, which is expected since $\beta$ is the corner parameter, and it would be impossible to close a loop in its absence. 

\subsection{Analytical example}
Let us set, for simplicity, $\gamma=0$. Consider $\tau_0$ (\ref{tau0z2}) and the $\hat M_{\mu}$ operators derived from it (\ref{Mz2}). Let us set $\gamma=0$; then we simply have
\begin{equation}
		\hat{M}_{1} = \Pi_{\uparrow}, \quad
		\hat{M}_{2} = \Pi_{\downarrow},\quad
		\hat{M}_{3} = \frac{1}{\sqrt{2}}\sigma_x,\quad
		\hat{M}_{4} = -\frac{i}{\sqrt{2}}\sigma_y.
		\label{Mz2g0}
\end{equation}
as well as 
\begin{equation}
	\lambda_{1}=\left|\alpha\right|^2,
	\quad
	\lambda_{2}=\left|\delta\right|^2,
	\quad
	\lambda_3 = 2\left|\beta\right|^2,
	\quad
	\lambda_4 = 0
\end{equation}

The choice of $\gamma=0$ sets all the zeroth block $\hat{M}_{\mu}$ operators to projectors onto orthogonal states, and the flux-free transfer matrix from (\ref{Edef}) takes the form 
\begin{equation}
	\begin{aligned}
	\hat{E} &= \left|\alpha\right|^{2\mathcal{N}} \Pi_{\uparrow} \otimes \cdots \otimes \Pi_{\uparrow}
+ \left|\delta\right|^{2\mathcal{N}} \Pi_{\downarrow}  \otimes \cdots \otimes \Pi_{\downarrow} +\\
&+ \left|\beta\right|^4 \overset{\mathcal{N}-1}{\underset{n=1}{\sum}}\underset{m}{\sum}\left|\alpha\right|^{2\left(\mathcal{N}-n-1\right)}\left|\delta\right|^{2\left(n-1\right)} \Pi_{\uparrow}  \otimes \cdots \otimes  \Pi_{\uparrow} \otimes \underbrace{\sigma_x}_{m} \otimes  \Pi_{\downarrow} \otimes \cdots \otimes  \Pi_{\downarrow} \otimes \underbrace{\sigma_x}_{m+n} \otimes \Pi_{\uparrow} \otimes \cdots \otimes  \Pi_{\uparrow} +\\
&+ \left|\beta\right|^4 \overset{\mathcal{N}-1}{\underset{n=1}{\sum}}\underset{m}{\sum}\left|\delta\right|^{2\left(\mathcal{N}-n-1\right)}\left|\alpha\right|^{2\left(n-1\right)} \Pi_{\downarrow} \otimes \cdots \otimes  \Pi_{\downarrow} \otimes \underbrace{\sigma_x}_{m} \otimes  \Pi_{\uparrow} \otimes \cdots \otimes  \Pi_{\uparrow} \otimes \underbrace{\sigma_x}_{m+n} \otimes \Pi_{\downarrow} \otimes \cdots \otimes  \Pi_{\downarrow} + O\left(\left|\beta\right|^8\right)
\end{aligned}
\label{Ez2}
\end{equation}
If we further assume that $\left|\beta\right| \ll \left|\alpha\right|,\left|\delta\right|$ we find ourselves in the perturbative case discussed above, and may use perturbation theory for finding the eigenvectors of $\hat{E}$. The zeroth, unperturbed part is in the first row of (\ref{Ez2}), from which we find two approximate, zeroth order eigenvectors,
\begin{equation}
	\left\langle w_1\right| = \left\langle \uparrow \right| \otimes \cdots \otimes \left\langle \uparrow \right|, \quad\quad \left\langle w_2 \right|= \left\langle \downarrow \right| \otimes \cdots \otimes \left\langle \downarrow \right|
\end{equation}
with zeroth order eigenvalues
\begin{equation}
	\rho_1 = \left|\alpha\right|^{2\mathcal{N}}, \quad\quad \rho_2 = \left|\delta\right|^{2\mathcal{N}}
\end{equation}
- which are the two highest ones. Let us set, without losing generality,  $\left|\alpha\right| > \left|\delta\right|$ (one can easily invert that in the following discussion).
The leading order corrections to the eigenvalues will be second order ($\propto\left|\beta\right|^8$) and to the eigenvectors will be of the first order ($\propto\left|\beta\right|^4$); we shall neglect them both. The norm of the state is then
\begin{equation}
	\left\langle \psi | \psi \right\rangle = \text{Tr}\left[E^{\mathcal{N}}\right] \underset{\mathcal{N} \gg 1} {\longrightarrow }  \rho_1^\mathcal{N}  = \left|\alpha\right|^{2\mathcal{N}^2}
\end{equation}
That is, the torus is tiled with $\mathcal{N}^2$ sites, each contributing a factor of $\left|\alpha\right|^{2}$ to the norm.

Let us now move on to the flux carrying transfer matrices. Looking at the straight flux ones $\hat{\tau}_-$ (\ref{tauh}) and $\hat{\tau}_|$ (\ref{tauv}), we see that our choice of $\gamma=0$ sets the zeroth block to zero. This implies that they will flip the local incoming spins in both directions - in particular in the direction orthogonal to the flux; i.e., the eigenspace of $\hat{\tau}_0$ out of the loop will be connected to the orthogonal one within the loop, eventually to give rise to an area law, unless $\left|\alpha\right| = \left|\delta\right|$. We see that
\begin{equation}
\hat{\tau}_- = \hat{\tau}_| = \left|\beta\right|^2 \sigma_x \otimes \sigma_x 
\end{equation}
(ignoring the $n,s$ space for the reasons explained above) - inverting the spins in the orthogonal direction to the flux lines, that is, changing indeed from the $\alpha$ to the $\delta$ sector and vice versa.

For the corners we get
\begin{equation}
\begin{aligned}
	\hat{\tau}_{\llcorner} & = \alpha \overline{\beta} \Pi_{\uparrow} \otimes \Pi_{\uparrow} + \delta \overline{\beta} \Pi_{\downarrow} \otimes \Pi_{\downarrow} + \beta\overline{\alpha} \sigma_+ \otimes \sigma_+ + \beta\overline{\delta} \sigma_- \otimes \sigma_- 
	\equiv \underset{\mu}{\sum}\xi_{\llcorner,\mu}\hat{H}_{\llcorner,\mu}\otimes\hat{V}_{\llcorner,\mu} \\
	\hat{\tau}_{\lrcorner} & = \alpha \overline{\beta} \Pi_{\uparrow} \otimes \Pi_{\uparrow} + \delta \overline{\beta} \Pi_{\downarrow} \otimes \Pi_{\downarrow} + \beta\overline{\delta} \sigma_+ \otimes \sigma_- + \beta\overline{\alpha} \sigma_- \otimes \sigma_+ \equiv \underset{\mu}{\sum}\xi_{\lrcorner,\mu}\hat{H}_{\lrcorner,\mu}\otimes\hat{V}_{\lrcorner,\mu}   \\
	\hat{\tau}_{\urcorner} & = \alpha \overline{\beta} \Pi_{\uparrow} \otimes \Pi_{\uparrow} + \delta \overline{\beta} \Pi_{\downarrow} \otimes \Pi_{\downarrow} + \beta\overline{\delta} \sigma_+ \otimes \sigma_+ + \beta\overline{\alpha} \sigma_- \otimes \sigma_- \equiv \underset{\mu}{\sum}\xi_{\urcorner,\mu}\hat{H}_{\urcorner,\mu}\otimes\hat{V}_{\urcorner,\mu}   \\
	\hat{\tau}_{\ulcorner} & = \alpha \overline{\beta} \Pi_{\uparrow} \otimes \Pi_{\uparrow} + \delta \overline{\beta} \Pi_{\downarrow} \otimes \Pi_{\downarrow} + \beta\overline{\alpha} \sigma_+ \otimes \sigma_-  +\beta\overline{\delta} \sigma_- \otimes \sigma_+ \equiv \underset{\mu}{\sum}\xi_{\ulcorner,\mu}\hat{H}_{\ulcorner,\mu}\otimes\hat{V}_{\ulcorner,\mu} 
\end{aligned}
\end{equation}

Let us consider the action of the lower row of the Wilson loop, $\hat{E}_b\left(R_1\right)$ on the input state $\left\langle w_1\right|$ with the highest eigenvalue, identifying without loss of generality, as usual, the origin of the torus  with the lower left corner of the loop. We get
\begin{equation}
\hat{E}_b\left(R_1\right) = \left|\alpha\right|^{2\left(\mathcal{N}-R_1-1\right)}
\left|\beta\right|^{2\left(R_1-1\right)}
\underset{\mu,\nu}{\sum}\xi_{\llcorner,\mu}\xi_{\lrcorner,\nu}
\text{Tr}\left[\hat{H}_{\llcorner,\mu} \sigma_x^{R_1-1} \hat{H}_{\lrcorner,\nu} \Pi_{\uparrow}\right]  
 \hat{V}_{\llcorner,\mu} \otimes \underbrace{\sigma_x \otimes \cdots \otimes \sigma_x}_{R_1-1} \otimes \hat{V}_{\lrcorner,\nu} \otimes \underbrace{\Pi_{\uparrow} \otimes \cdots \otimes \Pi_{\uparrow}}_{\mathcal{N}-R_1-1}+ ...
\end{equation}
where the omitted terms either annihilate $\left\langle w_1\right|$ or are of negligible magnitude.

Some of the $\mu,\nu$ configurations give rise to a zero trace. Others annihilate the input vector $\left\langle w_i\right|$. There are only four possible valid configurations: 
\begin{enumerate}
	\item $R_1$ is even, $\hat{H}_{\llcorner,\mu} = \sigma_{+}$, $\hat{H}_{\lrcorner,\nu} = \Pi_{\uparrow}$ and thus $\hat{V}_{\llcorner,\mu} = \sigma_{+}$, $\hat{V}_{\lrcorner,\nu} = \Pi_{\uparrow}$ and $\xi_{\llcorner,\mu}\xi_{\lrcorner,\nu} = \left|\alpha\overline{\beta}\right|^2$.
	\item $R_1$ is even, $\hat{H}_{\llcorner,\mu} = \Pi_{\uparrow}$, $\hat{H}_{\lrcorner,\nu} = \sigma_-$ and thus $\hat{V}_{\llcorner,\mu} = \Pi_{\uparrow}$, $\hat{V}_{\lrcorner,\nu} = \sigma_+$ and $\xi_{\llcorner,\mu}\xi_{\lrcorner,\nu} = \left|\alpha\overline{\beta}\right|^2$.
	\item $R_1$ is odd, $\hat{H}_{\llcorner,\mu} = \Pi_{\uparrow}$, $\hat{H}_{\lrcorner,\nu} = \Pi_{\uparrow}$ and thus $\hat{V}_{\llcorner,\mu} = \Pi_{\uparrow}$, $\hat{V}_{\lrcorner,\nu} = \Pi_{\uparrow}$ and $\xi_{\llcorner,\mu}\xi_{\lrcorner,\nu} = \left(\alpha\overline{\beta}\right)^2$.
    \item $R_1$ is odd, $\hat{H}_{\llcorner,\mu} = \sigma_+$, $\hat{H}_{\lrcorner,\nu} = \sigma_-$ and thus $\hat{V}_{\llcorner,\mu} = \sigma_{+}$, $\hat{V}_{\lrcorner,\nu} = \sigma_+$ and $\xi_{\llcorner,\mu}\xi_{\lrcorner,\nu} = \left(\beta\overline{\alpha}\right)^2$.
\end{enumerate}
The leading terms of the output vector $\left\langle w_1 \right|\hat{E}_b\left(R\right)$ are product vectors, with $\left\langle \downarrow\right|$ entering the loop and $\left\langle \uparrow \right|$ out of it. The two spins which are on the loop's boundaries are either flipped or not, depending on the particular configuration from the list above. We get for an even $R_1$
\begin{equation}
\left\langle w_1 \right|\hat{E}_b\left(R_1\right) =
		\left|\alpha\right|^{2\mathcal{N}}
		\left|\frac{\beta}{\alpha}\right|^{2R_1}\left(
		\left\langle \downarrow\right| \otimes \underbrace{\left\langle \downarrow\right| \otimes \cdots \otimes \left\langle \downarrow\right|}_{R_1-1} \otimes \left\langle \uparrow \right|\otimes \underbrace{\left\langle \uparrow \right| \otimes \cdots \otimes \left\langle \uparrow \right|}_{\mathcal{N}-R_1-1} + 
		\left\langle \uparrow\right| \otimes \underbrace{\left\langle \downarrow\right| \otimes \cdots \otimes \left\langle \downarrow\right|}_{R_1-1} \otimes \left\langle \downarrow \right|\otimes \underbrace{\left\langle \uparrow \right| \otimes \cdots \otimes \left\langle \uparrow \right|}_{\mathcal{N}-R_1-1}		
		\right)
		\label{Ebeven}
\end{equation}
and for an odd $R_1$
\begin{equation}
	\begin{aligned}
	\left\langle w_1 \right|\hat{E}_b\left(R_1\right) =
		\left|\alpha\right|^{2\mathcal{N}}
\left|\frac{\beta}{\alpha}\right|^{2R_1}\Bigg[&
\left(\frac{\alpha\overline\beta}{\left|\alpha\overline\beta\right|}\right)^2
	\left\langle \uparrow\right| \otimes \underbrace{\left\langle \downarrow\right| \otimes \cdots \otimes \left\langle \downarrow\right|}_{R_1-1} \otimes \left\langle \uparrow \right|\otimes \underbrace{\left\langle \uparrow \right| \otimes \cdots \otimes \left\langle \uparrow \right|}_{\mathcal{N}-R_1-1} + 
	\\&+\left(\frac{\beta\overline\alpha}{\left|\beta\overline\alpha\right|}\right)^2\left\langle \downarrow\right| \otimes \underbrace{\left\langle \downarrow\right| \otimes \cdots \otimes \left\langle \downarrow\right|}_{R_1-1} \otimes \left\langle \downarrow \right|\otimes \underbrace{\left\langle \uparrow \right| \otimes \cdots \otimes \left\langle \uparrow \right|}_{\mathcal{N}-R_1-1}	\Bigg]
		\label{Ebodd}
 	\end{aligned}
\end{equation}

We move on to the intermediate rows, with 
\begin{equation}
\hat{E}_{\parallel}\left(R_1\right)=	
\left|\alpha\right|^{2\left(\mathcal{N}-R_1-1\right)}
\left|\delta\right|^{2\left(R_1-1\right)}
\left|\beta\right|^4
 \sigma_x \otimes \underbrace{\Pi_{\downarrow} \otimes \cdots \otimes \Pi_{\downarrow}}_{R_1-1} \otimes \sigma_x \otimes \underbrace{\Pi_{\uparrow} \otimes \cdots \otimes \Pi_{\uparrow}}_{\mathcal{N}-R_1-1}+ ...
\end{equation}
where, once again, the terms not included are either small enough or annihilate the input vector. The highest eigenvalue (in absolute value) is
\begin{equation}
	\rho'_1\left(R\right) = \left|\alpha\right|^{2\left(\mathcal{N}-1\right)}
	\left|\frac{\beta^2}{\delta}\right|^2 
	\left|\frac{\delta}{\alpha}\right|^{2R_1}
\end{equation}
- exponential in the distance $R_1$, just as speculated in (\ref{rhogammakappa}), with $\Gamma = \left|\alpha\right|^{2\left(\mathcal{N}-1\right)}
\left|\frac{\beta^2}{\delta}\right|^2 $, and string tension $\kappa=-2\log\left|\frac{\delta}{\alpha}\right|$ - predicting an area law behaviour.

This eigenvalue is four-fold degenerate (in absolute value). Denoting by $\left|x=\pm 1\right\rangle$ the eigenvectors of $\sigma_x$, with eigenvalues $\pm 1$,
we get the four eigenvectors,
\begin{equation}
	\left\langle w'^{x,x'}_1\left(R\right)\right| = \left\langle x\right| \otimes \underbrace{\left\langle \downarrow\right| \otimes \cdots \otimes \left\langle \downarrow\right|}_{R_1-1} \otimes \left\langle x' \right|\otimes \underbrace{\left\langle \uparrow \right| \otimes \cdots \otimes \left\langle \uparrow \right|}_{\mathcal{N}-R_1-1}, \quad \text{ s.t. }\quad \left\langle w'^{x,x'}_1\left(R\right)\right| \hat{E}_{\parallel}\left(R_1\right) = xx'\rho'_1\left(R\right)\left\langle w'^{x,x'}_1\left(R\right)\right|
\end{equation}
Note that since the transfer matrices $\hat{E}$ and $\hat{E}_{\parallel}\left(R_1\right)$ are hermitian, $\left|v_i\right\rangle = \left|w_i\right\rangle$ and $\left|v'_i\left(R\right)\right\rangle = \left|w'_i\left(R\right)\right\rangle$. 

Connecting with the inputs (\ref{Ebeven}) and (\ref{Ebodd}) and using $\left\langle \uparrow | x \right\rangle = 1/\sqrt{2}$ and $\left\langle \downarrow | x \right\rangle = x/\sqrt{2}$  we obtain for an even $R_1$
\begin{equation}
	\begin{aligned}
\left\langle w_1 \right|\hat{E}_b\left(R_1\right) \left|v'^{x,x'}_1\left(R_1\right)\right\rangle &= \left|\alpha\right|^{2\mathcal{N}}\left|\frac{\beta}{\alpha}\right|^{2R_1}\left(\left\langle \downarrow | x \right\rangle  \left\langle \uparrow | x' \right\rangle + \left\langle \uparrow | x \right\rangle  \left\langle \downarrow | x' \right\rangle \right) \\&= \frac{1}{2}\left|\alpha\right|^{2\mathcal{N}}\left|\frac{\beta}{\alpha}\right|^{2R_1}\left(x+x'\right)
\equiv \frac{1}{2}\left|\alpha\right|^{2\mathcal{N}}\left|\frac{\beta}{\alpha}\right|^{2R_1}f_{\text{even}}\left(x,x'\right)
\end{aligned}
\end{equation}
and for an odd one
\begin{equation}
	\begin{aligned}
\left\langle w_1 \right|\hat{E}_b\left(R_1\right) \left|v'^{x,x'}_1\left(R_1\right)\right\rangle & = 
\left|\alpha\right|^{2\mathcal{N}}\left|\frac{\beta}{\alpha}\right|^{2R_1}
\left(\left(\frac{\alpha\overline\beta}{\left|\alpha\overline\beta\right|}\right)^2\left\langle \uparrow | x \right\rangle  \left\langle \uparrow | x' \right\rangle + 
\left(\frac{\beta\overline\alpha}{\left|\beta\overline\alpha\right|}\right)^2\left\langle \downarrow | x \right\rangle  \left\langle \downarrow | x' \right\rangle \right) \\&= \frac{1}{2}\left|\alpha\right|^{2\mathcal{N}}\left|\frac{\beta}{\alpha}\right|^{2R_1}\left(\left(\frac{\alpha\overline\beta}{\left|\alpha\overline\beta\right|}\right)^2+ xx'\left(\frac{\beta\overline\alpha}{\left|\beta\overline\alpha\right|}\right)^2\right)
\equiv \frac{1}{2}\left|\alpha\right|^{2\mathcal{N}}\left|\frac{\beta}{\alpha}\right|^{2R_1} f_{\text{odd}}\left(x,x'\right)
\end{aligned}
\end{equation}

We close the Wilson loop with $\hat{E}_t\left(R_1\right)$, where we consider the leading terms which do not annihilate the input vectors $\left\langle w'^{x,x'}_1\left(R\right)\right|$ or the output vector $\left\langle w_1\right|$, 
\begin{equation}
	\hat{E}_t\left(R_1\right) = \left|\alpha\right|^{2\left(\mathcal{N}-R_1-1\right)}
	\left|\beta\right|^{2\left(R_1-1\right)}
	\underset{\mu,\nu}{\sum}\xi_{\ulcorner,\mu}\xi_{\urcorner,\nu}
	\text{Tr}\left[\hat{H}_{\ulcorner,\mu} \sigma_x^{R_1-1} \hat{H}_{\urcorner,\nu} \Pi_{\uparrow}\right]  
	\hat{V}_{\ulcorner,\mu} \otimes \underbrace{\sigma_x \otimes \cdots \otimes \sigma_x}_{R_1-1} \otimes \hat{V}_{\urcorner,\nu} \otimes \underbrace{\Pi_{\uparrow} \otimes \cdots \otimes \Pi_{\uparrow}}_{\mathcal{N}-R_1-1}+ ...
\end{equation} 
Once again there are four possible cases:
\begin{enumerate}
	\item $R_1$ is even, $\hat{H}_{\ulcorner,\mu} = \sigma_{+}$, $\hat{H}_{\urcorner,\nu} = \Pi_{\uparrow}$ and thus $\hat{V}_{\ulcorner,\mu} = \sigma_{-}$, $\hat{V}_{\urcorner,\nu} = \Pi_{\uparrow}$ and $\xi_{\ulcorner,\mu}\xi_{\urcorner,\nu} = \left|\alpha\overline{\beta}\right|^2$.
	\item $R_1$ is even, $\hat{H}_{\ulcorner,\mu} = \Pi_{\uparrow}$, $\hat{H}_{\urcorner,\nu} = \sigma_-$ and thus $\hat{V}_{\ulcorner,\mu} = \Pi_{\uparrow}$, $\hat{V}_{\urcorner,\nu} = \sigma_-$ and $\xi_{\ulcorner,\mu}\xi_{\urcorner,\nu} = \left|\alpha\overline{\beta}\right|^2$.
	\item $R_1$ is odd, $\hat{H}_{\ulcorner,\mu} = \Pi_{\uparrow}$, $\hat{H}_{\urcorner,\nu} = \Pi_{\uparrow}$ and thus $\hat{V}_{\ulcorner,\mu} = \Pi_{\uparrow}$, $\hat{V}_{\urcorner,\nu} = \Pi_{\uparrow}$ and $\xi_{\ulcorner,\mu}\xi_{\urcorner,\nu} = \left(\alpha\overline{\beta}\right)^2$.
	\item $R_1$ is odd, $\hat{H}_{\ulcorner,\mu} = \sigma_+$, $\hat{H}_{\urcorner,\nu} = \sigma_-$ and thus $\hat{V}_{\ulcorner,\mu} = \sigma_{-}$, $\hat{V}_{\urcorner,\nu} = \sigma_-$ and $\xi_{\ulcorner,\mu}\xi_{\urcorner,\nu} = \left(\beta\overline{\alpha}\right)^2$.
\end{enumerate}
Implying that for an even $R_1$
\begin{equation}
	\hat{E}_t\left(R_1\right) \left|v_1\right\rangle=
	\left|\alpha\right|^{2\mathcal{N}}
	\left|\frac{\beta}{\alpha}\right|^{2R_1}\left(
	\left| \downarrow\right\rangle \otimes \underbrace{\left| \downarrow\right\rangle \otimes \cdots \otimes \left| \downarrow\right\rangle}_{R_1-1} \otimes \left| \uparrow \right\rangle\otimes \underbrace{\left| \uparrow \right\rangle \otimes \cdots \otimes \left| \uparrow \right\rangle}_{\mathcal{N}-R_1-1} + 
	\left| \uparrow\right\rangle \otimes \underbrace{\left| \downarrow\right\rangle \otimes \cdots \otimes \left| \downarrow\right\rangle}_{R_1-1} \otimes \left| \downarrow \right\rangle\otimes \underbrace{\left| \uparrow \right\rangle \otimes \cdots \otimes \left| \uparrow \right\rangle}_{\mathcal{N}-R_1-1}		
	\right)
	\label{Eteven}
\end{equation}
and for an odd $R_1$
\begin{equation}
	\begin{aligned}
	\hat{E}_t\left(R_1\right) \left|v_1\right\rangle=
		\left|\alpha\right|^{2\mathcal{N}}
	\left|\frac{\beta}{\alpha}\right|^{2R_1}\Big[&
	\left(\frac{\alpha\overline\beta}{\left|\alpha\overline\beta\right|}\right)^2
	\left| \uparrow\right\rangle \otimes \underbrace{\left| \downarrow\right\rangle \otimes \cdots \otimes \left| \downarrow\right\rangle}_{R_1-1} \otimes \left| \uparrow \right\rangle\otimes \underbrace{\left| \uparrow \right\rangle \otimes \cdots \otimes \left| \uparrow \right\rangle}_{\mathcal{N}-R_1-1} \\ &+ 
	 \left(\frac{\beta\overline\alpha}{\left|\beta\overline\alpha\right|}\right)^2
	\left| \downarrow\right\rangle \otimes \underbrace{\left| \downarrow\right\rangle \otimes \cdots \otimes \left| \downarrow\right\rangle}_{R_1-1} \otimes \left| \downarrow \right\rangle\otimes \underbrace{\left| \uparrow \right\rangle \otimes \cdots \otimes \left| \uparrow \right\rangle}_{\mathcal{N}-R_1-1}	\Big]	
	\label{Etodd}
	\end{aligned}
\end{equation}
Giving rise to, for an even $R_1$
\begin{equation}
	\begin{aligned}
	\left\langle w'^{x,x'}_1\left(R_1\right) \right|\hat{E}_t\left(R_1\right) \left|v_1\right\rangle &=
	 \left|\alpha\right|^{2\mathcal{N}}\left|\frac{\beta}{\alpha}\right|^{2R_1}
	 \left(\left\langle x | \downarrow \right\rangle  \left\langle x' | \uparrow \right\rangle + \left\langle x | \uparrow \right\rangle  \left\langle x' | \downarrow  \right\rangle \right) \\&= \frac{1}{2}\left|\alpha\right|^{2\mathcal{N}}\left|\frac{\beta}{\alpha}\right|^{2R_1}\left(x+x'\right) = \frac{1}{2}\left|\alpha\right|^{2\mathcal{N}}\left|\frac{\beta}{\alpha}\right|^{2R_1} f_{\text{even}}\left(x,x'\right)
\end{aligned}
\end{equation}
and for an odd one
\begin{equation}
	\begin{aligned}
	\left\langle w'^{x,x'}_1\left(R_1\right) \right|\hat{E}_t\left(R_1\right) \left|v_1\right\rangle & = 
		\left|\alpha\right|^{2\mathcal{N}}\left|\frac{\beta}{\alpha}\right|^{2R_1}
		\left(\left(\frac{\alpha\overline\beta}{\left|\alpha\overline\beta\right|}\right)^2\left\langle x| \uparrow \right\rangle  \left\langle x' | \uparrow \right\rangle + 
		\left(\frac{\beta\overline\alpha}{\left|\beta\overline\alpha\right|}\right)^2\left\langle x| \downarrow  \right\rangle  \left\langle x'| \downarrow \right\rangle \right) \\&= \frac{1}{2}\left|\alpha\right|^{2\mathcal{N}}\left|\frac{\beta}{\alpha}\right|^{2R_1}\left(\left(\frac{\alpha\overline\beta}{\left|\alpha\overline\beta\right|}\right)^2+ xx'\left(\frac{\beta\overline\alpha}{\left|\beta\overline\alpha\right|}\right)^2\right) =  \frac{1}{2}\left|\alpha\right|^{2\mathcal{N}}\left|\frac{\beta}{\alpha}\right|^{2R_1} f_{\text{odd}}\left(x,x'\right)
	\end{aligned}
\end{equation}

We are finally ready to obtain the Wilson loop expectation value using the procedure of section \ref{ssc}. We will have to slightly modify it, since in our case the highest eigenvalue of $\hat E_{\parallel}$ is only degenerate in absolute value; for large loops in the thermodynamic limit we thus modify Eq. (\ref{formula}) to
\begin{equation}
	\begin{aligned}
	\left\langle W\left(R_1,R_2\right)\right\rangle
	&=
	\frac{\rho'^{R_2-1}_1\left(R_1\right)}{\rho_1^{R_2+1}}
	\underset{x,x'}{\sum} \left(xx'\right)^{R_2-1}
\left\langle w_1\right| \hat  E_b \left(R_1\right) \left|v'^{x,x'}_1\left(R_1\right)\right\rangle
\left\langle w'^{x,x'}_1 \left(R_1\right) \right| \hat E_t \left(R_1\right)\left| v_1 \right\rangle 
\\&=
	\frac{1}{4}
	\left|\frac{\alpha\delta}{\beta^2}\right|^2 \left|\frac{\delta}{\alpha}\right|^{2R_1R_2}  \left|\frac{\beta^2}{\alpha\delta}\right|^{2\left(R_1+R_2\right)}
\underset{x,x'}{\sum} \left(xx'\right)^{R_2-1}
f^2_p\left(x,x'\right)
\end{aligned}
\end{equation}
where $p=\text{even,odd}$ is the parity of $R_1$.

One can already clearly see the area and perimeter dependent parts. The only thing left to do is to complete the computation of the sum, 
 where four different cases have to be considered, corresponding to the parities of $R_1,R_2$. It is straightforward to see that if the area is even (three of the four cases), the resulting number is $8$, while if the area is odd, the result is
 $8\text{Re}\left(\frac{\alpha\overline\beta}{\left|\alpha\overline\beta\right|}\right)^4$. Altogether we obtain, for large loops in the thermodynamic limit, for the $\left|\beta\right| \ll \left|\delta\right| < \left|\alpha\right|$ and $\gamma=0$ case, that
 \begin{equation}
 	\left\langle W\left(R_1,R_2\right) \right\rangle =\left\{
 	\begin{array}{ll}
 		2
 		\left|\frac{\alpha\delta}{\beta^2}\right|^2 \left|\frac{\delta}{\alpha}\right|^{2R_1R_2}  \left|\frac{\beta^2}{\alpha\delta}\right|^{2\left(R_1+R_2\right)}, &R_1R_2\text{ is even}\\
 		2\left|\frac{\alpha\delta}{\beta^2}\right|^2 \left|\frac{\delta}{\alpha}\right|^{2R_1R_2}  \left|\frac{\beta^2}{\alpha\delta}\right|^{2\left(R_1+R_2\right)}\text{Re}\left(\frac{\alpha\overline\beta}{\left|\alpha\overline\beta\right|}\right)^4, &R_1R_2\text{ is odd}
 	\end{array}\right. 
 \end{equation}
\end{widetext}
The Creutz parameter (\ref{Creutz}) is nothing but the string tension,
\begin{equation}
	\chi = \kappa = -2\log\left|\frac{\delta}{\alpha}\right|
	\end{equation}

We see that we have an area law, or a confining phase, as long as $\left|\delta\right| \neq \left|\alpha\right|$. While we excluded an equality in our arguments above, indeed we will have no area law if these two parameters are equal: then, the eigenvectors of $\hat{E}$ between which the fluxes transfer will have the same eigenvalue which does not allow for an area law, in full accordance with our general discussion. If we switch $\gamma$ on, it will have two effects: one will contaminate the eigenvectors of the transfer matrix $\hat{E}$, taking them farther from product vectors until the area law is broken, as well introduce terms in the flux-carrying transfer matrices that do not change the eigenvalue sector of $\hat {\tau}_0$ - violating another area law criterion.

\subsection{Numerical examples}

We will now present a few more examples which are computed numerically, using exact contraction, on a torus with size $\mathcal{N}_1 = 8 \times \mathcal{N}_2 = 100$. We considered different choices of parameters to demonstrate different behaviours; for each, we computed expectation value of the Wilson loop for several large loops.
We extracted the parameters $\kappa_A$ and $\kappa_P$ as follows: using the expression (\ref{Wlarge}) for a Wilson loop, we may define a function of $R_2$ depending on $R_1$ as a parameter,
\begin{equation}
f\left(R_2\right) = -\log\left\langle W\left(R_1,R_2\right) \right\rangle = f_1\left(R_1\right) R_2 + f_0\left(R_1\right)
\end{equation}
It is a linear function, which intersects with the vertical axis at
\begin{equation}
	f_0\left(R_1\right) = \kappa_P R_1  - \log W_0
\end{equation}
whose slope is
\begin{equation}
f_1\left(R_1\right) = \kappa_A R_1 + \kappa_P
\end{equation}
In the case of a perimeter law, the slope function will be constant, $f_1\left(R_1\right) = \kappa_P$ and when plotting $f\left(R_2\right)$ for different $R_1$ values, parallel lines will be obtained. In the case of an area law, the lines will have different slopes. Thus, $\kappa_A$ and $\kappa_P$ may be extracted by performing linear fits to the functions $f_{1,2}\left(R_1\right)$. Moreover, we have extracted the Creutz parameter too.

\begin{figure}
	\includegraphics[width=0.4\textwidth]{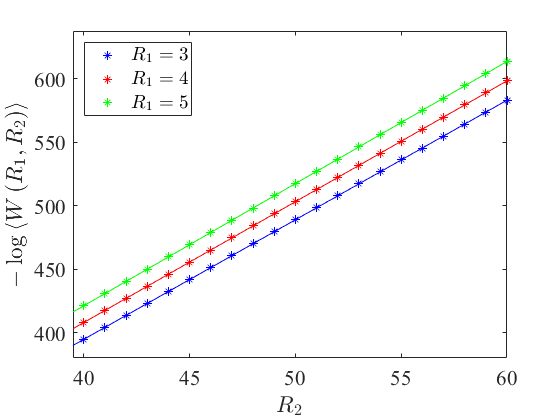}
	\includegraphics[width=0.4\textwidth]{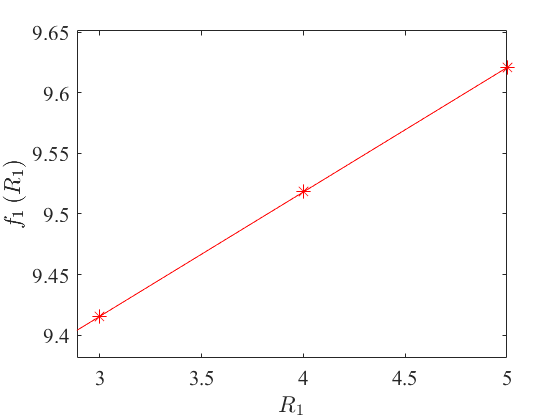}
	\includegraphics[width=0.4\textwidth]{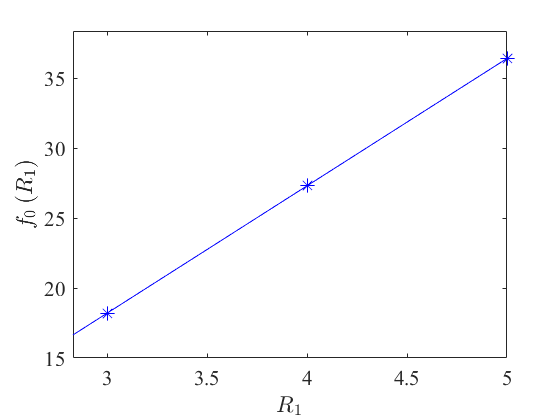}
	\caption{The $\alpha=1,\beta=0.1,\gamma=0,\delta=0.95$, which lies within the perturbative class discussed above, clearly shows an area law. It can be seen qualitatively on the top, where $-\log\left\langle W\left(R_1,R_2\right)\right\rangle$ is plotted as a function of $R_2$ for three different values of $R_1$ - resulting in three non-parallel lines. And if it is hard to detect the different slopes on the top, the middle figure shows it more quantitatively: the slope function $f_1\left(R_1\right) \approx 0.126 R_1 + 9.1078$ has a nonzero slope $\kappa_A \approx 9.1078$, and its intersection with the vertical axis is $\kappa_P \approx 9.1078$, the slope of the function plotted on the bottom, $f_0\left(R_1\right)$.  }
	\label{num10}
\end{figure}

\begin{figure}
	\includegraphics[width=\columnwidth]{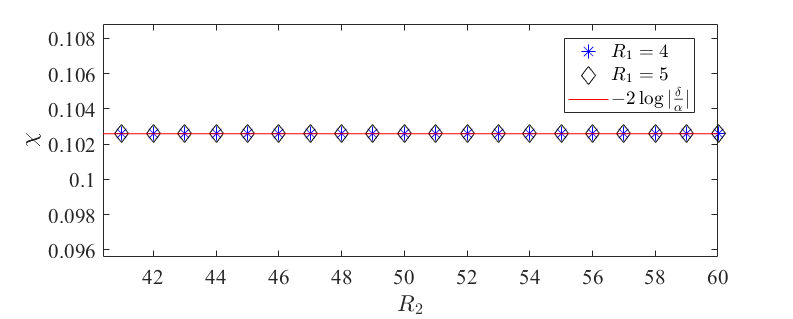}
	\caption{Computation of the Creutz Parameter $\chi\left(R_1,R_2\right)$ for $\alpha=1,\beta=0.1,\gamma=0,\delta=0.95$, for different values of $R_1$ and $R_2$. As can be seen, the values converge to the predicted value (thanks to the validity of the perturbative treatment in this parameter regime) of $-2\log\left|\frac{\delta}{\alpha}\right|\approx 0.1025$.}
	\label{num11}
\end{figure}

\begin{figure}
	\includegraphics[width=0.8\columnwidth]{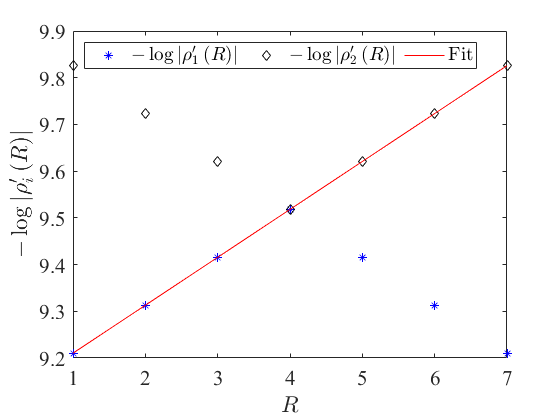}
	\caption{In the perturbative case worked out analytically, $\alpha=1,\beta=0.1,\gamma=0,\delta=0.95$, the highest eigenvalue of the intermediate transfer matrix $\hat{E}_{\parallel}\left(R\right)$ depend exponentially on the width $R$, as can be seen from the logarithmic plot given above, where the two highest eigenvalues (in absolute value, both degenerate in this case) are plotted for all values of $R$. The symmetric shape is due to the finiteness of the system ($\mathcal{N}=8$ in this case). For $R \leq 4$, the eigenvectors corresponding to highest eigenvalue connects with the right input state, while for $R \geq 4$ the next ones are relevant - all due to the symmetry. Also shown is a linear fit, computed with respect to the parameters predicted using the perturbative treatment.}
	\label{num12}
\end{figure}

The first set of parameters we examine is $\alpha=1,\beta=0.1,\gamma=0,\delta=0.95$. This choice is within the perturbative class studied above. It shows an area law, as can be seen from Fig. \ref{num10} and the Creutz parameter  $\chi = \kappa = -2\log\left|\frac{\delta}{\alpha}\right| \approx 0.1025$ (as shown in Fig. \ref{num11}). The expected exponential dependence of the eigenvalues of $\hat{E}_{\parallel}\left(R\right)$ is demonstrated in Fig. \ref{num12}.

\begin{figure*}
	\includegraphics[width=0.3\textwidth]{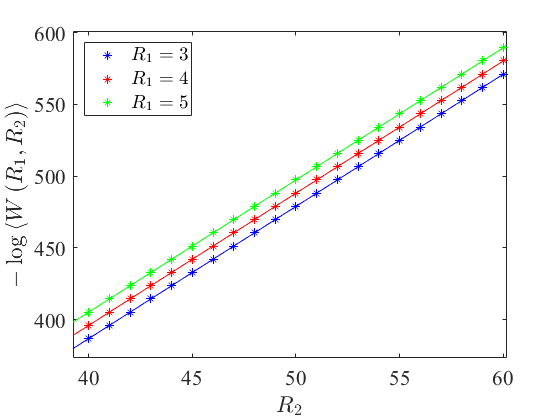}
	\includegraphics[width=0.3\textwidth]{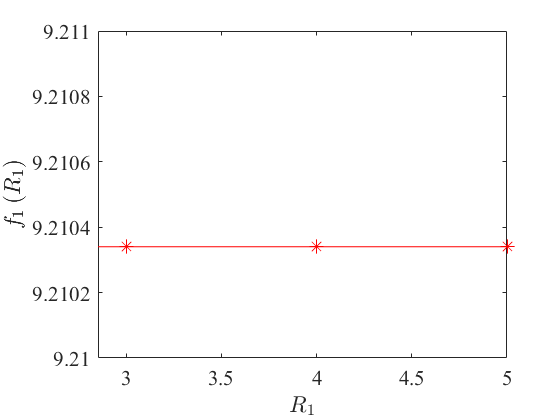}
	\includegraphics[width=0.3\textwidth]{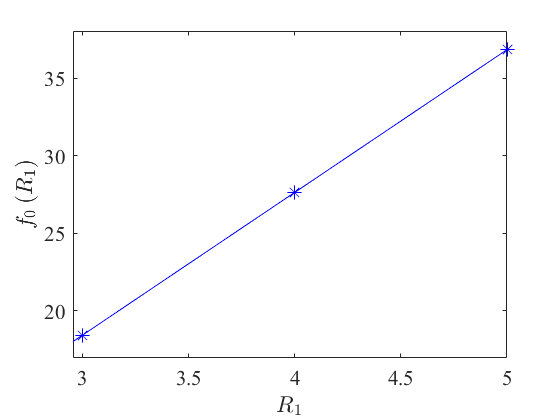}
	\caption{The $\alpha=1,\beta=0.1,\gamma=0,\delta=1$ case does not allow for an area law because of the degeneracy in the eigenvalues of $\hat{\tau}_0$ corresponding to projection operators. The state shows a perimeter law, which can be seen qualitatively on the left, where $-\log\left\langle W\left(R_1,R_2\right)\right\rangle$ is plotted as a function of $R_2$ for three different values of $R_1$ - resulting in three parallel lines. Quantitatively we see in the middle, where the three slopes of the three lines are plotted, that they are equal: $f_1\left(R_1\right) = \kappa_P \approx 9.2103$ is a constant function ($\kappa_A =0$). On the right we see the fit of $f_0\left(R_1\right) \approx 9.2103 \left(R_1-1\right)$. }
	\label{num21}
\end{figure*}

\begin{figure}
	\includegraphics[width=0.8\columnwidth]{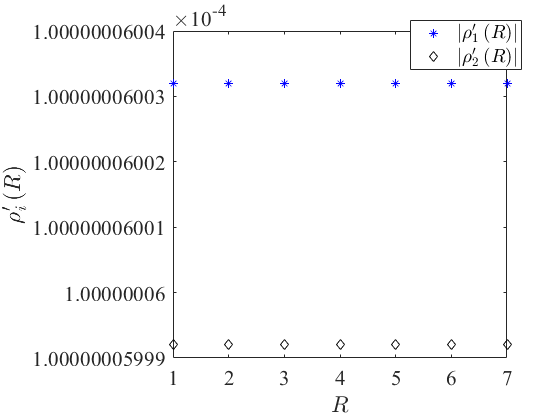}
	\caption{The $\alpha=1,\beta=0.1,\gamma=0,\delta=1$ case does not allow for an area law because of the degeneracy in the eigenvalues of $\hat{\tau}_0$ corresponding to projection operators. This is also manifested by the fact that the eigenvalues of the intermediate transfer matrix, $\hat{E}_{\parallel}\left(R\right)$, are completely independent of the distance $R$, as illustrated here.}
	\label{num22}
\end{figure}

Next, let us consider another example which lies within the perturbative regime: $\alpha=1,\beta=0.1,\gamma=0,\delta=1$. Here still $\gamma=0$ and $\beta$ is very small, so the eigenvectors of $\hat{E}$ would be product vectors, hence satisfying the first criterion for an area law. However, the eigenvalues of $\hat{\tau}_0$ are degenerate, implying no area law (the second criterion is violated). The perimeter law is clearly shown in Fig. \ref{num21}, and, as as one can see in Fig. \ref{num22}, the eigenvalues of $\hat{E}_{\parallel}\left(R\right)$ have no dependence on $R$.

Finally, we consider a completely different case, where $\alpha=0.1,\beta=0.1,\gamma=1,\delta=0.3$. For this choice of parameters, the previous perturbative treatment is not valid.
The eigenvalues associated with $\hat{\tau}_0$ are 
\begin{equation}
\lambda_1 \approx 1.05, \quad \lambda_2 \approx -0.95, \quad \lambda_3 = 0.02, \quad \lambda_4=0
\end{equation} associated with the operators
\begin{equation}
	\begin{aligned}
	&\hat{M}_{1} \approx \left( 
	\begin{matrix}
		&0.6928 &0 \\
		& 0 & 0.7211
	\end{matrix}
\right), 
	\quad\hat{M}_{2} \approx \left( 
	\begin{matrix}
		&-0.7211 &0 \\
		& 0 & 0.6928
	\end{matrix} \right), \\&
\hat{M}_{3} =\frac{1}{\sqrt{2}}\sigma_x, \quad\quad
	\hat{M}_{4} = -\frac{i}{\sqrt{2}}\sigma_y
	\end{aligned}
\end{equation}
- here, too, the most significant contributions are from the zeroth block with diagonal operators (the first two); however, they are far away from being projectors, hence we do not expect the eigenvectors of $\hat{E}$ to be anywhere close to product vectors.

\begin{figure}
	\includegraphics[width=0.4\textwidth]{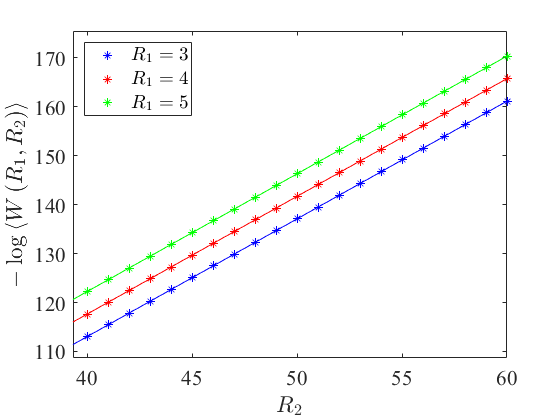}
	\includegraphics[width=0.4\textwidth]{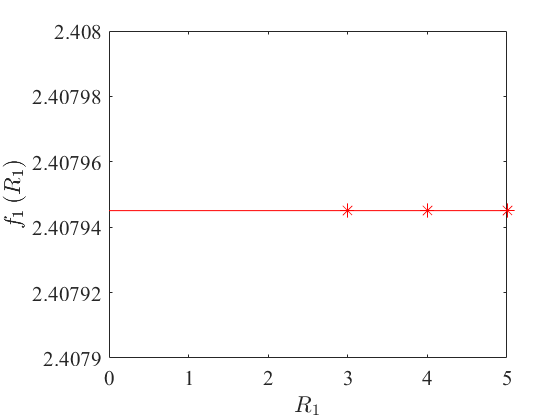}
	\caption{The $\alpha=0.1,\beta=0.1,\gamma=1,\delta=0.3$ case shows a perimeter law, which can be seen qualitatively on the top, where $-\log\left\langle W\left(R_1,R_2\right)\right\rangle$ is plotted as a function of $R_2$ for three different values of $R_1$ - resulting in three parallel lines. Quantitatively we see on the bottom, where the three slopes of the three lines are plotted, that they are equal. }
	\label{num31}
\end{figure}

\begin{figure}
	\includegraphics[width=0.8\columnwidth]{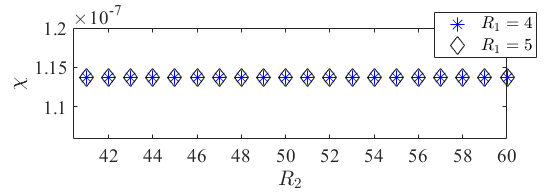}
	\caption{Another probe for the perimeter law of the $\alpha=0.1,\beta=0.1,\gamma=1,\delta=0.3$ case is the zero Creutz parameter, as plotted here (the plotted results are not exactly zero due to the fact our loops are not very large).}
	\label{num32}
\end{figure}

\begin{figure}
	\includegraphics[width=0.8\columnwidth]{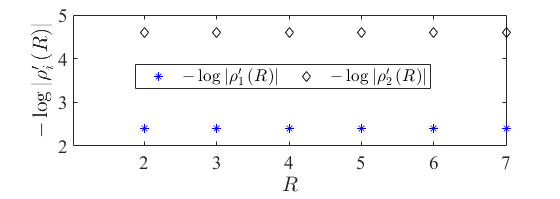}
	\caption{In the perimeter law case of $\alpha=0.1,\beta=0.1,\gamma=1,\delta=0.3$ case, as expected, the eigenvalues of the intermediate transfer matrix $\hat{E}_{\parallel}\left(R\right)$ are independent of $R$.}
	\label{num33}
\end{figure}

Let as also consider the straight flux carrying transfer operator $\hat{\tau}_{-}$ and to $\hat{\tau}_{|}$. We find the singular values 
\begin{equation}
	\eta_1 \approx 0.4472, \quad \eta_2 =0.02, \quad \eta_3 = \eta_4=0,
\end{equation}
associated, in the flux direction, with the operators
\begin{equation}
	\begin{aligned}
		&\hat{K}_{1} = -\frac{1}{\sqrt{2}} \mathbf{1}
		, \quad
		\hat{K}_{2} = - \frac{1}{\sqrt{2}}\sigma_x
		, \\&
		\hat{K}_{3} =-\frac{1}{\sqrt{2}}\sigma_z, \quad
		\hat{K}_{4} = -\frac{i}{\sqrt{2}}\sigma_y
		\end{aligned}
\end{equation}
and, in the direction orthogonal to the flux, with the operators
\begin{equation}
	\begin{aligned}
		&\hat{L}_{1} \approx \left( 
		\begin{matrix}
			&-0.3162 &0 \\
			& 0 & -0.9487
		\end{matrix}
		\right), \quad\quad
		\hat{L}_{2} =  - \frac{1}{\sqrt{2}}\sigma_x, \\&
		\hat{L}_{3}  \approx \left( 
		\begin{matrix}
			&-0.9487 &0 \\
			& 0 & 0.3162
		\end{matrix}
		\right), \quad\quad
		\hat{L}_{4} = -\frac{i}{\sqrt{2}}\sigma_y
			\end{aligned}
\end{equation}
which imply that even if our eigenvectors were product vectors (which they are not), the most prominent contribution, coming from $\eta_1$, would be diagonal in the subsector (as seen from $\hat{L}_1$). Therefore all our area law criteria are violated. Indeed, this set of parameters show a perimeter law decay of the Wilson loop, as can be seen in Fig. \ref{num31}, in the zero Creutz parameter (see Fig. \ref{num32}) and in the eigenvalues of $\hat{E}_{\parallel}\left(R\right)$ which are independent of $R$ (as shown in Fig. \ref{num33}).

\section{Summary}

In this work we have seen how local properties of two dimensional lattice gauge theory PEPS, manifested in their transfer operators (on-site) and matrices (rows) simplify their contraction and dictate their long-range, Wilson loop behaviour. We have related the area law  with transfer matrices whose eigenvectors are product vectors - that is, a product of local contributions of the transfer operators on each side, manifesting the lack of long-range order, as expected for a disordered, confining phase. The perimeter law, appearing in ordered phases, has to do with non-product eigenvectors, where the separate sites contribute in a correlated, long-ordered manner. 
These results may be used for detecting phases of PEPS used for pure gauge theory studies, and for the design of PEPS used as ansatz states for such scenarios. 

One possible extension is the inclusion of dynamical matter - which is different from the current work both in the mathematical sense (different structure of the tensors, implying different symmetry properties) and the physical one (in that case, at least with fermionic matter as in conventional standard model scenarios, the Wilson loop does not serve as an order parameter for confinement any more). This could possibly connected with the formalism of gauged Gaussian fermionic PEPS \cite{zohar_fermionic_2015,zohar_projected_2016} which can be contracted using sign-problem free Monte-Carlo techniques \cite{zohar_combining_2018,emonts_variational_2020} both for the study of further examples and application to physical models of interest. 

Another important and relevant generalization is the extension to higher dimensions, where further geometry arguments have to be taken into account, potentially containing many further interesting physical and mathematical properties.

\section*{Acknowledgements}
I would like to thank J.I. Cirac, P. Emonts, A. Molnar and T. B. Wahl for fruitful discussions and N. C. Hallakoun for her technical support. This research was supported by the Israel Science Foundation (grant No. 523/20).

\bibliography{ref}
\end{document}